\documentclass[twocolumn]{aastex631}  % for arXiv
\usepackage{natbib}
\usepackage{booktabs}
\usepackage{amsmath}
\usepackage[figuresright]{rotating}
\usepackage{url}
\usepackage{booktabs}
\usepackage{multirow}
\usepackage{tablefootnote}

\usepackage{hyperref}
\usepackage{soul}

\begin{document}

\title{Two-Component $\gamma$-ray Emission Spectrum and X-Ray Polarization of the Radio Galaxy Pictor A}

\correspondingauthor{Hai-Ming Zhang \& Jin Zhang}
\email{hmzhang@gxu.edu.cn; j.zhang@bit.edu.cn}

\author[0009-0003-9471-4724]{Jia-Xuan Li}
\affiliation{School of Physics, Beijing Institute of Technology, Beijing 100081, People's Republic of China; j.zhang@bit.edu.cn}

\author{Xin-Ke Hu}
\affiliation{School of Physics, Beijing Institute of Technology, Beijing 100081, People's Republic of China; j.zhang@bit.edu.cn}

\author[0000-0003-2547-1469]{Ji-Shun Lian}
\affiliation{School of Physics, Beijing Institute of Technology, Beijing 100081, People's Republic of China; j.zhang@bit.edu.cn}

\author[0009-0000-6577-1488]{Yu-Wei Yu}
\affiliation{School of Physics, Beijing Institute of Technology, Beijing 100081, People's Republic of China; j.zhang@bit.edu.cn}

\author{Wei Deng}
\affiliation{Guangxi Key Laboratory for Relativistic Astrophysics, School of Physical Science and Technology, Guangxi University, Nanning 530004, People's Republic of China; hmzhang@gxu.edu.cn}

\author{Kuan Liu}
\affiliation{Guangxi Key Laboratory for Relativistic Astrophysics, School of Physical Science and Technology, Guangxi University, Nanning 530004, People's Republic of China; hmzhang@gxu.edu.cn}

\author[0000-0001-6863-5369]{Hai-Ming Zhang\dag}
\affiliation{Guangxi Key Laboratory for Relativistic Astrophysics, School of Physical Science and Technology, Guangxi University, Nanning 530004, People's Republic of China; hmzhang@gxu.edu.cn}

\author{Liang Chen}
\affiliation{Key Laboratory for Research in Galaxies and Cosmology, Shanghai Astronomical Observatory, Chinese Academy of Sciences, 80 Nandan Road, Shanghai 200030, People's Republic of China}

\author[0000-0003-3554-2996]{Jin Zhang\dag}
\affiliation{School of Physics, Beijing Institute of Technology, Beijing 100081, People's Republic of China; j.zhang@bit.edu.cn}

\begin{abstract}

Pictor A is a $\gamma$-ray emitting radio galaxy and has a bright hotspot called WHS, located $\sim$4 arcmin away from the nucleus. In this work, we present an analysis of its 16-year Fermi-LAT data and report the Imaging X-ray Polarimetry Explorer (IXPE) observations for this source. Our analysis of the Fermi-LAT observations reveals evidence of two components in the average $\gamma$-ray spectrum of Pictor A, exhibiting a statistically significant hardening from $\Gamma_{\rm \gamma,1}=3.25\pm0.15$ to $\Gamma_{\rm \gamma,2}=1.81\pm0.07$ at a break energy of $2.46\pm0.09$ GeV. Notably, variability of $\gamma$-rays is evident in Pictor A, predominantly driven by the component below the break energy, while the component above the break energy remains stable. Furthermore, our analysis reveals that a power-law function provides an adequate fit for the high-flux-state spectrum, while a broken power-law function remains necessary to accurately model the low-flux-state spectrum. We suggest that the low-energy component originates from the nucleus, while the high-energy component primarily stems from WHS. The broadband spectral energy distributions of both nucleus and WHS can be well represented by a simple leptonic model, with both $\gamma$-ray components attributed to the synchrotron-self-Compton (SSC) process. Analysis of IXPE data provides upper limits on the polarization degree of $\Pi_{\rm X}<$6.6\% for the nucleus and $\Pi_{\rm X}<$56.4\% for the WHS within the 2--8 keV band. For the nucleus, this result aligns with X-ray emission originating from the SSC process. However, the upper limit of $\Pi_{\rm X}<$56.4\% for WHS is insufficient to conclusively determine the X-ray emission mechanism in this region.

\end{abstract}


\section{Introduction}

Radio galaxies (RGs), a subset of radio-loud Active Galactic Nuclei (RL-AGNs), are distinguished by the presence of jet substructures that extend from the radio core to scales of kiloparsecs to megaparsecs (kpc--Mpc). Hotspots, characterized by high surface brightness at the edges of extended radio lobes in RGs, are considered to represent the jet termination, where magnetic fields are amplified and particles are accelerated to high energies \citep{1989A&A...219...63M, 2022ApJ...941..204T}. Thanks to Chandra observations, numerous hotspots and knots within RG jets have been detected in the X-ray band (\citealt{2006ARA&A..44..463H} for a review), however, ongoing debate persists regarding their X-ray radiation mechanisms (e.g., \citealt{2005ApJ...622..797K, 2010ApJ...710.1017Z, 2018ApJ...858...27Z}). {For substructures characterized by a hard X-ray spectrum, the X-ray emission is likely produced by inverse Compton scattering (IC) from relativistic electrons that are responsible for the radio-optical emission via the synchrotron process (e.g.,\citealt{2000ApJ...544L..23T,2005ApJ...630..721T,Georganopoulos_2003,Stawarz_2007,2010ApJ...710.1017Z, 2018ApJ...858...27Z}), or alternatively by synchrotron radiation from a second population of high-energy particles (e.g., \citealt{2020ApJ...903..109T, 2020ApJ...893...41W, 2022PASJ...74..602S, 2023MNRAS.525.5298H}).}

In addition, both the IC and second synchrotron radiation models of X-rays predict the observable $\gamma$-ray emission for some large-scale jet substructures \citep{2010ApJ...710.1017Z, 2018ApJ...858...27Z, 2020ApJ...893...41W, 2023MNRAS.525.5298H}. The detection of $\gamma$-rays from large-scale radio lobes of RGs Cen A \citep{2010ApJ...719.1433A, 2016A&A...595A..29S, 2020Natur.582..356H}, Fornax A \citep{2016ApJ...826....1A}, and NGC 6251 \citep{2024ApJ...965..163Y} confirms that the large-scale jet substructures are acceleration sites of high-energy particles. To date, the Fermi Large Area Telescope (Fermi-LAT) has detected numerous RGs \citep{2022ApJS..260...53A, 2023arXiv230712546B}; however, the limited spatial resolution of Fermi-LAT makes it difficult to judge the location of the $\gamma$-ray emission for the majority of RGs. It has been proposed that some large-scale jet substructures may have the detectable $\gamma$-ray emission through the search for a steady $\gamma$-ray emission component and modeling of the broadband spectral energy distributions (SEDs; \citealt{2018RAA....18...70G, 2018ApJ...858...27Z}). 

Polarization measurements serve as a valuable tool for evaluating the radiation mechanisms of X-rays. The launch of the Imaging X-ray Polarimetry Explorer \citep[IXPE;][]{2022HEAD...1930101W} has enabled this assessment to become feasible. Due to the high brightness of RL-AGNs, they constitute one of the primary observation targets for IXPE, particularly blazars \citep[e.g.,][]{2022Natur.611..677L,2022ApJ...938L...7D,2023ApJ...942L..10M,2023ApJ...953L..28M,2023ApJ...959...61E,2024ApJ...963....5E,2024A&A...689A.119K,2024ApJ...972...74M}. The first RG observed by IXPE was Cen A \citep{2022ApJ...935..116E}; however, no statistically significant degree of polarization was detected. Only an upper limit of $6.5\%$ was established for the core region, and insufficient counts were obtained to measure statistically significant X-ray polarization from the X-ray jet. Recently, IXPE has observed the second RG, Pictor A.

Pictor A is a $\gamma$-ray emitting RG, located at a redshift of $z=0.035$ \citep{1965ApJ...141....1S}. It has a bright primary hotspot, known as the western hotspot (WHS), which is situated about 4 arcmin ($\sim$165 kpc) away from the radio core \citep{1997A&A...328...12P}. Its nucleus is a strong X-ray source and has been widely studied; a power-law (PL) spectrum is observed \citep{1998ApJ...505..577E,1999ApJ...526...60S}, maybe with a weak, narrow Fe K$\alpha$ line \citep{2016MNRAS.455.3526H}. Extensive research has also been conducted on this bright WHS \citep{1987ApJ...314...70R, 1995ApJ...446L..93T, 1997A&A...328...12P, 2008AJ....136.2473T, 1999ApJS..123..447S, 2001ApJ...547..740W, 2017ApJ...850..193I, 2022PASJ...74..602S, 2022ApJ...941..204T, 2023MNRAS.521.2704G, 2024arXiv240206218S}. The observed flat spectrum and flux variability in the X-ray band suggest that the X-rays from WHS are produced by synchrotron radiation \citep{2020ApJ...903..109T, 2022PASJ...74..602S, 2023MNRAS.521.2704G, 2024arXiv240206218S}. X-ray polarization observations can serve as a valuable tool to verify its radiation mechanisms. 

The $\gamma$-rays detected by Fermi-LAT for Pictor A are generally attributed to the contribution of the nucleus; adding the X-ray emission from the nucleus, they are produced through the synchrotron-self-Compton (SSC) process \citep{2012MNRAS.421.2303B,2017RAA....17...90X,2023MNRAS.521.2704G}. However, it is also possible that the $\gamma$-rays stem from the WHS \citep{2009ApJ...701..423Z}. In this paper, we perform a comprehensive analysis of the $\sim$16-year Fermi-LAT observation data to investigate the origins of $\gamma$-ray emission from Pictor A, and we examine the X-ray emission mechanisms of the nucleus and WHS using the IXPE observations for Pictor A in conjunction with the broadband SED modeling. Observations and data analysis are given in Section \ref{sec2}, followed by the presentation of results in Section \ref{sec3}. The SEDs of the nucleus and WHS are constructed and modeled in Section \ref{sec4}. A discussion and conclusions are presented in Section \ref{sec5}. $H_0=71$ km s$^{-1}$ Mpc$^{-1}$, $\Omega_{\rm m}=0.27$, and $\Omega_{\Lambda}=0.73$ are adopted in this paper.

\section{Observations and Data Analysis}\label{sec2}

\subsection{Fermi-LAT}\label{sec2.1}

Pictor A is associated with $\gamma$-ray source 4FGL J0519.6-4544 in the 4FGL-DR4 \citep{2022ApJS..260...53A, 2023arXiv230712546B}. We select the data within 15$^\circ$ region of interest (ROI) centered on the radio position of Pictor A (R.A.=79$^\circ$.957, Decl.=-45$^\circ$.779) and download the PASS 8 data covering from 2008 August 4 to 2024 July 10 (MJD 54682--60501) within the energy range of 0.1--500 GeV from the Fermi Science Support Center\textbf{\footnote{\url{https://fermi.gsfc.nasa.gov/cgi-bin/ssc/LAT/LATDataQuery.cgi}}}. 

The publicly available software \texttt{fermitools} (v.2.2.0)  and \texttt{Fermipy} \citep[v.1.1;][]{Wood2017} are used in our analysis. We use event class ``SOURCE'' (evclass=128) and event type ``FRONT+BACK'' (evtype=3) for the binned likelihood analysis based on LAT data selection recommendations\textbf{\footnote{\url{https://fermi.gsfc.nasa.gov/ssc/data/analysis/documentation/Cicerone/Cicerone_Data_Exploration/Data_preparation.html}}} and set the maximum zenith angle of 90$^\circ$ in order to eliminate the contamination of $\gamma$-ray emission from the earth limb. A standard filter expression ``(DATA\_QUAL\textgreater0)\&\&(LAT\_CONFIG==1)'' and the instrument response function of $\rm P8R3\_SOURCE\_V3$ are used in our analysis. All of the $\gamma$-ray source models included in the 4FGL-DR4 \citep{2020ApJS..247...33A} within the ROI and two background model including the isotropic emission ("$\rm iso\_P8R3\_SOURCE\_V3\_V1.txt$") and the diffuse galactic interstellar emission ("$\rm gll\_iem\_v07.fits$") are added to the model. Only the parameters of the sources within $6.5^\circ$ centered on Pictor A and the normalization of two background models are left free. 

\begin{figure}
    \includegraphics[angle=0, scale=0.47]{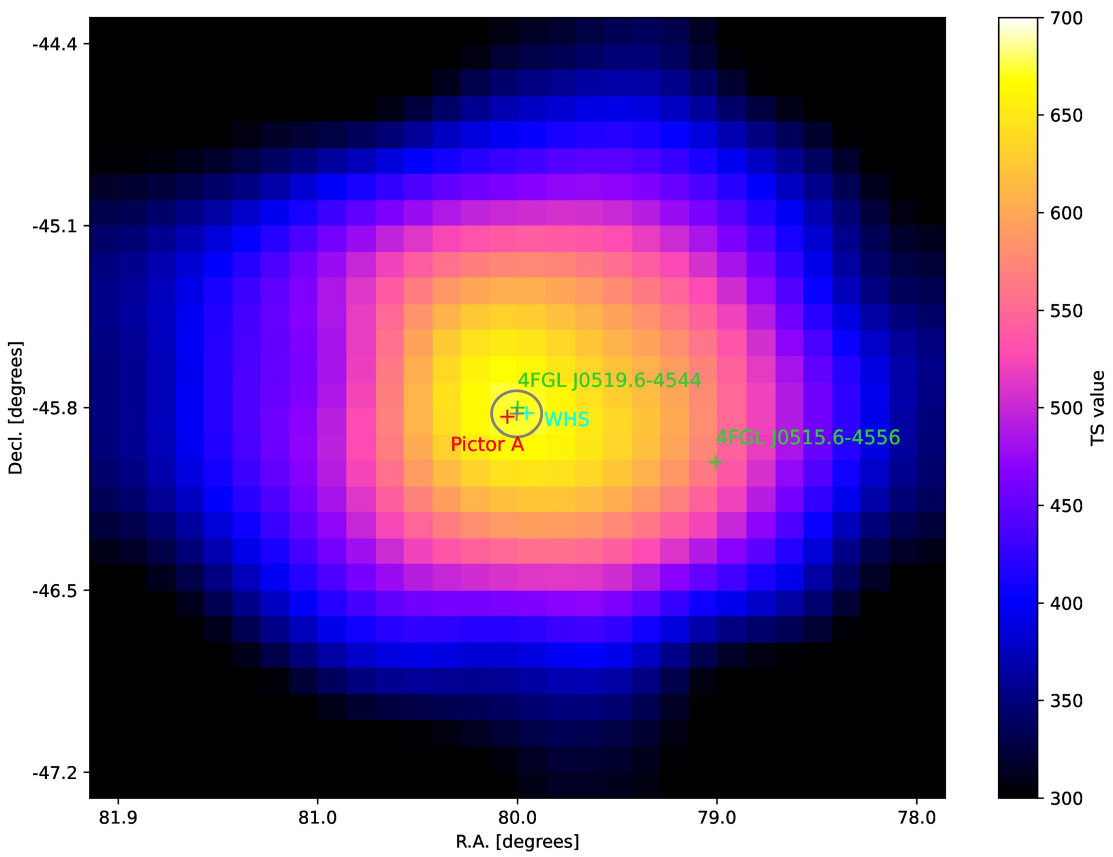}
    \includegraphics[angle=0, scale=0.47]{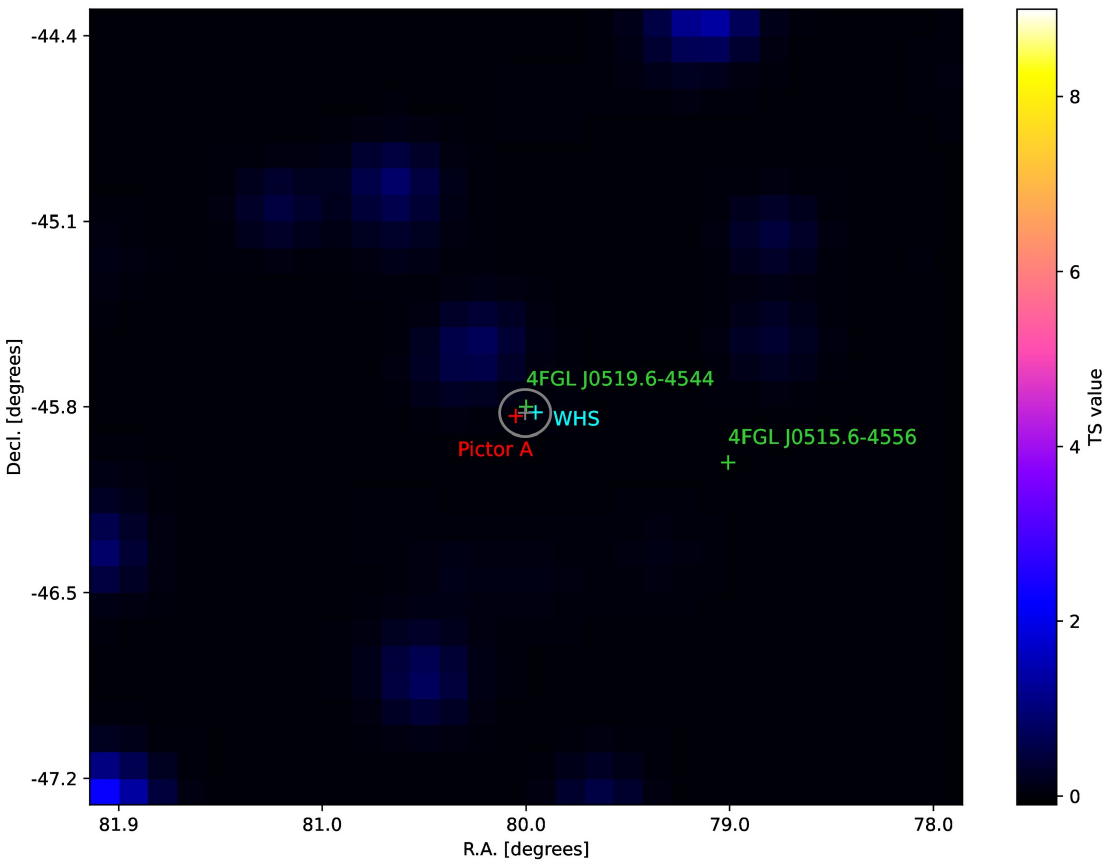}
    \caption{3$^\circ$ $\times$ 3$^\circ$ TS map (left panel) and residual TS map (right panel)
    of Pictor A in the 0.1--500 GeV band. The green crosses represent the positions of 4FGL J0519.6-4544 and 4FGL J0515.6-4556 in the 4FGL-DR4. The red and cyan crosses represent the radio position of Pictor A and the Chandra pointing position \citep{2001ApJ...547..740W} of WHS, respectively. The grey cross and grey circle represent the best-fit position of $\gamma$-ray source in this work along with its corresponding 95\% containment region.}
    \label{TS-map}
\end{figure}

We use the test statistic (TS) to evaluate the significance of a $\gamma$-ray source signal, $\rm TS=2(log\mathcal{L}_{\rm src}-log\mathcal{L}_{\rm null})$, where $\mathcal{L}_{\rm src}$ and $\mathcal{L}_{\rm null}$ are the likelihood values of the background with and without the target source, respectively. The TS map of the Pictor A region is shown in the left panel of Figure \ref{TS-map}, where the grey cross represents the best-fit position of the $\gamma$-ray source in this work, i.e., $\rm R.A.=79.91\degr\pm0.02\degr$ and $\rm Decl.=-45.77\degr\pm0.02\degr$. This best-fit position is consistent with that of 4FGL J0519.6--4544 and spatially associated with Pictor A. As shown in the right panel of Figure \ref{TS-map}, the maximum TS value of the residual TS map is $\sim$4, indicating that the $\gamma$-ray emission from this source is well fitted by the model and no new $\gamma$-ray source is found. It should be noted that a bright flat-spectrum radio quasar (FSRQ), PKS J0515--4556, associated with 4FGL J0515.6--4556, is located $0.75\degr$ away from Pictor A. We have performed an extensive series of tests and conclusively ruled out any potential signal contamination from PKS J0515--4556. Further details are provided {in Section \ref{sec3.2}.}

The spectrum of Pictor~A exhibits a broken power-law (BPL) shape, as shown in Figure \ref{spec_LAT}. However, Pictor A is identified as a point-like source with a PL spectrum in 4FGL-DR4 \citep{2022ApJS..260...53A, 2023arXiv230712546B}. In order to evaluate which function provides a more accurate representation of the spectrum, We calculate the value of $\Delta \rm TS$. $\Delta \rm TS$ is identified as
\begin{equation}
\label{equ_deltaTS}
\Delta \rm TS=2(log\mathcal{L}_{\rm BPL}-log\mathcal{L}_{\rm PL}),
\end{equation}
where $\mathcal{L}_{\rm BPL}$ and $\mathcal{L}_{\rm PL}$ are the obtained likelihood values utilizing BPL and PL functions, respectively. The PL function is defined as 
\begin{equation}
	\label{equ_PL}
	\frac{dN(E)}{dE}=N_0\times\left(\frac{E}{E_0}\right)^{-\Gamma_\gamma},
\end{equation}
and the BPL function is defined as
\begin{equation}
	\label{equ_BPL}
	\frac{dN(E)}{dE}=N_0\times\left\{
	\begin{array}{rcl}
		\left(\frac{E}{E_{\rm b}}\right)^{-\Gamma_{\rm \gamma,1}}& & {E<E_{\rm b}}\\
		\left(\frac{E}{E_{\rm b}}\right)^{-\Gamma_{\rm \gamma,2}}& & {E\geqslant E_{\rm b}}\\
	\end{array},\right.
\end{equation}
where $E_{\rm b}$ represents the break energy, $\Gamma_{\rm \gamma,1}$ and $\Gamma_{\rm \gamma,2}$ are the indices below and above the break energy, respectively.

The $\sim$16-year light curves in different energy bands are generated by freeing the normalizations of Pictor A and 4FGL J0515.6--4556, while keeping all other spectral parameters fixed at their best-fit values obtained from the $\sim$16-year data analysis. A likelihood-based statistic is the predominant approach for quantifying variability {(\citealt{2012ApJS..199...31N,2019ApJ...884...91P,2020ApJS..247...33A}). To assess the variability of Pictor A, we adhere to the definition provided in \citet{2012ApJS..199...31N}} and calculate the variability index (TS$_{\rm var}$) as
\begin{equation}
	\label{eqTSvar}	\mathrm{TS}_{\mathrm{var}}=2\sum_{i=0}^{N}\left[\log\left(\mathcal{L}_i\left(F_i\right)\right)-\log\left(\mathcal{L}_i\left(F_{\text{glob}}\right)\right)\right],
\end{equation}
where N is the number of time bins, $F_i$ is the fitting flux for bin $i$, $\mathcal{L}_{i}(F_i)$ is the likelihood corresponding to bin $i$, and $F_{\rm glob}$ is the best fit flux for the glob time by assuming a constant flux.

\subsection{IXPE}\label{sec2.2}

Pictor A was observed by IXPE in two different periods: first from 2024 June 15 to 2024 July 5 with a total exposure of $\sim$ 960 ks (Obs.1), and second from 2024 November 13 to 2024 December 18 with a total exposure of $\sim$ 934 ks (Obs.2). The WHS is the primary target of these IXPE observations. However, the nucleus of Pictor A was also captured within the field of view (FOV) of IXPE, albeit in close proximity to the edge of the FOV.

The nucleus and WHS are both identified as point-like sources under a $\sim30^{\prime\prime}$ angular resolution of IXPE. Prior to conducting the IXPE data analysis, several corrections need to be applied to the publicly available Level-2 event files. Firstly, coordinate corrections are implemented to address detector pointing misalignment. Additionally, instrumental background is filtered using the \texttt{filter$\_$background.py} script \footnote{\url{https://heasarc.gsfc.nasa.gov/docs/ixpe/analysis/contributed.html}} to enhance the significance of results for faint sources with extended features \citep{2023AJ....165..143D}. Given the long exposure time, good time intervals are filtered to mitigate particle events caused by solar activity \citep{2024ApJ...967L..38F,2024ApJ...962...92X}.

We calculate the X-ray polarization parameters from the IXPE observations using two different methods: (1) a model-independent method \citep{2015APh....68...45K}; and (2) a spectropolarimetric analysis, as described in \citet{2017ApJ...838...72S}. The source region of the nucleus is defined as a circle with a radius of $60^{\prime\prime}$ centered on its radio position. Similarly, the source region for the WHS is also defined as a circle on the Chandra pointing position in \cite{2001ApJ...547..740W}, but with a smaller radius of $30^{\prime\prime}$ due to its faintness compared to the nucleus. The background photons for both the nucleus and WHS are extracted from nearby circular regions with radii of $90^{\prime\prime}$ and $45^{\prime\prime}$, respectively, taking into account the effects of the FOV edges.

We first estimate the X-ray polarization of Pictor A for Obs.1 and Obs.2 separately; however, no significant differences are observed between these two observation periods. Therefore, to enhance photon statistics, we combine the data from Obs.1 and Obs.2 to estimate the polarization for both the nucleus and the WHS.

The polarization parameters of the nucleus and WHS are first calculated using the \texttt{PCUBE} algorithm within the \texttt{xpbin} task of the software \texttt{ixpeobssim} \citep[v.31.0.3;][]{2022SoftX..1901194B} in the 2--8 keV energy range with the unweighted analysis method \footnote{weights=False and irfname=`ixpe:obssim20240701:v013' being set within \texttt{xpbin}.}. The influence of the sky background is eliminated following the background subtraction procedure presented in \citet{2022SoftX..1901194B}.

The spectropolarimetric analysis is performed using \texttt{Xspec} \citep[v.12.14.1;][]{1999ascl.soft10005A}  in the HEASoft (v.6.34) package. Stokes parameter spectra ($I$, $Q$ and $U$) for both the source and background are generated using the \texttt{PHA1}, \texttt{PHA1Q}, and \texttt{PHA1U} algorithms within \texttt{xpbin}. The $I$, $Q$ and $U$ spectra are regrouped with a constant energy binning of 0.2 keV via the \texttt{ftgrouppha} task. Specifically, we utilize a weighted analysis method \footnote{weights=True and irfname=`ixpe:obssim20240701$\_$alpha075:v013' being set within \texttt{xpbin}.} on the $I$, $Q$ and $U$ spectra to enhance the significance of the measurements using the \texttt{alpha075} response matrices. The spectropolarimetric fits are performed in a two-step procedure \citep[e.g.,][]{2024ApJ...963....5E,2024ApJ...962...92X,2024ApJ...970L..22H}. Firstly, the $I$ spectra are fitted with an absorbed PL model of the form \texttt{CONSTANT} $\times$ \texttt{TBABS} $\times$ \texttt{POWERLAW}. The PL function is
\begin{equation}\label{PL} 
\frac{dN(E)}{dE}=N_{0}\times\left(\frac{E}{E_0}\right)^{-\Gamma_{\rm X}}\, , 
\end{equation}
where $E_{0}=1$ keV is the scale parameter of photon energy, $N_0$ is the PL normalization, and $\Gamma_{\rm X}$ is the photon spectral index \citep{2004A&A...413..489M}. The \texttt{CONSTANT} and \texttt{TBABS} models account for the uncertainties in the absolute effective area of the three detector units (DUs) and Galactic photoelectric absorption, respectively. The column densities are fixed to the Galactic values, i.e., $N_{\rm H}=3.62\times10^{20}$ cm$^{-2}$ for the nucleus and $N_{\rm H} =3.57\times10^{20}$ cm$^{-2}$ for the WHS \citep{2016A&A...594A.116H}. During the fits of the $I$ spectra, both $N_{0}$ and $\Gamma_{\rm X}$ are allowed to vary. Secondly, we simultaneously fit the $I$, $Q$ and $U$ spectra of three DUs using an absorbed PL model with a constant polarization of the form \texttt{CONSTANT} $\times$ \texttt{TBABS} $\times$ \texttt{POLCONST} $\times$ \texttt{POWERLAW}. The polarization model \texttt{POLCONST} assumes constant polarization parameters within the operating energy range and has only two free parameters: the X-ray polarization degree ($\Pi_{\rm X}$) and angle ($\psi_{\rm X}$). For these spectropolarimetric fits, the values of $N_{0}$ and $\Gamma_{\rm X}$ are fixed at their best-fit values obtained from the $I$ spectra fitting, leaving only $\Pi_{\rm X}$ and $\psi_{\rm X}$ as free parameters. 

The analysis using the \texttt{PCUBE} algorithm within software \texttt{ixpeobssim} can yield a minimum detectable polarization at the 99\% confidence level (MDP$_{99}$). If the estimated value of $\Pi_{\rm X}$ is lower than the corresponding MDP$_{99}$, we will determine the 99\% confidence level upper limit using the \texttt{error} task within the \texttt{Xspec} software package, following the recommendation provided in the IXPE statistics guide document\footnote{\url{https://heasarc.gsfc.nasa.gov/docs/ixpe/analysis/}}.

\section{Results}\label{sec3}

\subsection{Complex $\gamma$-ray Behaviors of Pictor A}\label{sec3.1}

\begin{figure}
    \includegraphics[angle=0, scale=0.45]{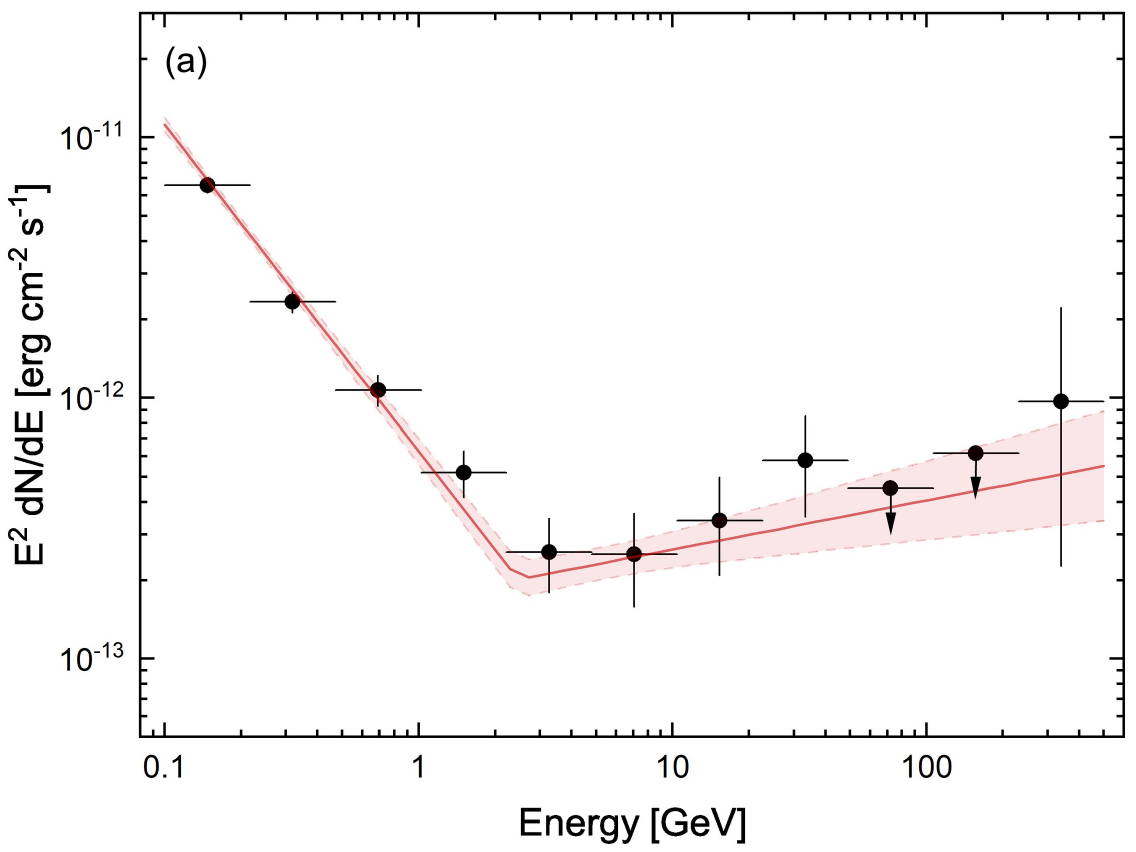}\hspace{0.3cm}
    \includegraphics[angle=0, scale=0.45]{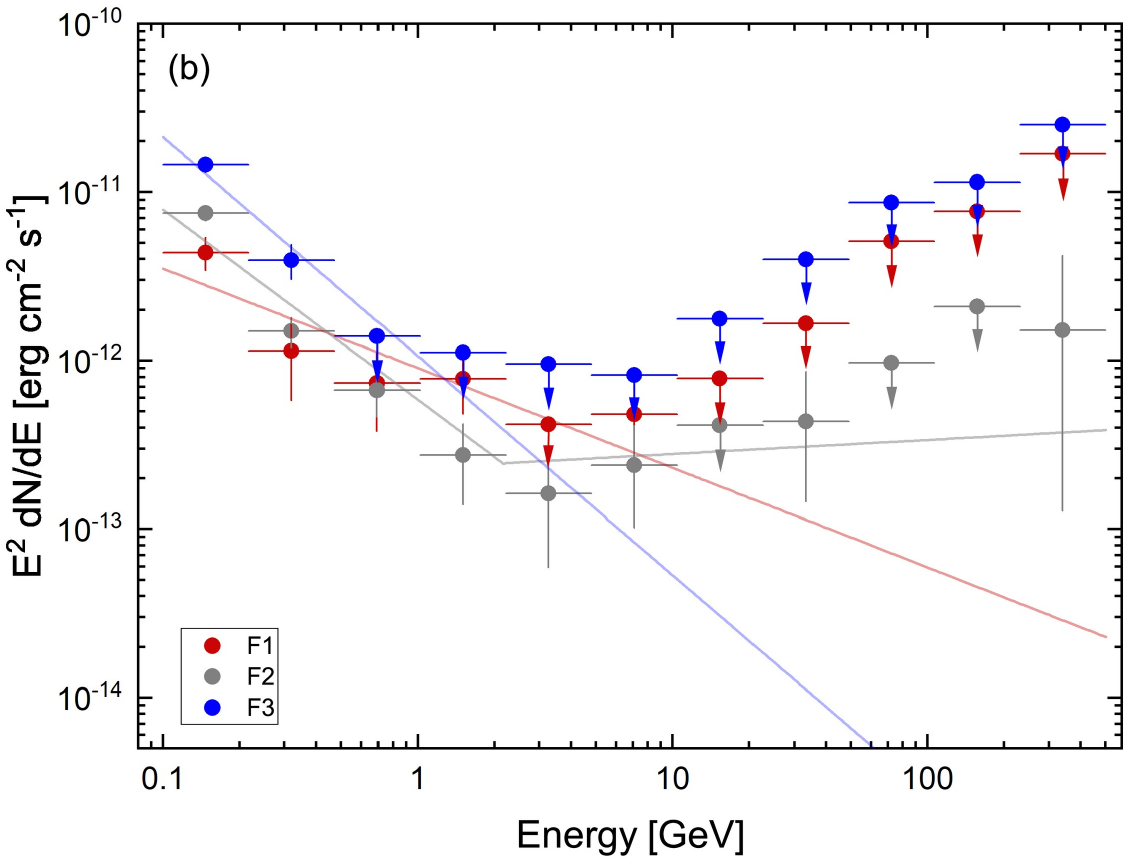}
    \caption{The Fermi-LAT spectra for Pictor A in the 0.1--500 GeV band. Panel (a): the $\sim$16-year average spectrum. Panel (b): the spectra of three high/low states, corresponding to the three regions shaded in different colors in Figure \ref{LC}(a). The $2\sigma$ upper limit (points with a downward arrow) is reported when $\rm TS\leq4$ for that energy bin. The color lines represent the corresponding best-fit results using either a BPL or a PL function. The pink shaded region in Panel (a) indicates the 1$\sigma$ uncertainty of the fitting result. }
    \label{spec_LAT}
\end{figure}

By analyzing the $\sim$16-year Fermi-LAT observation data for Pictor A, we obtain TS$\sim$773.10 and an average flux of  $(1.06\pm0.08)\times10^{-11}~\rm erg~cm^{-2}~s^{-1}$ in the 0.1--500~GeV energy band. As illustrated in Figure \ref{spec_LAT}, the average spectrum of Pictor A over the full time shows a noticeable break at $2.46\pm0.09$ GeV, with the spectral index hardening from $\Gamma_{\rm \gamma,1}=3.25\pm0.15$ to $\Gamma_{\rm \gamma,2}=1.81\pm0.07$. The BPL spectral model is preferred over the single PL spectral model for fitting this spectrum at a confidence level of 6.2$\sigma$ ($\Delta \rm TS=38.88$). Moreover, both energy bands (0.1--2.46 GeV and 2.46--500 GeV) can be well described by a PL function, yielding photon spectral indices of $3.12\pm0.06$ and $1.97\pm0.18$, respectively, which are consistent with those obtained from fitting the entire energy band using a BPL model. These results are summarized in Table \ref{table_LAT}. These findings suggest the presence of statistically significant hardening in the observed $\gamma$-ray spectrum of Pictor A, similar to what has been observed in the other two RGs: Cen A \citep{2013ApJ...770L...6S, 2017PhRvD..95f3018B} and M87 \citep{2019A&A...623A...2A}. Such an ``unusual” break could most naturally be explained by a superposition of different spectral components. In order to further investigate the properties of GeV emission from Pictor A, additional data analysis on Fermi-LAT observations is conducted and the main findings are summarized below.

\begin{figure}
    \centering \includegraphics[angle=0, scale=0.9]{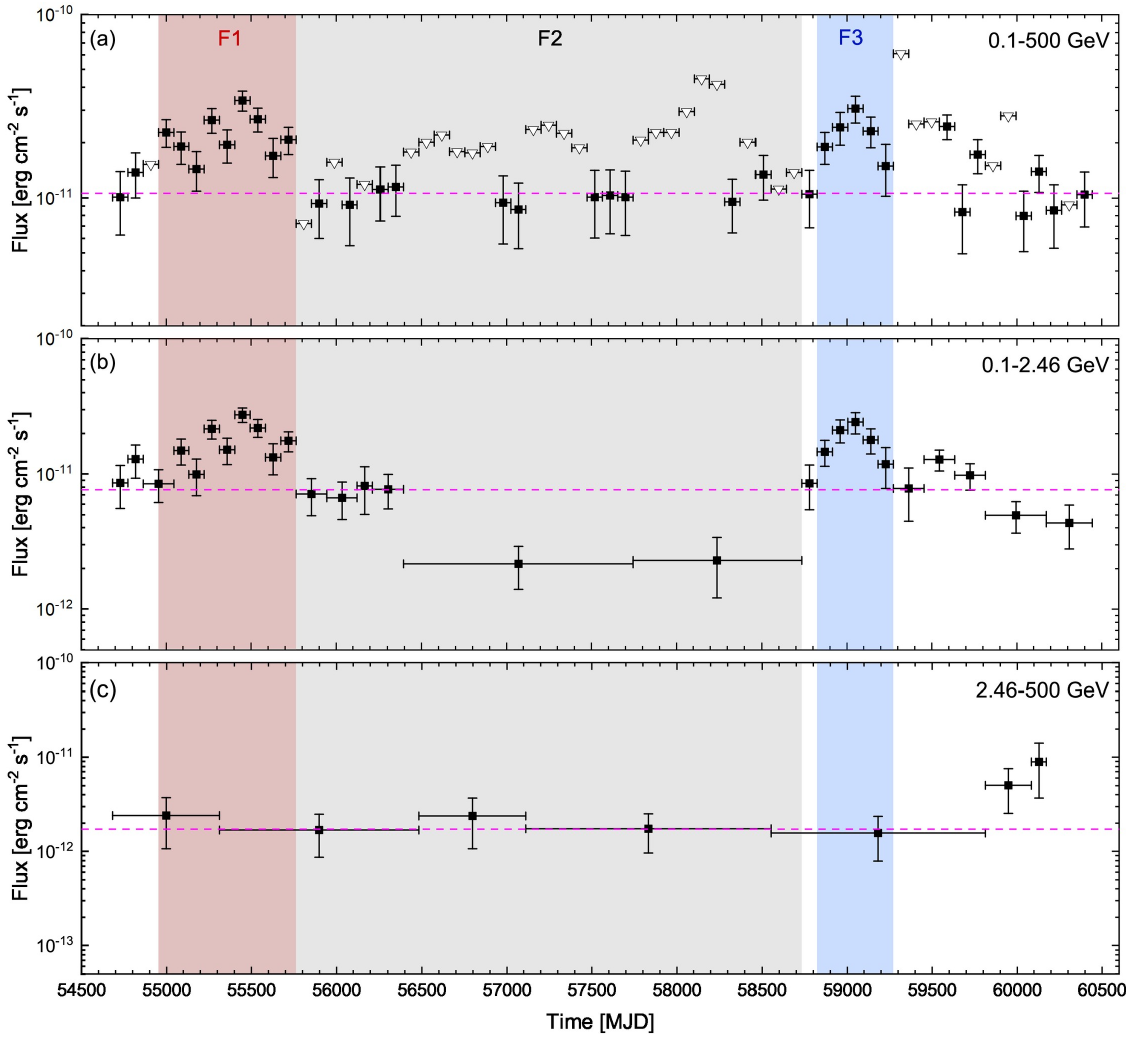}
    \caption{The long-term light curves of Pictor A, derived with the $\sim$16-year Fermi-LAT observation data in different energy bands, including the 0.1--500 GeV band (Panel (a)) together with the energy bands below (0.1--2.46 GeV in Panel (b)) and above (2.46--500 GeV in Panel (c)) the spectral break. The horizontal magenta dashed lines represent the average flux in that energy band. In panel (a), each time bin is 90 days, and the $3\sigma$ upper limit, denoted by open inverted triangles, is reported for time bins where $\rm TS\leq9$. In panels (b) and (c), the light curves are derived with an adaptive-binning method based on a criterion of TS$\geq$9 for each time bin, where the minimum time bin is 90 days.}
    \label{LC}
\end{figure}

\begin{itemize}

\item \textit{The full energy light curve.} We construct the $\sim$16-year Fermi-LAT light curve of Pictor A in the 0.1--500 GeV band using a uniform time-bin size of 90 days, as shown in Figure \ref{LC}(a). To quantify the variability of the $\gamma$-ray light curve, we compute the variability index using Equation \ref{eqTSvar}, yielding $\rm{TS}_{var}=2715.7$ with $\rm{N}=64$. Given $\rm{N}-1=63$ degrees of freedom, a value of $\rm{TS}_{var}\geq135.9$ is required to identify variable sources at a confidence level of 5$\sigma$. So, the result indicates that the $\gamma$-ray light curve of Pictor A exhibits significant flux variations at a confidence level substantially exceeding 5$\sigma$. Based on a comparison of these fluxes with the average value, we tentatively identify two periods (F1 and F3) as high-flux states and one period (F2) as a low-flux state for Pictor A, as illustrated in Figure \ref{LC}(a).
 
\item \textit{The light curves below and above the spectral break.} To investigate the temporal characteristics of the two spectral components, we also construct the light curves in the 0.1--2.46 GeV band (below the spectral break) and the 2.46--500 GeV band (above the spectral break), as displayed in Figures \ref{LC}(b) and \ref{LC}(c). Given the presence of numerous upper-limit points in the 0.1--500 GeV band light curve (Figure \ref{LC}(a)), we use an adaptive-binning method with a criterion of TS$\geq9$ for each time bin to derive the light curves in the two energy bands. The minimum time-bin size is set to 90 days, consistent with the time-bin size used for the 0.1--500 GeV light curve. Additionally, we compute the $\rm{TS}_{var}$ values for both light curves, obtaining $\rm{TS}_{var}=2306.3$ with $\rm{N}=28$ for the low-energy light curve (a value of $\rm{TS}_{var}\geq80.8$ with $\rm{N}-1=27$ degrees of freedom is required to identify variable sources at a confidence level of 5$\sigma$) and $\rm{TS}_{var}=5.5$ with $\rm{N}=7$ (corresponding to $0.05\sigma$) for the high-energy light curve, respectively. Notably, the low-energy light curve demonstrates significant flux variability substantially exceeding a confidence level of $5\sigma$, whereas no significant flux variability is observed in the high-energy light curve. This indicates that the dominant statistical contribution to the full energy light curve comes from the low-energy band component. Detailed information for each time bin of the two light curves is provided in Table \ref{LC_detail}.

\item \textit{The spectra of low-flux and high-flux states}. The spectrum shown in Figure \ref{spec_LAT} is a $\sim$16-year average spectrum, covering the different flux states of source. We thus extract the time-integrated spectra in the 0.1--500 GeV band for three different stages: MJD 55762--58732 (period F2) for the low-flux state, together with MJD 54952--55762 (period F1) and MJD 58822--59272 (period F3) for the two high-flux states, respectively. Using Equation \ref{equ_deltaTS}, we obtain $\Delta \rm TS=0.63$ for the F1 state and $\Delta \rm TS=1.55$ for the F3 state, respectively. These values suggest that the BPL spectral shape is not significant in either spectrum. As illustrated in Figure \ref{spec_LAT}, both spectra from the high-flux states can be well fitted using a PL function. However, they exhibit different spectral indices: $\Gamma_{\gamma}=2.59\pm0.21$ for the F1 state and $\Gamma_{\gamma}=3.30\pm0.17$ for the F3 state. Nonetheless, to account for the low-flux spectrum (F2 state), a BPL function is still necessary ($\Delta \rm TS=11.51$, corresponding to a significance of $3.4\sigma$), characterized by $\Gamma_{\rm \gamma,1}=3.13\pm0.08$, $\Gamma_{\rm \gamma,2}=1.92\pm0.16$, and a break energy of $2.15\pm0.27$ GeV. 
\end{itemize}

The aforementioned findings provide support for the idea that the spectrum depicted in Figure \ref{spec_LAT} results from the superposition of two distinct physical components. One component dominates the low-energy band, particularly during periods of high-flux emission, while another component primarily contributes to the high-energy band, which is more prominent during low-flux state of the source. The contrasting variability patterns above and below the spectral break could potentially serve as additional evidence for the existence of these two separate components in the $\gamma$-ray emission of Pictor A. 

\subsection{Assessing the Contamination from 4FGL J0515.6--4556}\label{sec3.2}

To investigate whether the broken $\gamma$-ray spectrum of Pictor A is influenced by contamination from the radiation of 4FGL J0515.6--4556 (associated with the FSRQ PKS J0515--4556), we conduct a series of tests as follows. 

\begin{figure}
    \includegraphics[angle=0, scale=0.47]{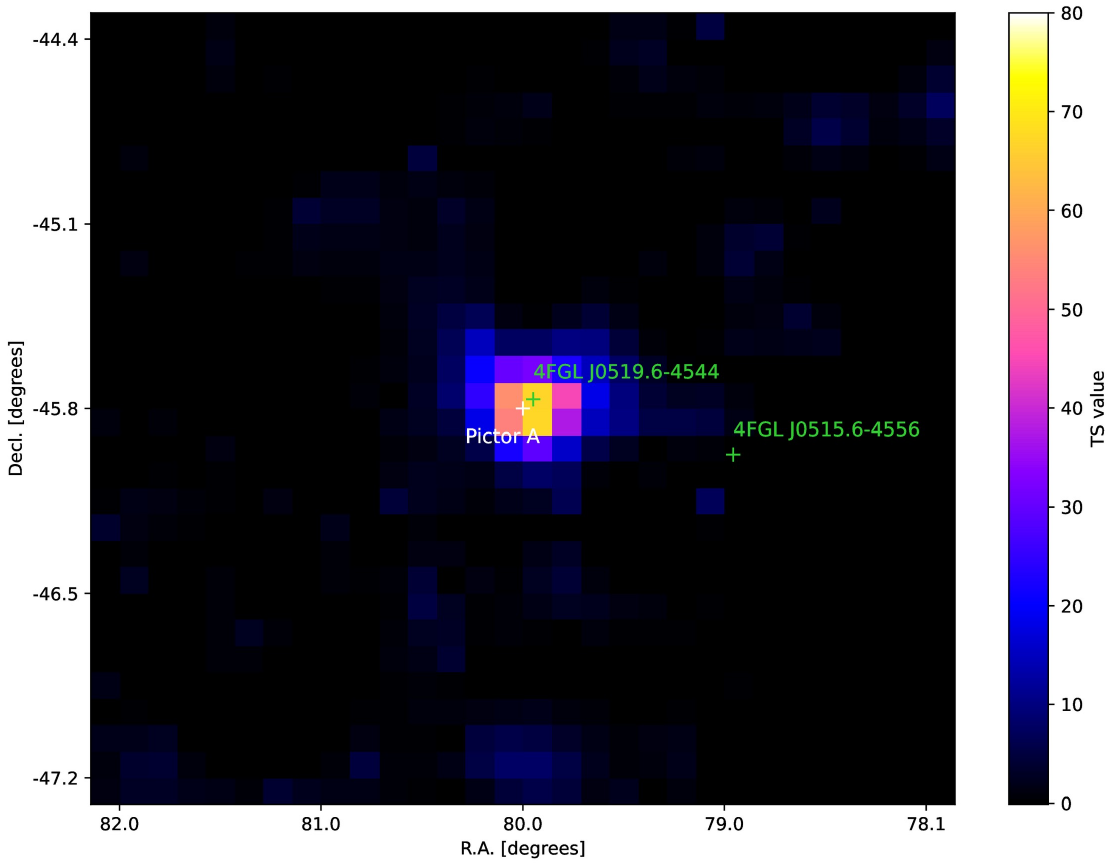}
    \includegraphics[angle=0, scale=0.47]{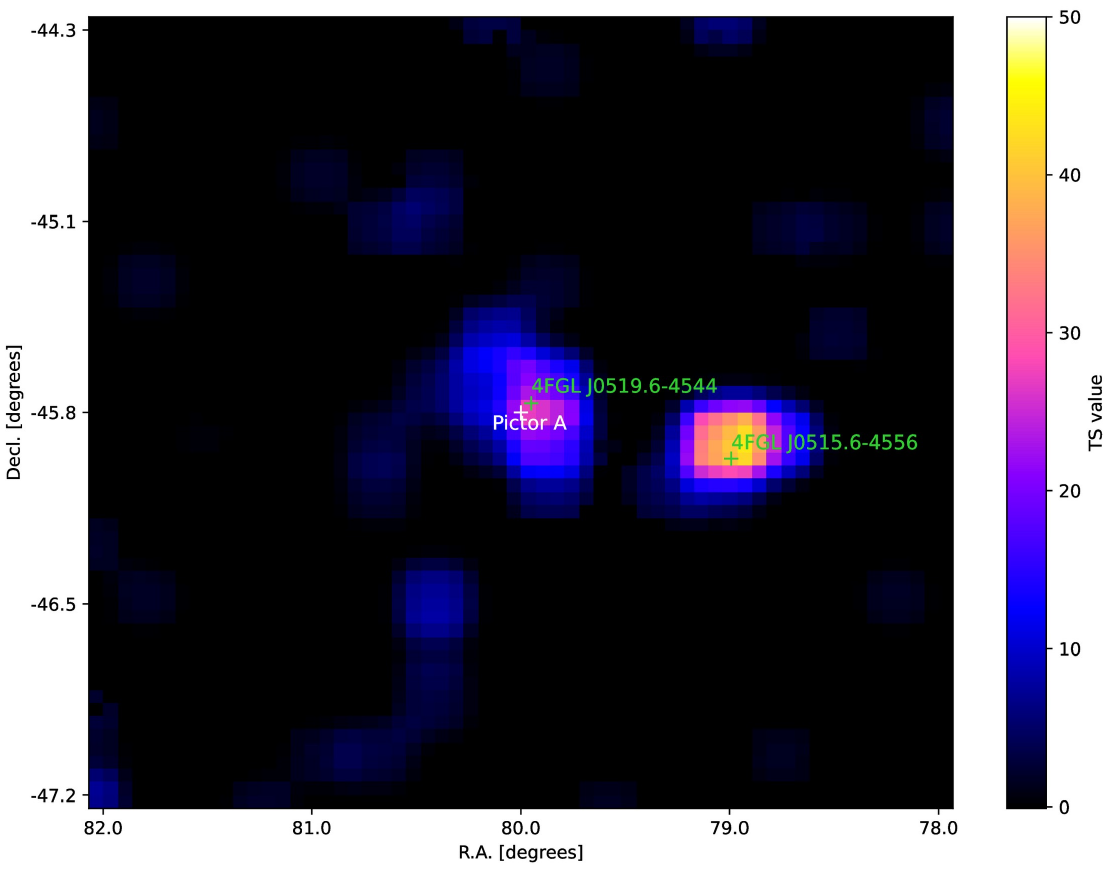}
    \caption{Left panel: 3$^\circ$ $\times$ 3$^\circ$ residual TS map in the 2.0--500 GeV band by removing 4FGL J0519.6--4544 (associated with Pictor A) from the \textit{model.xml}. Right panel: 3$^\circ$ $\times$ 3$^\circ$ TS map of Pictor A and PKS J0515--4556 in the 20.0--500 GeV band. The green crosses represent the positions of 4FGL J0519.6-4544 and 4FGL J0515.6-4556 in the 4FGL-DR4 \citep{2022ApJS..260...53A, 2023arXiv230712546B}. The white crosses represent the radio position of Pictor A.}
    \label{check-TSmap}
\end{figure}

\begin{itemize}

\item \textit{Contributions of $\gamma$-rays above 2 GeV.} We first consider an extreme scenario wherein the observed $\gamma$-ray emission is attributed solely to 4FGL J0515.6--4556, by removing 4FGL J0519.6--4544 from the \textit{model.xml}. As the angular resolution of the LAT is less than 0.6$\degr$ at 2 GeV \citep{2009ApJ...697.1071A}, we generate a 3$^\circ$ $\times$ 3$^\circ$ residual TS map centered on the radio position of Pictor A by re-analyzing 16-year Fermi-LAT data in the 2.0--500 GeV band. As shown in the left panel of Figure \ref{check-TSmap}, a significant residual signal with a maximum TS value of 67 is clearly observed at the location of Pictor A. Therefore, it is evident that the model considering only the single source 4FGL J0515.6--4556 cannot fully account for all observed emission. There are substantial contributions of $\gamma$-rays above 2 GeV originating from Pictor A.  

\item \textit{The spectra above 2.46 GeV.} As shown in Figure \ref{spec_LAT}, the $\gamma$-ray spectrum of Pictor A displays a spectral break at 2.46 GeV. We thus re-analyze the 16-year Fermi-LAT data above 2.46~GeV (break energy) to simultaneously generate the spectra of Pictor A and PKS J0515--4556, as illustrated in Figure \ref{check-spec}. Notably, Pictor A displays a flat power-law spectrum, while PKS J0515--4556 follows a log-parabolic spectral shape. The derived spectrum of PKS J0515--4556 is consistent with that reported in the 4FGL-DR4 \citep{2022ApJS..260...53A, 2023arXiv230712546B}. Furthermore, the flux of PKS J0515--4556 decreases logarithmically, becoming comparable to that of Pictor A around 30 GeV, while the $\gamma$-ray flux of Pictor A exhibits a gradual increase extending beyond 100 GeV.

\item \textit{TS map above 20 GeV.} To further investigate whether the high-energy emission in the energy band 20--500 GeV primarily originates from the bright PKS J0515--4556, {we re-generate the TS map of Pictor A and PKS J0515--4556, as illustrated in the right panel of Figure \ref{check-TSmap}.} The maximum TS values obtained are 25 for Pictor A and 40 for PKS J0515--4556, respectively. Considering the angular resolution of the LAT above 20~GeV is much better than 0.75 degrees, these results indicate that Pictor A also contributes significantly to the high-energy $\gamma$-ray emission, which is consistent with our previous spectral analysis.
\end{itemize}

The results from the above analysis are consistent, indicating that the spectral break at 2.46 GeV observed in Pictor A is not attributable to contamination from emission by PKS J0515--4556. It should be noted that the systematic uncertainty associated with the Point Spread Function (PSF) was not incorporated into the analysis, as the Fermi team is still conducting an ongoing study of the LAT PSF using in-orbit data. Nevertheless, even considering the maximum uncertainty of the PSF within the energy band above 10 GeV, the PSF at 10 GeV expands only to 0.19$\degr$ (with an uncertainty of 25\% and a PSF value of 0.15$\degr$). Given that the angular separation between the two sources is 0.75$\degr$, we conclude that this level of PSF uncertainty does not significantly affect the spectral results.

\subsection{Possible $\gamma$-ray Emission from WHS}

The analysis results of the Fermi-LAT data clearly indicate that the $\gamma$-rays detected from Pictor A have two distinct origins: one component dominates the observed flux and exhibits significant variability, primarily contributing to the emission at low energies (below the spectral break) with a steep spectrum; while another component contributes to the high-energy emission (above the spectral break) without evident variability, displaying a hard spectrum. Through broadband SED fitting of the nucleus and WHS in Pictor A, \citet{2009ApJ...701..423Z} proposed that there is detectable $\gamma$-ray emission in WHS. The combined spectrum from both the nucleus and WHS in the GeV band would manifest as a BPL feature, with a harder spectrum at higher energy ranges, as illustrated in Figure 3 of their study. Therefore, we suggest that the high-energy component observed above the spectral break in Pictor A is predominantly due to the emission from WHS, while radiation from the nucleus primarily accounts for low-energy emission below this break and exhibits evident variability in Pictor A.

\subsection{X-Ray Polarization of the nucleus and WHS}\label{sec3.4}

The \texttt{PCUBE} analysis yields an MDP$_{99}$ of 5.4\% for the nucleus and an MDP$_{99}$ exceeding 90\% for the WHS, based on the combined data from the three IXPE DUs, as shown in Figure \ref{pcube}. No significant detection of X-ray polarization at a confidence level greater than 99\% is observed in the 2--8 keV band for both the nucleus and the WHS. Due to the limited number of detected photons, the estimated MDP$_{99}$ value for the WHS surpasses the theoretical maximum polarization degree of $\sim$ 70\%. Based on the spectropolarimetric fitting results presented in Figure \ref{specpol}, upper limits for the polarization degree are derived: $\Pi_{\rm X}<6.6\%$ for the nucleus and $\Pi_{\rm X}<56.4\%$ for the WHS, at a 99\% confidence level within the 2--8 keV band. Figure \ref{contour} illustrates the confidence levels of the measurements for both the nucleus and WHS through polarization contour plots. The detailed optimal parameters derived from the spectropolarimetric fitting are summarized in Table \ref{table_IXPE}.

We also perform an energy-resolved polarization analysis on the nucleus following the methodology provided in \citet{2024ApJ...963....5E} and \citet{2024ApJ...970L..22H}. Specifically, we divide the entire 2--8 keV energy range into finer energy bins (1 keV, 1.5 keV, 2 keV, and 3 keV) and estimate the X-ray polarization parameters for each bin using both the model-independent approach and the spectropolarimetric method. As displayed in Figure \ref{energy-pol}, the polarization parameters can be estimated with a confidence level exceeding 99\% only within the energy ranges of 4--6 keV and 5--6.5 keV using both the model-independent approach and the spectropolarimetric method. Based on the spectropolarimetric fit results, we obtain $\Pi_{\rm X}=9.5\%\pm2.4\%$ with $\psi_{\rm X}=107.0\degr\pm7.2\degr$ in the 4--6 keV band at a confidence level of $4.0\sigma$, and $\Pi_{\rm X}=13.8\%\pm3.9\%$ with $\psi_{\rm X}=119.7\degr\pm8.0\degr$ in the 5--6.5 keV band at a confidence level of $3.5\sigma$. Meanwhile, the \texttt{PCUBE} analysis yields $\Pi_{\rm X}=9.0\%\pm2.6\%$ with $\psi_{\rm X}=104.7\degr\pm8.4\degr$ and $\Pi_{\rm X}=14.2\%\pm4.3\%$ with $\psi_{\rm}=126.7\degr\pm8.7\degr$ for the respective energy bands. The derived values of $\psi_{\rm X}$ are slight larger than the jet position angle, which is approximately $100\degr$ \citep{1997A&A...328...12P,2000AJ....119.1695T}.

The radio \citep{1997A&A...328...12P} and optical bands \citep{1987ApJ...314...70R,1995ApJ...446L..93T} have detected high polarization in WHS, reaching up to the theoretical maximum of approximately 70\%. Furthermore, there is a remarkable similarity in structures between the radio and optical bands. Due to the limited number of detected photons, the IXPE observations for WHS do not yield significant polarization data. However, we can establish an upper limit on the polarization degree for WHS, i.e., $\Pi_{\rm X}<56.4\%$ at a 99\% confidence level in the 2--8 keV band.

\section{Constructing and Modeling the SED}\label{sec4}

The observed BPL spectral pattern in Pictor A, characterized by a harder spectrum at higher energy bands, suggests the existence of two distinct components contributing to its $\gamma$-ray emission. In order to further investigate the $\gamma$-ray emission characteristics of Pictor A, we construct broadband SEDs for both the nucleus and WHS regions, covering the energy band from radio to $\gamma$-rays, as illustrated in Figure \ref{SED}. The data for both the nucleus and WHS, ranging from radio to X-rays, are obtained from \cite{2023MNRAS.521.2704G} and references therein (for more details, please refer to the caption of Figure \ref{SED}), while the $\gamma$-ray spectrum is derived through our own analysis in this study. As described in Section \ref{sec3}, the low-energy and high-energy components are proposed to primarily originate from the nucleus and WHS, respectively. The results appear to be consistent with their broadband SEDs, as shown in Figure \ref{SED}; the combinations with their respective radio to X-ray data result in smooth features in their broadband SEDs. 

The leptonic model is commonly used to fit the broadband SEDs of the jet emission from AGNs. The radiation region is assumed to be a sphere with a radius $R$, magnetic field strength $B$, the bulk Lorentz factor $\Gamma$, and the Doppler boosting factor $\delta$. The electron distribution in the emission region is taken as a PL or a BPL, characterized by an electron density parameter ($N_0$), a break energy ($\gamma_{\rm b}$), and indices ($p_1$ and $p_2$) ranging from $\gamma_{\min}$ to $\gamma_{\max}$. In the case of a PL electron distribution, $p_1=p_2$ and $\gamma_{\rm b}=\gamma_{\max}$. Additionally, the synchrotron-self absorption, Klein--Nishina effect, and extragalactic background light absorption \citep{2022ApJ...941...33F} are considered during the SED modeling.

\subsection{Nucleus Region}

The synchrotron and SSC processes of relativistic electrons with a BPL distribution are commonly used to explain the observed SED of the nucleus in $\gamma$-ray emitting RGs \citep{2009ApJ...707...55A, 2011A&A...533A..72M, 2015ApJ...798...74F, 2017RAA....17...90X, 2024ApJ...965..163Y}, which is also used to represent the observed SED of the nucleus in Pictor A. The emission region radius is fixed at $R=10^{16}$~cm, as also mentioned in \cite{2009ApJ...701..423Z} and \cite{2023MNRAS.521.2704G}. We adopt $\Gamma=5$ with a viewing angle of $30^\circ$ \citep{2023MNRAS.521.2704G}, corresponding to $\delta=1.3$. The values of $p_1$ and $p_2$ are determined using the observed spectral indices in X-rays and optical--UV bands, respectively. We set $\gamma_{\min}=1$, while $\gamma_{\max}$ is approximately constrained by the observed GeV spectrum, showing a cutoff at the break frequency. 
By adjusting the parameters $B$, $N_0$, and $\gamma_{\rm b}$ to match the broadband SED of the nucleus, we obtain $B=7$ G, $\gamma_{\rm b}=1.6 \times 10^3$, and $\gamma_{\rm max}=1.6 \times 10^4$. The modeling parameters are also listed in Table \ref{table_SED}. 

The fitting result is presented in Figure \ref{SED}; the one-zone synchrotron+SSC model well represents the broadband SED of the nucleus. It is evident that this model fails to account for the high-energy component observed in the Fermi-LAT spectrum, even without considering the low-energy component of the $\gamma$-ray spectrum. If the high-energy component also stems from the inner jet of Pictor A, a more complex radiation model is required to reproduced the observed broadband SED of nucleus, such as a spine-layer model \citep{2008MNRAS.385L..98T} or hadronic processes \citep[e.g.,][]{2022ApJ...925L..19C}.

It should be noted that the derived parameter values are based on visual assessments, resulting in a set of model parameters that provide an acceptable fit. Furthermore, there exists a degeneracy between $B$ and $\delta$ during the SED modeling process (e.g., \citealt{2012ApJ...752..157Z}). For instance, with a similar value of $R\sim10^{16}$ cm, \cite{2023MNRAS.521.2704G} reported $\delta=1.3$ with $B=10$ G, whereas \citet{2009ApJ...701..423Z} found $\delta=1.6$ with $B=7$ G, and \citet{2017RAA....17...90X} obtained $\delta=2.5$ with $B=4.2$ G. Considering the dominance of the low-energy component in the high-flux $\gamma$-ray emission state of Pictor A, as illustrated in Figure \ref{spec_LAT}, we incorporate the X-ray spectrum obtained during the high-flux state \citep{2023MNRAS.521.2704G} into account in our SED modeling. Consequently, the obtained value of $B$ differs from theirs due to the different X-ray and $\gamma$-ray data.

\subsection{WHS}

Considering the complex radio structure \citep{1997A&A...328...12P,2008AJ....136.2473T} and flat X-ray spectrum of WHS (e.g., \citealt{2009ApJ...701..423Z,2023MNRAS.521.2704G}), we use a two-zone leptonic model to fit its broadband SED from radio to X-rays, same model as taken in \citet{2009ApJ...701..423Z} and \citet{2023MNRAS.521.2704G}. This model includes synchrotron and SSC radiations originating from both a diffuse region and a compact substructure, where both regions are treated as stationary blobs characterized by $\Gamma=\delta=1$. The energy density of the synchrotron radiation field in this scenario surpasses that of the cosmic microwave background (CMB) by more than one order of magnitude. Consequently, the contribution of the IC scattering of CMB photons by relativistic electrons can be considered negligible when compared to the SSC process. The emission of WHS from radio, optical to ultraviolet is primarily attributed to the synchrotron process occurring in the whole diffuse region, while the X-rays are produced through synchrotron radiation emitted by a compact substructure contained within the diffuse region. The $\gamma$-ray spectrum above the break in Figure \ref{spec_LAT} is taken into consideration in the SED modeling of the WHS.  

For the diffuse region of WHS, a radius of $R=500$ pc is adopted, which corresponds to half of the overall size (1.4 arcsec, \citealt{1995ApJ...446L..93T}) of WHS in the optical band. The electron distribution follows a BPL, with $p_1$ and $p_2$ being well constrained by observed data across radio to UV bands. $\gamma_{\min}$ cannot be determined and is fixed at 1, while $\gamma_{\max}$ is roughly constrained by the last data point in the UV band. Considering the equipartition condition ($U_{B}=U_{\rm e}$), where $U_{B}$ and $U_{\rm e}$ are the energy densities of the magnetic fields and electrons in the radiation region, it is found that the prediction flux from SSC process fails to explain the $\gamma$-ray spectrum at GeV energies. The fitting of the highest-energy point in the GeV band necessitates a significantly small magnetic field that deviates substantially from the equipartition condition, while also resulting in the model's prediction line surpassing the two data points above the break. Therefore, we abandon this hypothesis.

For the compact substructure of WHS, a radius of $R=30$ pc is taken, which is half the average size of the five compact substructures reported by \cite{2008AJ....136.2473T}. Considering the limited observational data, we adopt a PL electron distribution. $p_1$ is determined based on the spectral index between radio and soft X-rays. $\gamma_{\min}$ is also fixed at 1. When considering equipartition conditions, the predicted line from the SSC process is significantly lower than the $\gamma$-ray fluxes observed at the GeV band, similar to that of the diffuse region. We propose that the $\gamma$-rays primarily originate from the compact substructure; therefore, the equipartition condition in the compact substructure is not taken into account during SED modeling of WHS. 

We adjust the values of $N_0$ and $\gamma_{\rm b}$ for the diffuse region under the equipartition condition, along with adjusting the parameters $N_0$, $B$, and $\gamma_{\rm max}$ of the compact substructure to represent the observed SED of WHS from radio to $\gamma$-ray bands. The fitting parameter values are visually assessed and are provided in Table \ref{table_SED}. The fitting result is presented in Figure \ref{SED}. The parameter values for the diffuse region are found to be comparable to those reported in \cite{2023MNRAS.521.2704G}. However, the $B$ value for the compact substructure is one order of magnitude smaller, accompanied by a slightly higher value of $\gamma_{\rm max}$ in order to match the $\gamma$-ray spectrum. Additionally, we observe that the predicted flux of SED modeling is over the sensitivity curve of Cherenkov Telescope Array (CTA) south array (CTA-S, 50 hr), but remains below the sensitivity of High Energy Stereoscopic System (H.E.S.S.), where the sensitivity curves are obtained from \cite{2022ChPhC..46c0003W} and the CTA webpage\footnote{\url{https://www.ctao.org/for-scientists/performance/}}, respectively. Considering the hard spectrum in the GeV band, it could potentially be considered as a candidate for CTA.

\section{Discussion and Conclusions}\label{sec5}

\begin{figure}
    \centering
    \includegraphics[angle=0, scale=0.45]{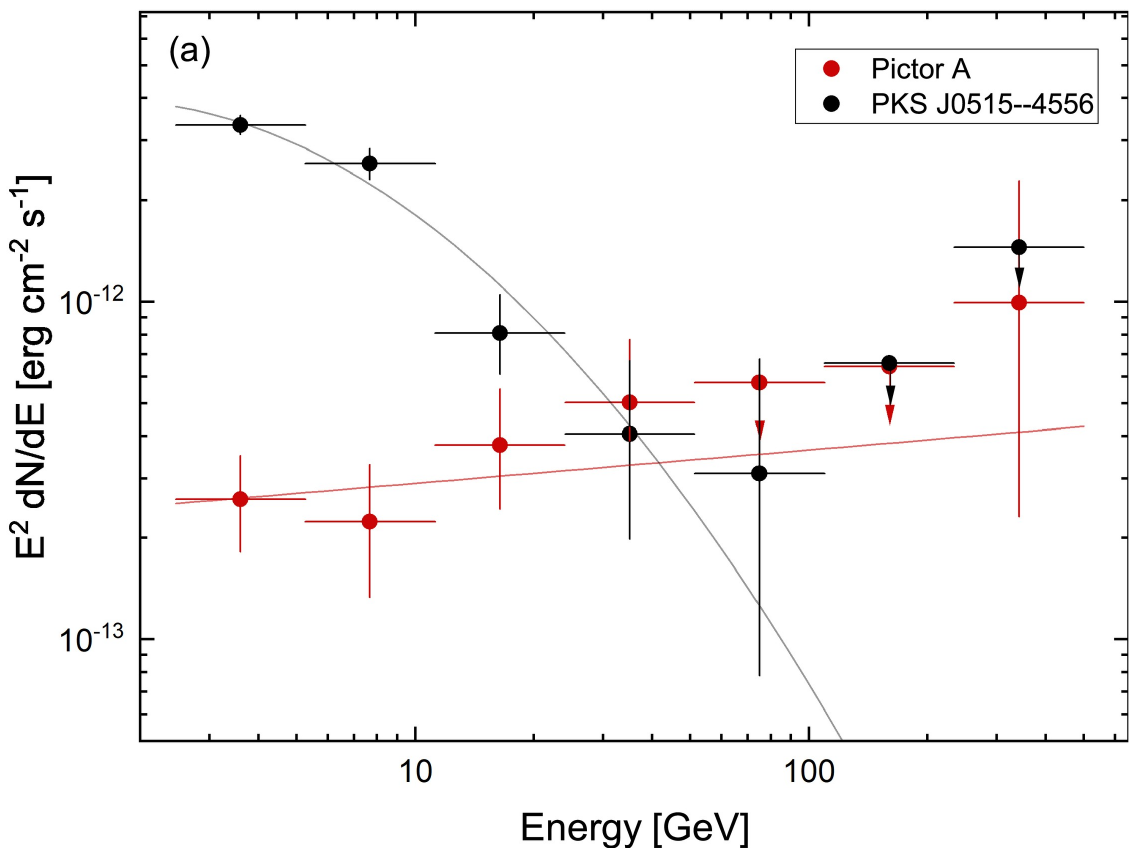}\hspace{0.3cm}
    \includegraphics[angle=0, scale=0.45]{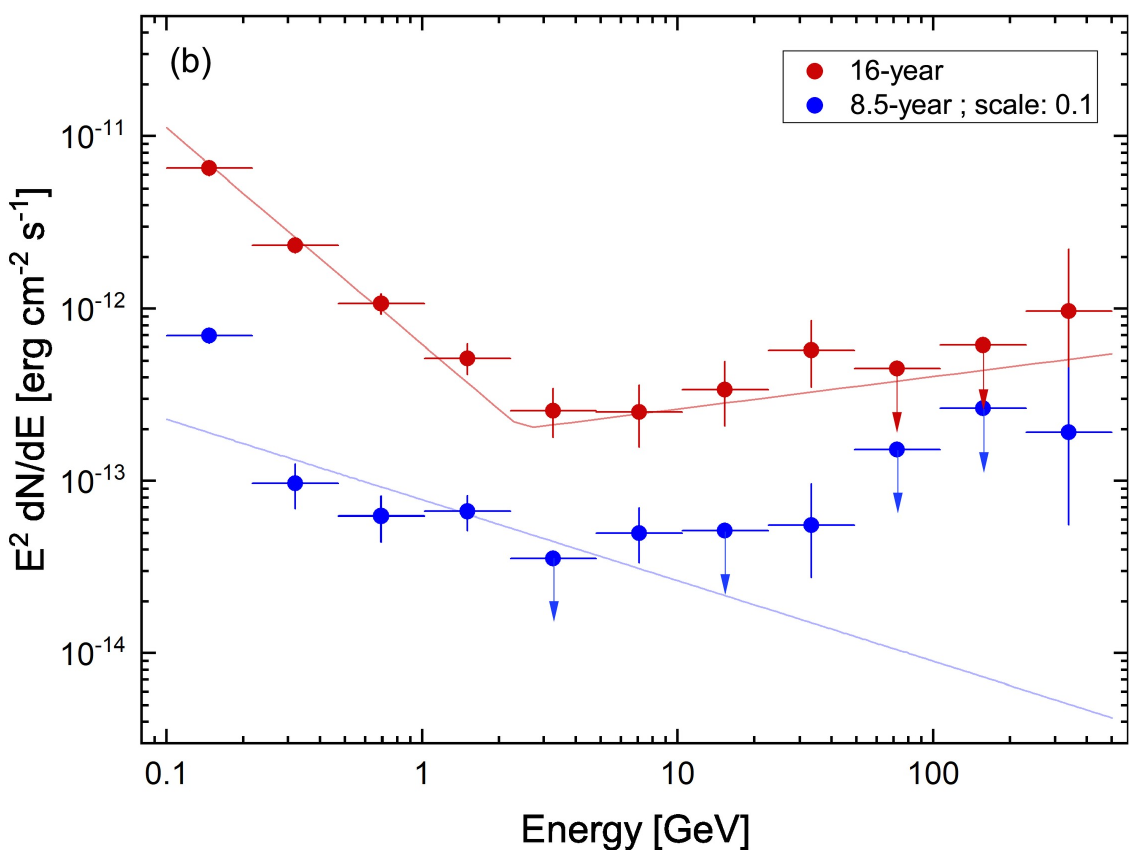}
    \caption{Panel (a): The spectra of Pictor A (red symbols) and PKS J0515--4556 (black symbols) in the 2.46--500 GeV band of 16-year dataset. Panel (b): The $\gamma$-ray spectra of Pictor A in the 0.1-500 GeV energy band for different observation durations: red symbols represent the 16-year dataset and blue symbols represent the first 8.5-year dataset. For clarity of comparison, the spectrum of the 8.5-year dataset has been scaled by a factor of 0.1. The $2\sigma$ upper limit (points with a downward arrow) is reported when $\rm TS\leq4$ for that energy bin in both panel (a) and (b).}    
    \label{check-spec}
\end{figure}

\begin{figure}
    \centering
    \includegraphics[angle=0, scale=0.35]{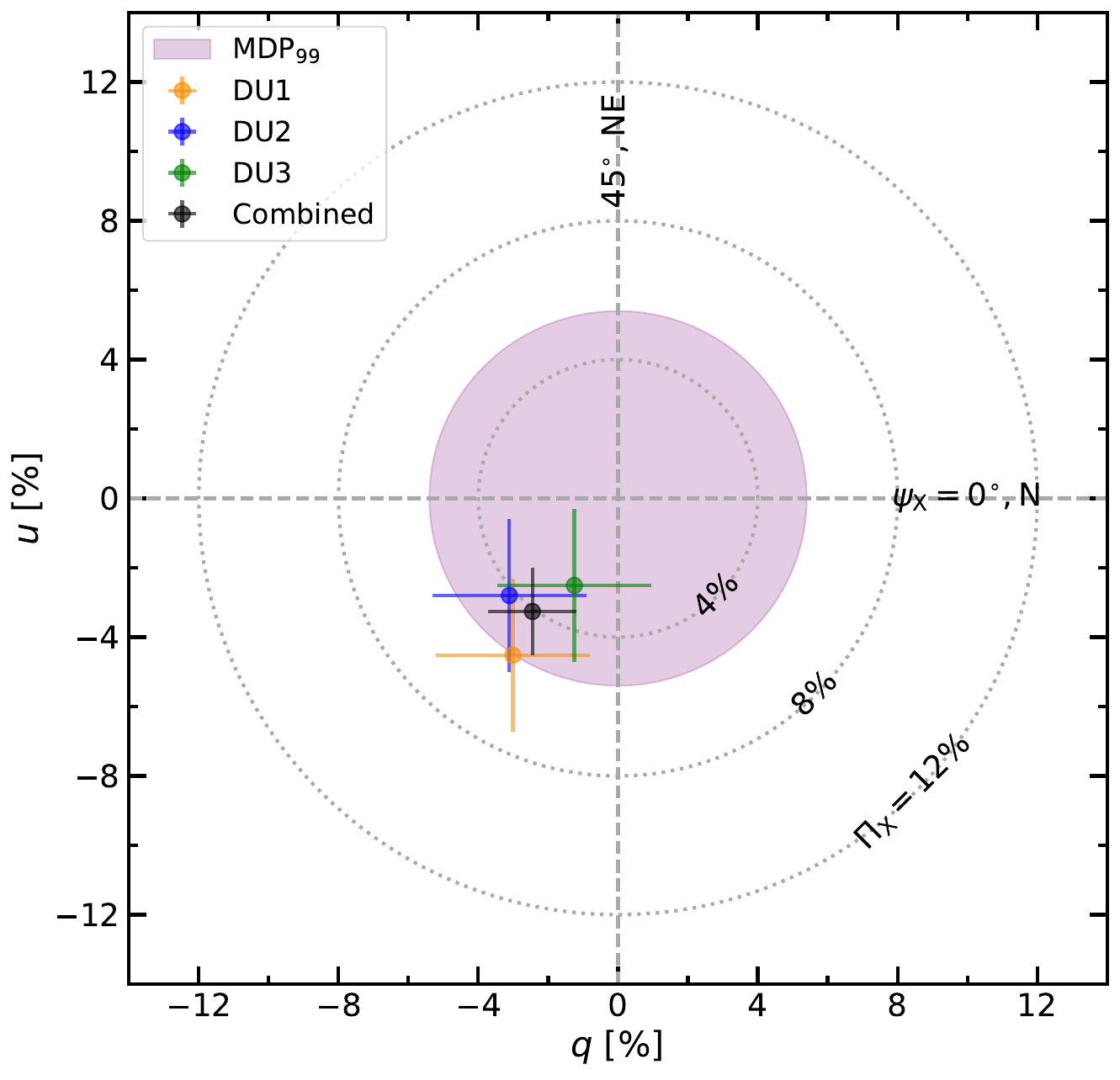}
    \includegraphics[angle=0, scale=0.35]{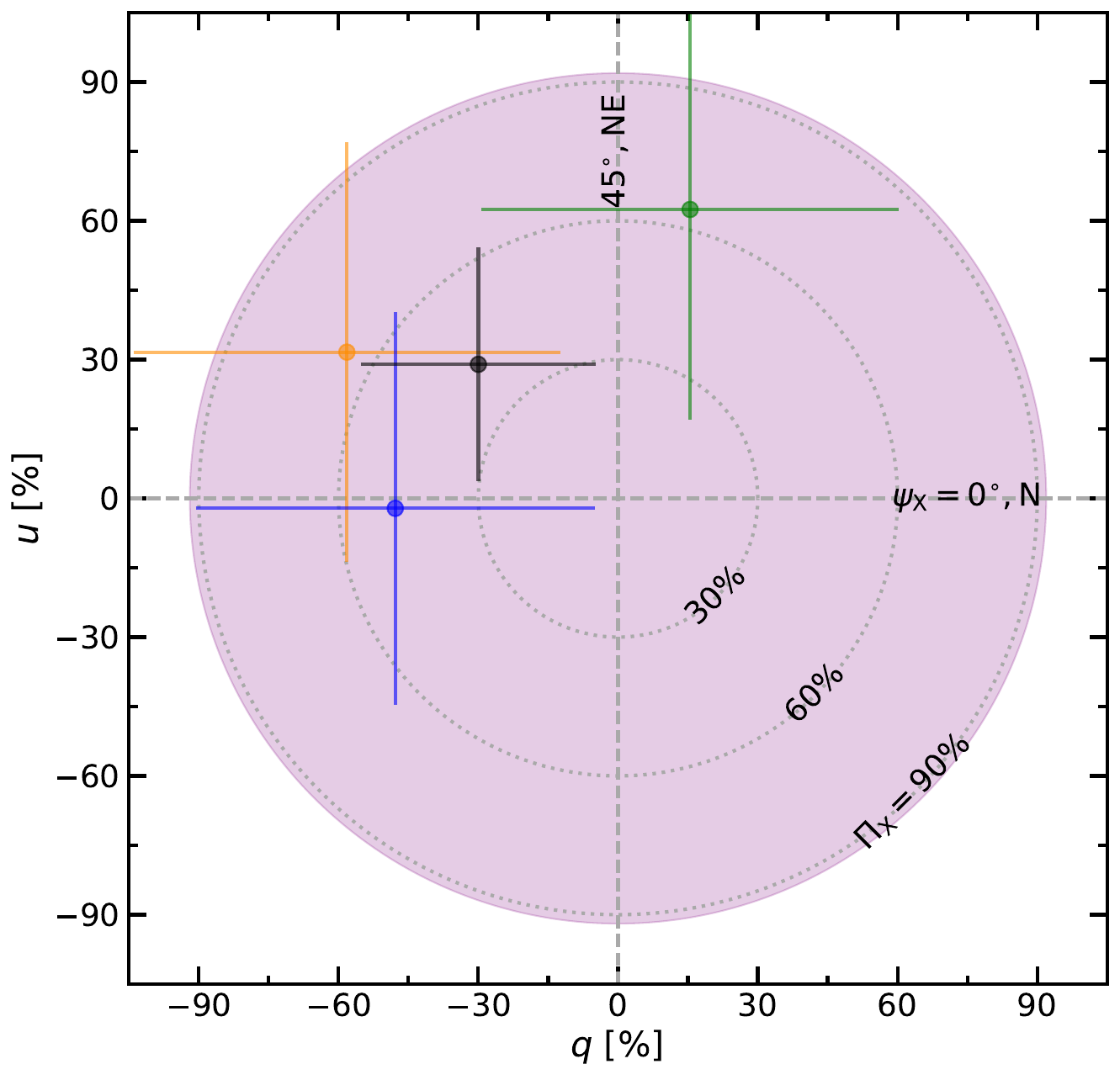}
    \caption{Normalized Stokes parameters $q$ and $u$ of the nucleus (left panel) and WHS (right panel) in the 2--8 keV band for different DUs, as measured using the \texttt{PCUBE} algorithm in software \texttt{ixpeobssim}.}
    \label{pcube}
\end{figure}

\begin{figure}
    \centering
    \includegraphics[angle=0, scale=0.35]{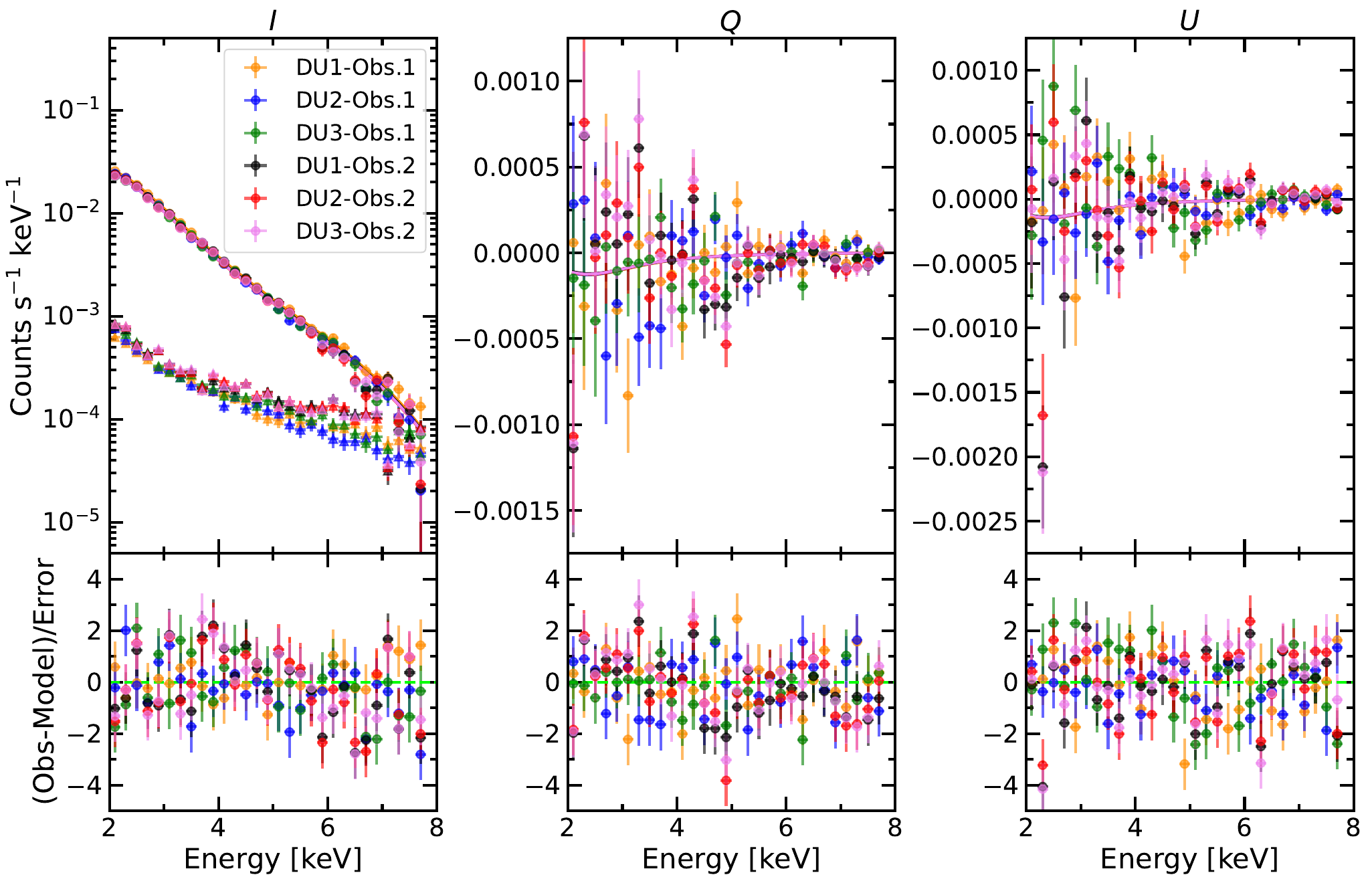}
    \includegraphics[angle=0, scale=0.35]{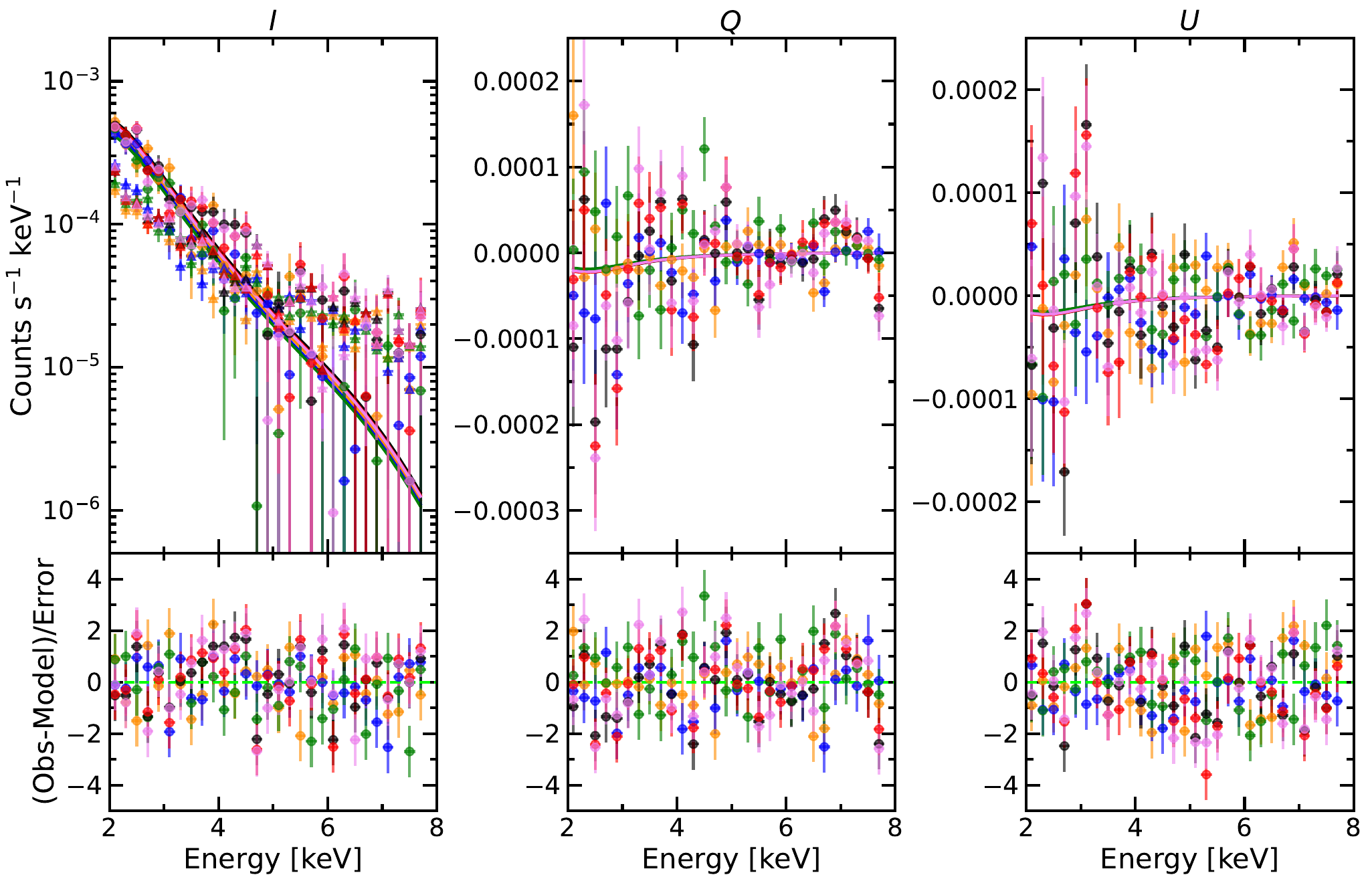}
    \caption{Spectropolarimetric fit results for the nucleus (top panel) and WHS (bottom). Panels represent the fits to the IXPE Stokes parameters $I$, $Q$, and $U$ with their associated residuals from left to right. In the $I$ spectra panels, the triangles indicate the background spectra for different DUs.}
    \label{specpol}
\end{figure}

\begin{figure}
    \centering
    \includegraphics[angle=0, scale=0.35]{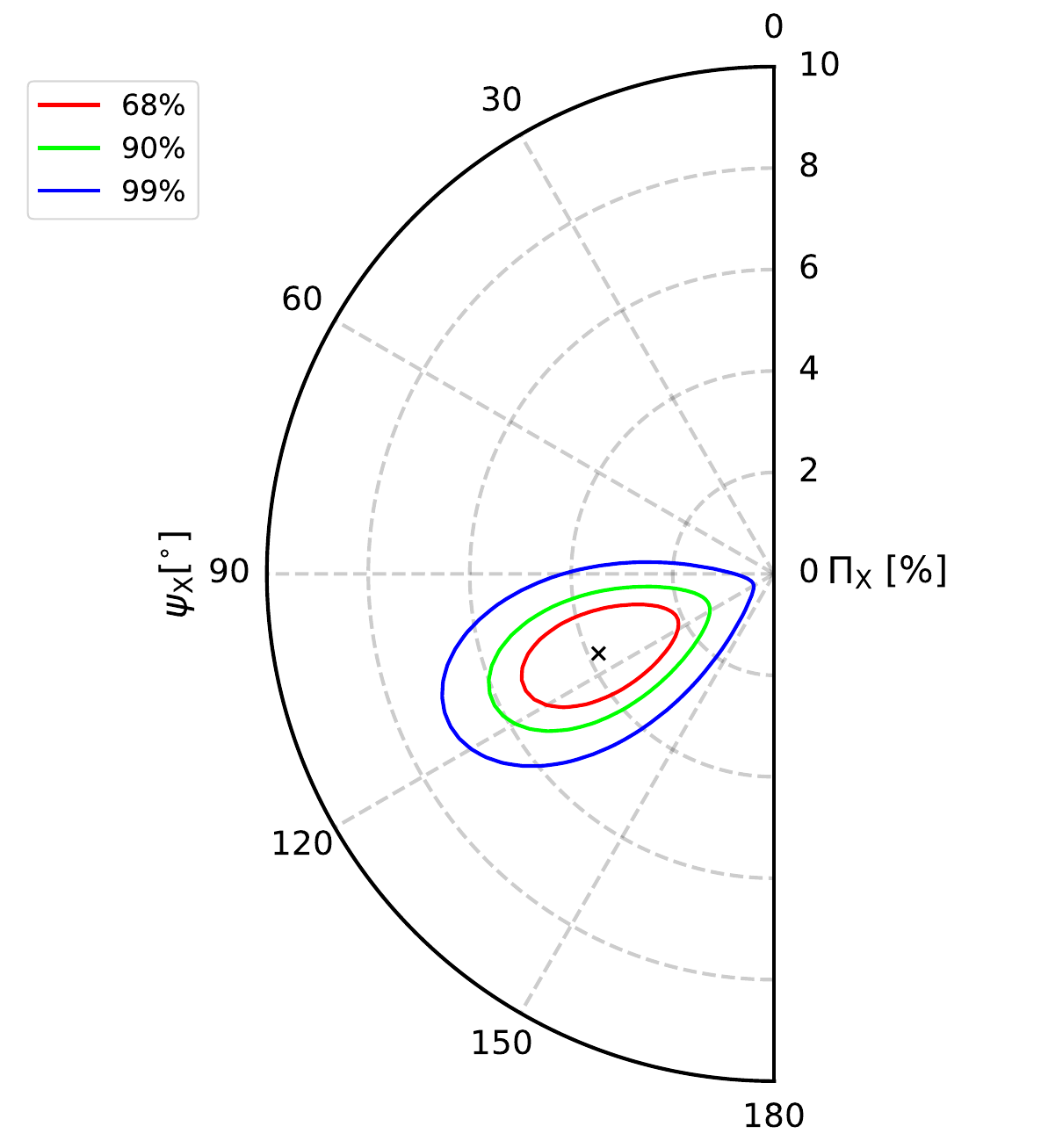}
    \includegraphics[angle=0, scale=0.35]{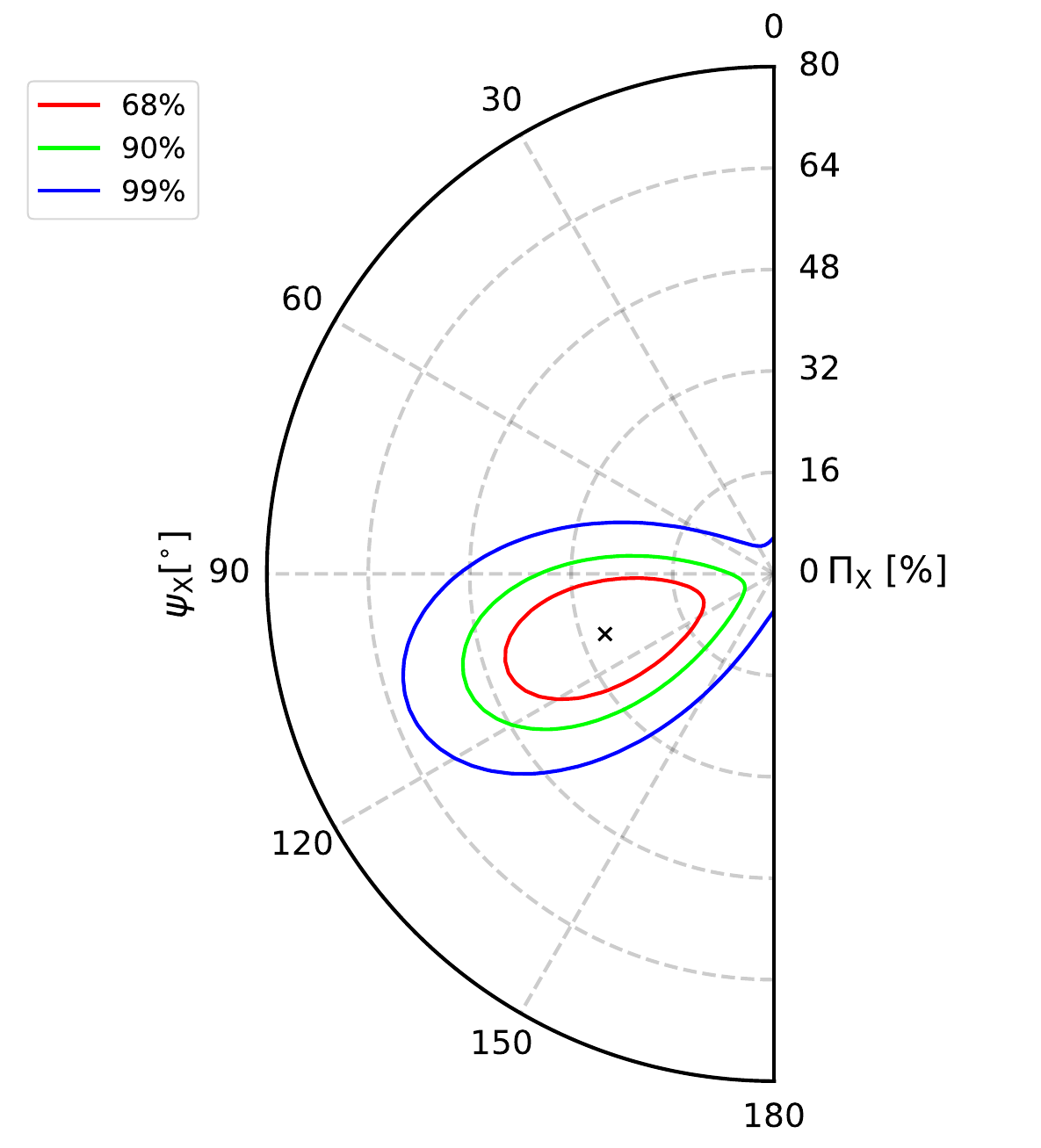}
    \caption{X-ray polarization contours in the 2--8 keV energy band for the nucleus (left panel) and WHS (right panel), derived from the spectropolarimetric fitting using \texttt{Xspec}. The black crosses represent the best-fit values of $\Pi_{\rm X}$ and $\psi_{\rm X}$. The red, green, and blue contours denote the 68\%, 90\%, and 99\% confidence intervals, respectively.}
    \label{contour}
\end{figure}

By analyzing the Fermi-LAT observation data, we find that the $\sim$ 16-year average spectrum of Pictor A in the GeV band exhibits statistically significant hardening feature. The BPL spectral form is more preferred than a simple PL function for describing the $\gamma$-ray spectrum of Pictor A with a confidence level of $6.2\sigma$. Pictor A was initially identified as a $\gamma$-ray source by \citet{2012MNRAS.421.2303B} using the first three years of Fermi-LAT data. Subsequently, \citet{2019A&A...627A.148A} analyzed $\sim$8.5 years of Fermi-LAT data in the 0.1--100 GeV energy range for Pictor A, while \citet{2020MNRAS.492.4666R} examined $\sim$10 years of Fermi-LAT observations spanning 0.1--300 GeV. Both studies reported a single PL spectrum for Pictor A with photon spectral indices $\Gamma_{\gamma}=2.63\pm0.08$ and $\Gamma_{\gamma}=2.37\pm0.10$, respectively. Notably, \citet{2020MNRAS.492.4666R} also reported the detection of a photon with an energy exceeding 100 GeV (in their figure 3). To quantitatively evaluate the impact of energy range and exposure time on the spectral break, we calculate the difference in TS values ($\Delta \rm TS$) between PL model and BPL model (as described in Section \ref{sec3.1}) for different observation durations. For the first 8.5-year dataset, consistent with the duration in \citet{2019A&A...627A.148A}, the $\Delta \rm TS$ values are 1.23 (corresponding to $1.1\sigma$) in the 0.1–100 GeV energy band and 6.85 ($2.6\sigma$) when extending to 0.1–500 GeV (Figure \ref{check-spec}). Within the full 16-year dataset, these values significantly increase to 20.17 ($4.5\sigma$) in the 0.1–100 GeV energy band and 38.88 ($6.2\sigma$) in the broader 0.1–500 GeV energy band. This progressive enhancement of significance demonstrates that increasing the exposure time improves the spectral break detection significance across all energy bands, while expanding the energy range to 500 GeV further enhances the statistical robustness of high-energy spectral features. 

Significant spectral hardening behavior in the GeV $\gamma$-ray band has also been observed in two other RGs: Cen A \citep{2013ApJ...770L...6S, 2017PhRvD..95f3018B} and M 87 \citep{2019A&A...623A...2A}. In both cases, it was proposed that the GeV emission consists of two distinct physical components. For M 87, the hardening $\gamma$-ray spectrum was observed at a significance level of $\sim2.66\sigma$, and both components were suggested to stem from the inner jet \citep{2019A&A...623A...2A}. For Cen A, the significance level of the hardening $\gamma$-ray spectrum increased from $2.3\sigma$  \citep{2013ApJ...770L...6S} based on the 4-year Fermi-LAT observation data to $>5\sigma$ \citep{2017PhRvD..95f3018B} when utilizing the 8-year Fermi-LAT observation data. They suggested that this additional high-energy component beyond the common SSC emission may be related with the interactions between relativistic protons and ambient gas in the large-scale jet \citep{2013ApJ...770L...6S}, or the dark matter around the black hole, or a population of millisecond pulsars in the core region \citep{2017PhRvD..95f3018B}.    

The low-energy component, characterized by a steep spectrum and significant variability, is likely to originate from the nucleus and accounts for the majority of the emission flux of Pictor A. Contrarily, the high-energy component, which exhibits a hard spectrum and remains in a steady emission state, is favored by the extended $\gamma$-ray production scenario. We propose that this high-energy component stems from the WHS, and that the observed BPL spectrum in the GeV band of Pictor A results from the combined emission originating from both its nucleus and WHS.

We use a single-zone leptonic model to explain the broadband SED of the nucleus, while adopting a two-zone leptonic model to represent the broadband SED of WHS. The two components in the GeV spectrum of Pictor A smoothly connect with the X-ray spectra of the nucleus and WHS, respectively; both can be naturally represented by the SSC process occurring in two different regions. Moreover, our SED fitting predicts that CTA-S will be able to detect very high-energy (VHE) emission from WHS. Notably, H.E.S.S. observations have unveiled VHE $\gamma$-ray emission along the large-scale jet of RG Cen A \citep{2020Natur.582..356H}, providing compelling evidence supporting large-scale jets as sites for accelerating ultrarelativistic electrons.

By analyzing the IXPE observational data for Pictor A, we have derived an upper limit of $\Pi_{\rm X}<6.6\%$ for the polarization degree in the 2--8 keV band of the nucleus at a 99\% confidence level. This finding is consistent with the low-polarization expectation that the X-ray emission of the nucleus is produced by the SSC process \citep{2012ApJ...744...30K,2019ApJ...885...76P}, similar to the IXPE observation results of RG Cen A \citep{2022ApJ...935..116E} and other low and intermediate spectral peak blazars \citep{2023arXiv231011510M}. We also note that the nucleus of Pictor A generally exhibits low polarization in the radio band, typically less than 5\% (\citealt{1997A&A...328...12P} and references therein). However, no optical polarimetry data is available in the archived records.

\begin{figure}
    \centering
    \includegraphics[angle=0, scale=0.25]{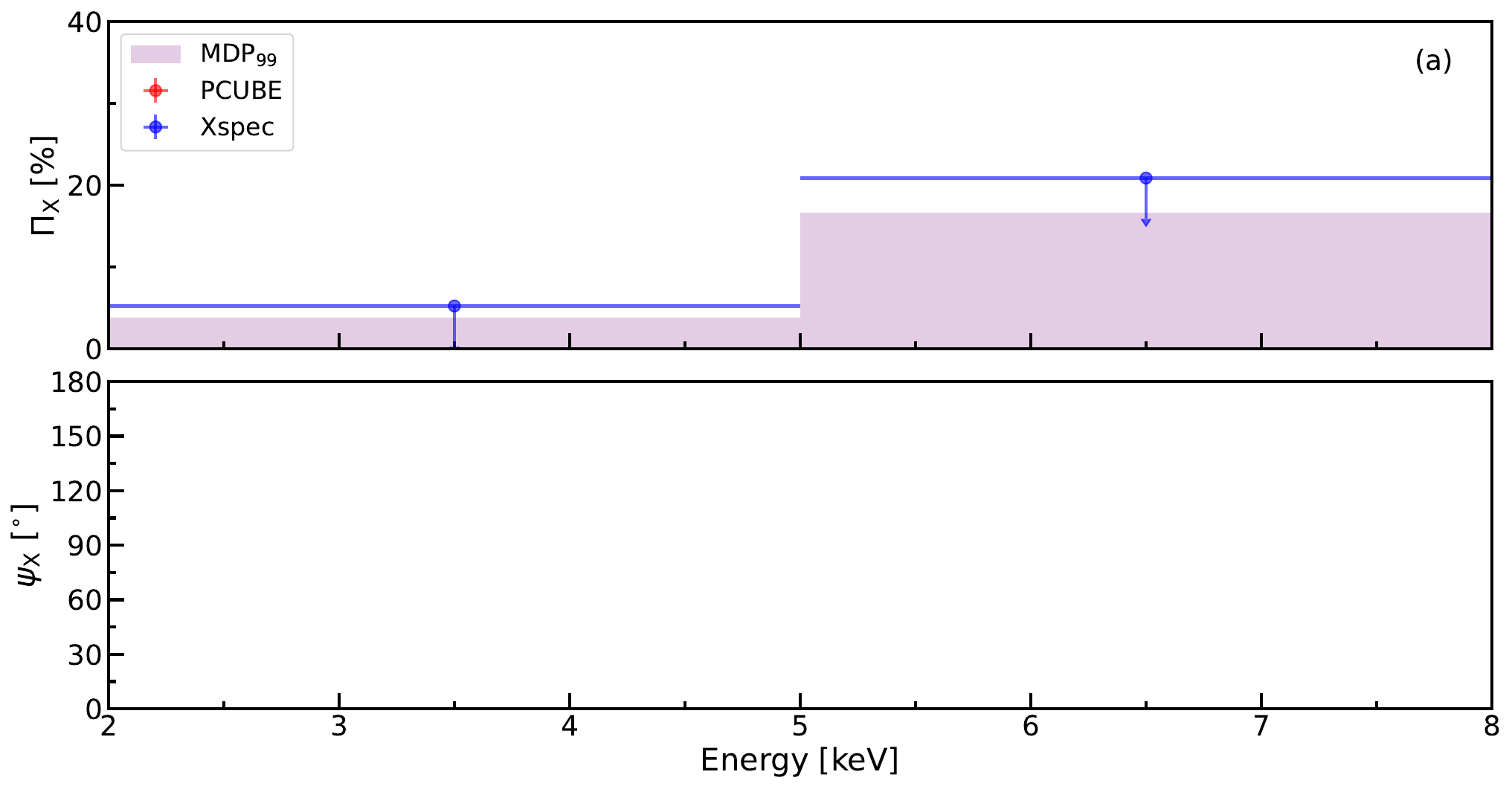}
    \includegraphics[angle=0, scale=0.25]{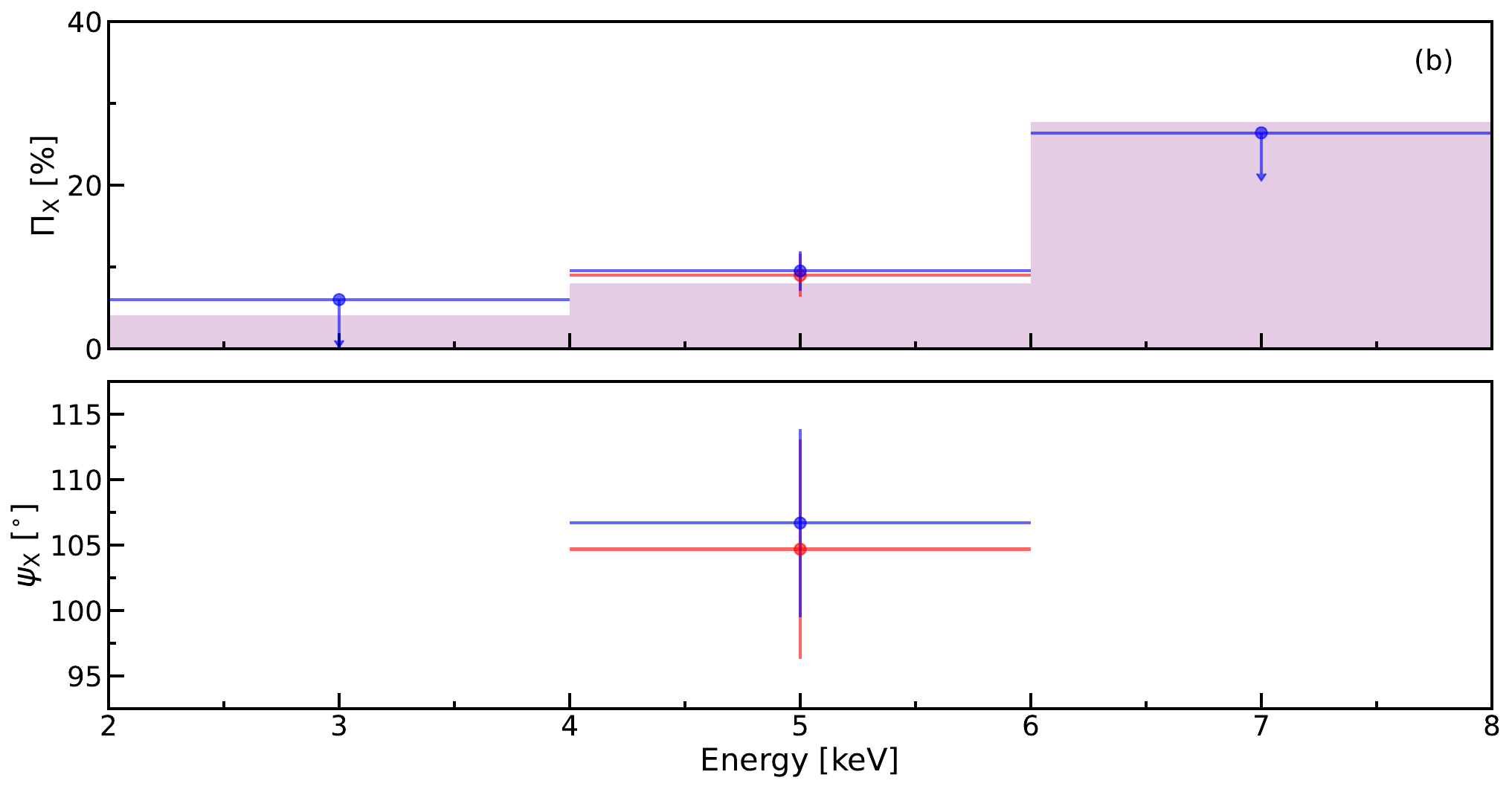}
    \includegraphics[angle=0, scale=0.25]{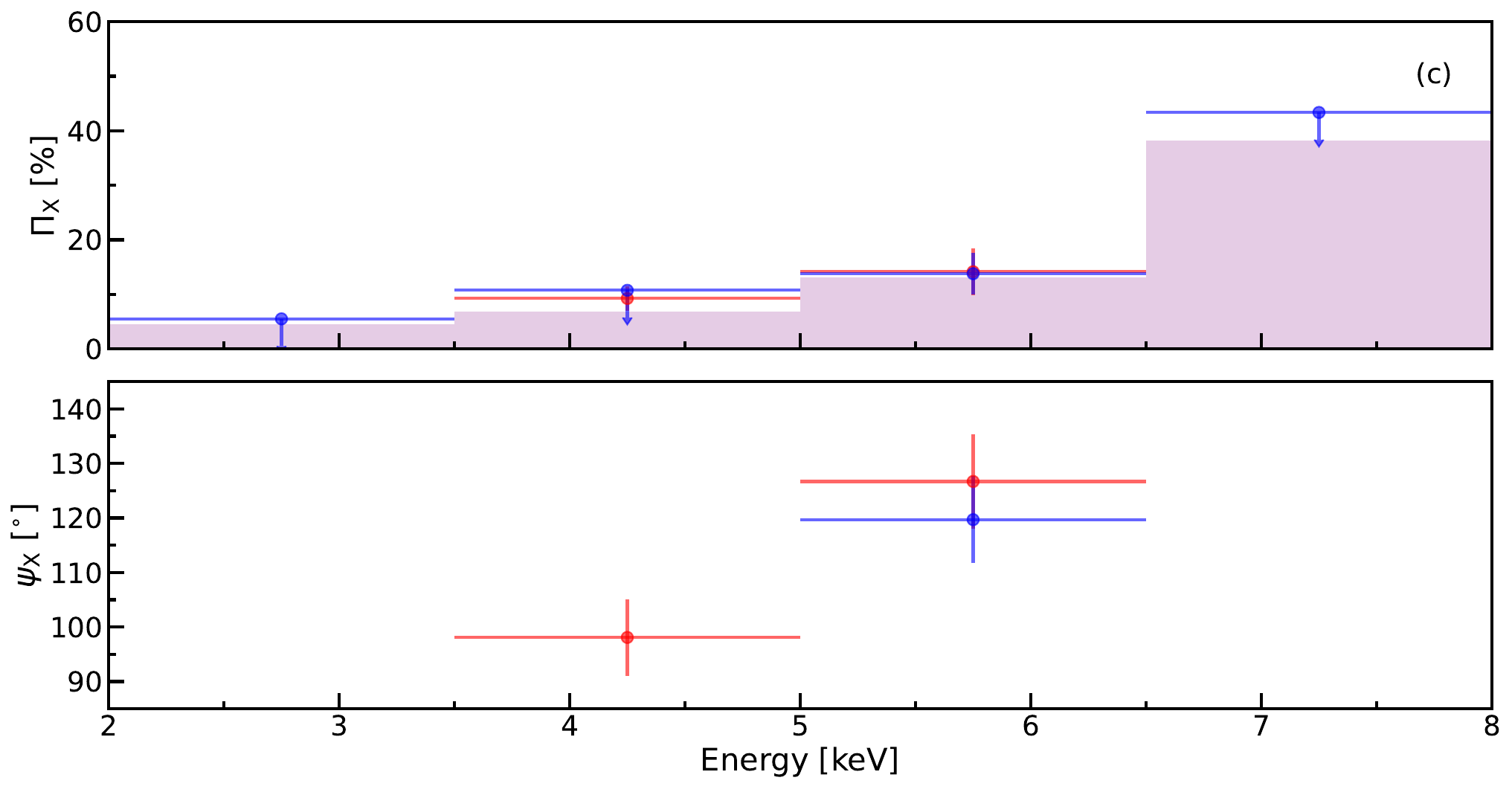}
    \includegraphics[angle=0, scale=0.25]{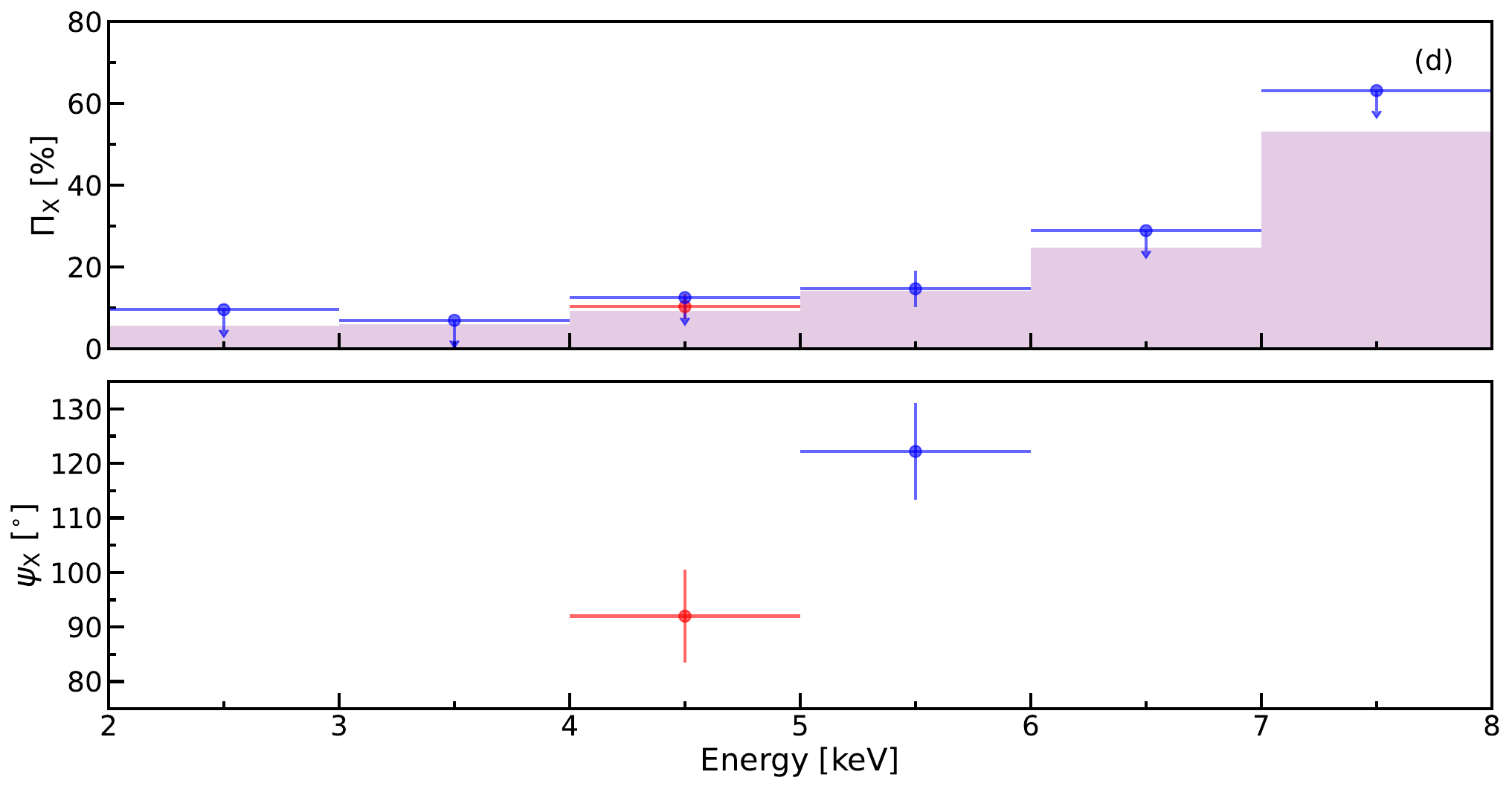}
    \caption{Results of energy-resolved analysis for the polarization of the nucleus in different energy bins: panel (a) for 3 keV bin$^{-1}$, panel (b) for 2 keV bin$^{-1}$, panel (c) for 1.5 keV bin$^{-1}$ and panel (d) for 1 keV bin$^{-1}$, respectively. In each panel, the red points represent the results estimated via \texttt{PCUBE} algorithm in \texttt{ixpeobssim}, the blue points represent the results estimated via spectropolarimetric fits in \texttt{Xspec}, and the violet shaded areas represent the MDP$_{99}$ values for different energy bins. Particularly, if the best-fit value of $\Pi_{\rm X}$ from spectropolarimetric fit is lower than the associated value of MDP$_{99}$, an upper limit at the 99\% confidence level will be provided, marked by blue points with down-arrows.}
    \label{energy-pol}
\end{figure}

\begin{figure}
    \centering \includegraphics[width=0.49\textwidth]{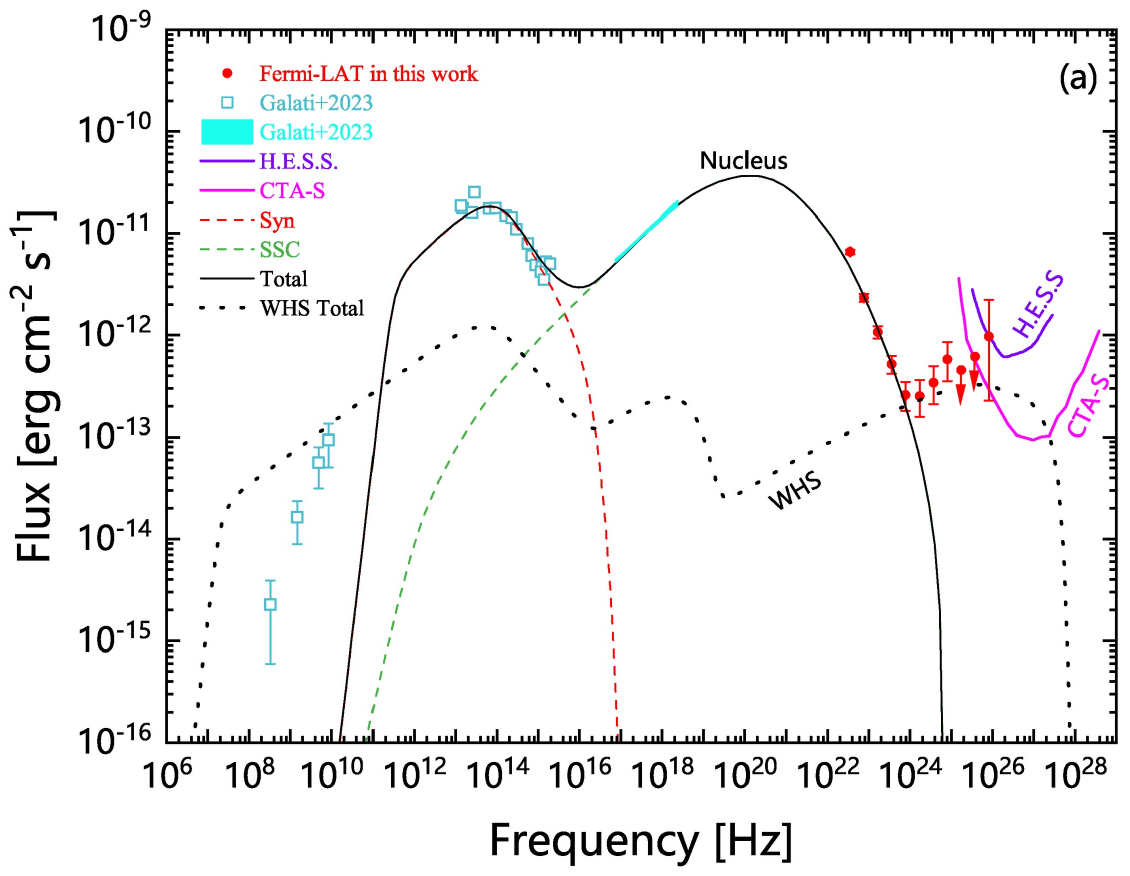} %\hspace{-0.8cm}
\includegraphics[width=0.49\textwidth]{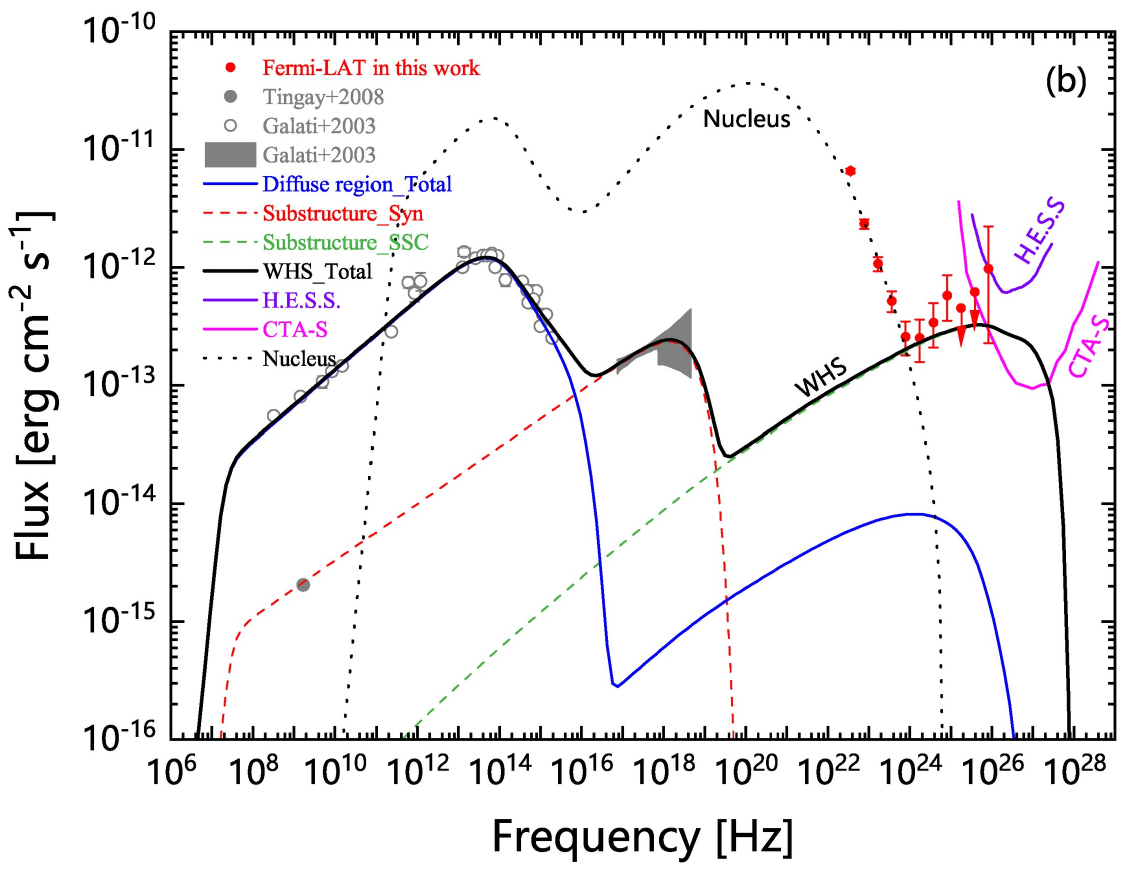}
    \caption{Observed SEDs with model fitting for the nucleus (Panel (a)) and the WHS (Panel (b)) in Pictor A. The data from radio to X-rays for both the nucleus and WHS are taken from \cite{2023MNRAS.521.2704G} and the references therein; for the WHS, including the radio and optical data from \cite{1997A&A...325...57M}, the IR and the substructure data from \cite{2008AJ....136.2473T}, the mid-infrared data from \cite{2017ApJ...850..193I}, two optical/near-ultraviolet points from \cite{2001ApJ...547..740W}, the far-infrared data from \cite{2020ApJ...899...17I}, and the NuSTAR spectrum from \cite{2022PASJ...74..602S}; for the nucleus, including the radio data from \cite{1997A&A...328...12P}, the IR data from \cite{1990MNRAS.246..706S}, and the mid-infrared data from \citep{2010AJ....140.1868W, 2013wise.rept....1C}. It should be noted that this work takes into consideration the X-ray spectrum of the nucleus in a high-flux state, which is different from that in \cite{2023MNRAS.521.2704G}. The Fermi-LAT spectrum is derived from the analysis conducted in this study, same as in Figure \ref{spec_LAT}(a).}
    \label{SED}
\end{figure}

While no significant polarization is detected in the 2--8 keV band for the nucleus, X-ray polarization signals are detected in several narrow energy bins with a confidence level exceeding 99\%, as illustrated in Figure \ref{energy-pol}. As described in Section \ref{sec2.2}, the X-ray spectrum of the nucleus can be adequately explained by a simple PL function. This is also consistent with recent studies employing other X-ray detectors (e.g., \citealt{2022PASJ...74..602S,2023MNRAS.521.2704G}). It is unlikely that this narrow-energy-band polarization arises from superimposed components originating from other regions or radiation processes. Given that the detection significance in these narrow energy bins is $\leq4\sigma$, we propose this to be a tentative detection, which may potentially be attributed to statistical fluctuations.

Numerous studies have suggested that the X-rays emitted by WHS originate from synchrotron radiation within a compact substructure \citep{2001ApJ...547..740W, 2008AJ....136.2473T, 2009ApJ...701..423Z, 2016MNRAS.455.3526H, 2023MNRAS.521.2704G}. Considering the significant polarization observed in both radio and optical bands, with some measurements reaching up to the theoretical maximum of approximately 70\% \citep{1987ApJ...314...70R, 1995ApJ...446L..93T, 1997A&A...328...12P}, it is expected that high polarization detection of WHS in the X-ray band  would provide a valuable tool for investigating radiation mechanisms of X-rays in WHS. Unfortunately, due to the limited number of detected photons in the IXPE observations, we have only been able to establish an upper limit of $\Pi_{\rm X}<56.4\%$ at a 99\% confidence level within the 2--8 keV band, which does not provide sufficient information to further examine the X-ray emission mechanism in WHS.

\begin{acknowledgments}

We sincerely appreciate the referee for the valuable suggestions, which have greatly enhanced the quality of the manuscript. We also appreciate helpful discussion with Fei Xie. This work reports observations obtained with the Imaging X-ray Polarimetry Explorer (IXPE), a joint US (NASA) and Italian (ASI) mission, led by Marshall Space Flight Center (MSFC). The research uses data products provided by the IXPE Science Operations Center (MSFC), using algorithms developed by the IXPE Collaboration, and distributed by the High-Energy Astrophysics Science Archive Research Center (HEASARC). This work is supported by the National Key R\&D Program of China (grant 2023YFE0117200) and the National Natural Science Foundation of China (grants 12203022, 12022305, 11973050, and 12373041).

\end{acknowledgments}

\clearpage 
\bibliography{reference}

\begin{thebibliography}{}
\expandafter\ifx\csname natexlab\endcsname\relax\def\natexlab#1{#1}\fi
\providecommand{\url}[1]{\href{#1}{#1}}
\providecommand{\dodoi}[1]{doi:~\href{http://doi.org/#1}{\nolinkurl{#1}}}
\providecommand{\doeprint}[1]{\href{http://ascl.net/#1}{\nolinkurl{http://ascl.net/#1}}}
\providecommand{\doarXiv}[1]{\href{https://arxiv.org/abs/#1}{\nolinkurl{https://arxiv.org/abs/#1}}}

\bibitem[{{Abdo} {et~al.}(2009){Abdo}, {Ackermann}, {Ajello}, {Atwood}, {Axelsson}, {Baldini}, {Ballet}, {Barbiellini}, {Bastieri}, {Bechtol}, {Bellazzini}, {Berenji}, {Blandford}, {Bloom}, {Bonamente}, {Borgland}, {Bregeon}, {Brez}, {Brigida}, {Bruel}, {Burnett}, {Caliandro}, {Cameron}, {Cannon}, {Caraveo}, {Casandjian}, {Cavazzuti}, {Cecchi}, {{\c{C}}elik}, {Charles}, {Cheung}, {Chiang}, {Ciprini}, {Claus}, {Cohen-Tanugi}, {Colafrancesco}, {Conrad}, {Costamante}, {Cutini}, {Davis}, {Dermer}, {de Angelis}, {de Palma}, {Digel}, {Donato}, {Silva}, {Drell}, {Dubois}, {Dumora}, {Edmonds}, {Farnier}, {Favuzzi}, {Fegan}, {Finke}, {Focke}, {Fortin}, {Frailis}, {Fukazawa}, {Funk}, {Fusco}, {Gargano}, {Gasparrini}, {Gehrels}, {Georganopoulos}, {Germani}, {Giebels}, {Giglietto}, {Giommi}, {Giordano}, {Giroletti}, {Glanzman}, {Godfrey}, {Grenier}, {Grondin}, {Grove}, {Guillemot}, {Guiriec}, {Hanabata}, {Harding}, {Hayashida}, {Hays}, {Horan}, {J{\'o}hannesson}, {Johnson}, {Johnson}, {Johnson}, {Johnson}, {Kamae},
  {Katagiri}, {Kataoka}, {Kawai}, {Kerr}, {Kn{\"o}dlseder}, {Kocian}, {Kuss}, {Lande}, {Latronico}, {Lemoine-Goumard}, {Longo}, {Loparco}, {Lott}, {Lovellette}, {Lubrano}, {Madejski}, {Makeev}, {Mazziotta}, {McConville}, {McEnery}, {Meurer}, {Michelson}, {Mitthumsiri}, {Mizuno}, {Moiseev}, {Monte}, {Monzani}, {Morselli}, {Moskalenko}, {Murgia}, {Nolan}, {Norris}, {Nuss}, {Ohsugi}, {Omodei}, {Orlando}, {Ormes}, {Ozaki}, {Paneque}, {Panetta}, {Parent}, {Pelassa}, {Pepe}, {Pesce-Rollins}, {Piron}, {Porter}, {Rain{\`o}}, {Rando}, {Razzano}, {Reimer}, {Reimer}, {Reposeur}, {Ritz}, {Rochester}, {Rodriguez}, {Romani}, {Roth}, {Ryde}, {Sadrozinski}, {Sambruna}, {Sanchez}, {Sander}, {Saz Parkinson}, {Scargle}, {Sgr{\`o}}, {Shaw}, {Smith}, {Smith}, {Spandre}, {Spinelli}, {Strickman}, {Suson}, {Tajima}, {Takahashi}, {Tanaka}, {Taylor}, {Thayer}, {Thompson}, {Tibaldo}, {Torres}, {Tosti}, {Tramacere}, {Uchiyama}, {Usher}, {Vasileiou}, {Vilchez}, {Waite}, {Wang}, {Winer}, {Wood}, {Ylinen}, {Ziegler}, {Harris}, {Massaro},
  \& {Stawarz}}]{2009ApJ...707...55A}
{Abdo}, A.~A., {Ackermann}, M., {Ajello}, M., {et~al.} 2009, \apj, 707, 55, \dodoi{10.1088/0004-637X/707/1/55}

\bibitem[{{Abdo} {et~al.}(2010){Abdo}, {Ackermann}, {Ajello}, {Atwood}, {Baldini}, {Ballet}, {Barbiellini}, {Bastieri}, {Baughman}, {Bechtol}, {Bellazzini}, {Berenji}, {Blandford}, {Bloom}, {Bonamente}, {Borgland}, {Bouvier}, {Brandt}, {Bregeon}, {Brez}, {Brigida}, {Bruel}, {Buehler}, {Buson}, {Caliandro}, {Cameron}, {Cannon}, {Caraveo}, {Carrigan}, {Casandjian}, {Cavazzuti}, {Cecchi}, {{\c{C}}elik}, {Charles}, {Chekhtman}, {Cheung}, {Chiang}, {Ciprini}, {Claus}, {Cohen-Tanugi}, {Colafrancesco}, {Cominsky}, {Conrad}, {Costamante}, {Davis}, {Dermer}, {de Angelis}, {de Palma}, {Silva}, {Drell}, {Dubois}, {Dumora}, {Falcone}, {Farnier}, {Favuzzi}, {Fegan}, {Finke}, {Focke}, {Fortin}, {Frailis}, {Fukazawa}, {Funk}, {Fusco}, {Gargano}, {Gasparrini}, {Gehrels}, {Georganopoulos}, {Germani}, {Giebels}, {Giglietto}, {Giommi}, {Giordano}, {Giroletti}, {Glanzman}, {Godfrey}, {Grandi}, {Grenier}, {Grondin}, {Grove}, {Guillemot}, {Guiriec}, {Hadasch}, {Harding}, {Hase}, {Hayashida}, {Hays}, {Horan}, {Hughes}, {Itoh},
  {Jackson}, {J{\'o}hannesson}, {Johnson}, {Johnson}, {Johnson}, {Kadler}, {Kamae}, {Katagiri}, {Kataoka}, {Kawai}, {Kishishita}, {Kn{\"o}dlseder}, {Kuss}, {Lande}, {Latronico}, {Lee}, {Lemoine-Goumard}, {Llena Garde}, {Longo}, {Loparco}, {Lott}, {Lovellette}, {Lubrano}, {Makeev}, {Mazziotta}, {McConville}, {McEnery}, {Michelson}, {Mitthumsiri}, {Mizuno}, {Moiseev}, {Monte}, {Monzani}, {Morselli}, {Moskalenko}, {Murgia}, {M{\"u}ller}, {Nakamori}, {Naumann-Godo}, {Nolan}, {Norris}, {Nuss}, {Ohno}, {Ohsugi}, {Ojha}, {Okumura}, {Omodei}, {Orlando}, {Ormes}, {Ozaki}, {Pagani}, {Paneque}, {Panetta}, {Parent}, {Pelassa}, {Pepe}, {Pesce-Rollins}, {Piron}, {Pl{\"o}tz}, {Porter}, {Rain{\`o}}, {Rando}, {Razzano}, {Razzaque}, {Reimer}, {Reimer}, {Reposeur}, {Ripken}, {Ritz}, {Rodriguez}, {Roth}, {Ryde}, {Sadrozinski}, {Sanchez}, {Sander}, {Scargle}, {Sgr{\`o}}, {Siskind}, {Smith}, {Spandre}, {Spinelli}, {Starck}, {Stawarz}, {Strickman}, {Suson}, {Tajima}, {Takahashi}, {Takahashi}, {Tanaka}, {Thayer}, {Thayer},
  {Thompson}, {Tibaldo}, {Torres}, {Tosti}, {Tramacere}, {Uchiyama}, {Usher}, {Vandenbroucke}, {Vasileiou}, {Vilchez}, {Vitale}, {Waite}, {Wang}, {Winer}, {Wood}, {Yang}, {Ylinen}, \& {Ziegler}}]{2010ApJ...719.1433A}
---. 2010, \apj, 719, 1433, \dodoi{10.1088/0004-637X/719/2/1433}

\bibitem[{{Abdollahi} {et~al.}(2020){Abdollahi}, {Acero}, {Ackermann}, {Ajello}, {Atwood}, {Axelsson}, {Baldini}, {Ballet}, {Barbiellini}, {Bastieri}, {Becerra Gonzalez}, {Bellazzini}, {Berretta}, {Bissaldi}, {Blandford}, {Bloom}, {Bonino}, {Bottacini}, {Brandt}, {Bregeon}, {Bruel}, {Buehler}, {Burnett}, {Buson}, {Cameron}, {Caputo}, {Caraveo}, {Casandjian}, {Castro}, {Cavazzuti}, {Charles}, {Chaty}, {Chen}, {Cheung}, {Chiaro}, {Ciprini}, {Cohen-Tanugi}, {Cominsky}, {Coronado-Bl{\'a}zquez}, {Costantin}, {Cuoco}, {Cutini}, {D'Ammando}, {DeKlotz}, {de la Torre Luque}, {de Palma}, {Desai}, {Digel}, {Di Lalla}, {Di Mauro}, {Di Venere}, {Dom{\'\i}nguez}, {Dumora}, {Fana Dirirsa}, {Fegan}, {Ferrara}, {Franckowiak}, {Fukazawa}, {Funk}, {Fusco}, {Gargano}, {Gasparrini}, {Giglietto}, {Giommi}, {Giordano}, {Giroletti}, {Glanzman}, {Green}, {Grenier}, {Griffin}, {Grondin}, {Grove}, {Guiriec}, {Harding}, {Hayashi}, {Hays}, {Hewitt}, {Horan}, {J{\'o}hannesson}, {Johnson}, {Kamae}, {Kerr}, {Kocevski}, {Kovac'evic'},
  {Kuss}, {Landriu}, {Larsson}, {Latronico}, {Lemoine-Goumard}, {Li}, {Liodakis}, {Longo}, {Loparco}, {Lott}, {Lovellette}, {Lubrano}, {Madejski}, {Maldera}, {Malyshev}, {Manfreda}, {Marchesini}, {Marcotulli}, {Mart{\'\i}-Devesa}, {Martin}, {Massaro}, {Mazziotta}, {McEnery}, {Mereu}, {Meyer}, {Michelson}, {Mirabal}, {Mizuno}, {Monzani}, {Morselli}, {Moskalenko}, {Negro}, {Nuss}, {Ojha}, {Omodei}, {Orienti}, {Orlando}, {Ormes}, {Palatiello}, {Paliya}, {Paneque}, {Pei}, {Pe{\~n}a-Herazo}, {Perkins}, {Persic}, {Pesce-Rollins}, {Petrosian}, {Petrov}, {Piron}, {Poon}, {Porter}, {Principe}, {Rain{\`o}}, {Rando}, {Razzano}, {Razzaque}, {Reimer}, {Reimer}, {Remy}, {Reposeur}, {Romani}, {Saz Parkinson}, {Schinzel}, {Serini}, {Sgr{\`o}}, {Siskind}, {Smith}, {Spandre}, {Spinelli}, {Strong}, {Suson}, {Tajima}, {Takahashi}, {Tak}, {Thayer}, {Thompson}, {Tibaldo}, {Torres}, {Torresi}, {Valverde}, {Van Klaveren}, {van Zyl}, {Wood}, {Yassine}, \& {Zaharijas}}]{2020ApJS..247...33A}
{Abdollahi}, S., {Acero}, F., {Ackermann}, M., {et~al.} 2020, \apjs, 247, 33, \dodoi{10.3847/1538-4365/ab6bcb}

\bibitem[{{Abdollahi} {et~al.}(2022){Abdollahi}, {Acero}, {Baldini}, {Ballet}, {Bastieri}, {Bellazzini}, {Berenji}, {Berretta}, {Bissaldi}, {Blandford}, {Bloom}, {Bonino}, {Brill}, {Britto}, {Bruel}, {Burnett}, {Buson}, {Cameron}, {Caputo}, {Caraveo}, {Castro}, {Chaty}, {Cheung}, {Chiaro}, {Cibrario}, {Ciprini}, {Coronado-Bl{\'a}zquez}, {Crnogorcevic}, {Cutini}, {D'Ammando}, {De Gaetano}, {Digel}, {Di Lalla}, {Dirirsa}, {Di Venere}, {Dom{\'\i}nguez}, {Fallah Ramazani}, {Fegan}, {Ferrara}, {Fiori}, {Fleischhack}, {Franckowiak}, {Fukazawa}, {Funk}, {Fusco}, {Galanti}, {Gammaldi}, {Gargano}, {Garrappa}, {Gasparrini}, {Giacchino}, {Giglietto}, {Giordano}, {Giroletti}, {Glanzman}, {Green}, {Grenier}, {Grondin}, {Guillemot}, {Guiriec}, {Gustafsson}, {Harding}, {Hays}, {Hewitt}, {Horan}, {Hou}, {J{\'o}hannesson}, {Karwin}, {Kayanoki}, {Kerr}, {Kuss}, {Landriu}, {Larsson}, {Latronico}, {Lemoine-Goumard}, {Li}, {Liodakis}, {Longo}, {Loparco}, {Lott}, {Lubrano}, {Maldera}, {Malyshev}, {Manfreda}, {Mart{\'\i}-Devesa},
  {Mazziotta}, {Mereu}, {Meyer}, {Michelson}, {Mirabal}, {Mitthumsiri}, {Mizuno}, {Moiseev}, {Monzani}, {Morselli}, {Moskalenko}, {Negro}, {Nuss}, {Omodei}, {Orienti}, {Orlando}, {Paneque}, {Pei}, {Perkins}, {Persic}, {Pesce-Rollins}, {Petrosian}, {Pillera}, {Poon}, {Porter}, {Principe}, {Rain{\`o}}, {Rando}, {Rani}, {Razzano}, {Razzaque}, {Reimer}, {Reimer}, {Reposeur}, {S{\'a}nchez-Conde}, {Saz Parkinson}, {Scotton}, {Serini}, {Sgr{\`o}}, {Siskind}, {Smith}, {Spandre}, {Spinelli}, {Sueoka}, {Suson}, {Tajima}, {Tak}, {Thayer}, {Thompson}, {Torres}, {Troja}, {Valverde}, {Wood}, \& {Zaharijas}}]{2022ApJS..260...53A}
{Abdollahi}, S., {Acero}, F., {Baldini}, L., {et~al.} 2022, \apjs, 260, 53, \dodoi{10.3847/1538-4365/ac6751}

\bibitem[{{Ackermann} {et~al.}(2016){Ackermann}, {Ajello}, {Baldini}, {Ballet}, {Barbiellini}, {Bastieri}, {Bellazzini}, {Bissaldi}, {Blandford}, {Bloom}, {Bonino}, {Brandt}, {Bregeon}, {Bruel}, {Buehler}, {Buson}, {Caliandro}, {Cameron}, {Caragiulo}, {Caraveo}, {Cavazzuti}, {Cecchi}, {Charles}, {Chekhtman}, {Cheung}, {Chiaro}, {Ciprini}, {Cohen}, {Cohen-Tanugi}, {Costanza}, {Cutini}, {D'Ammando}, {Davis}, {de Angelis}, {de Palma}, {Desiante}, {Digel}, {Di Lalla}, {Di Mauro}, {Di Venere}, {Favuzzi}, {Fegan}, {Ferrara}, {Focke}, {Fukazawa}, {Funk}, {Fusco}, {Gargano}, {Gasparrini}, {Georganopoulos}, {Giglietto}, {Giordano}, {Giroletti}, {Godfrey}, {Green}, {Grenier}, {Guiriec}, {Hays}, {Hewitt}, {Hill}, {Jogler}, {J{\'o}hannesson}, {Kensei}, {Kuss}, {Larsson}, {Latronico}, {Li}, {Li}, {Longo}, {Loparco}, {Lubrano}, {Magill}, {Maldera}, {Manfreda}, {Mayer}, {Mazziotta}, {McConville}, {McEnery}, {Michelson}, {Mitthumsiri}, {Mizuno}, {Monzani}, {Morselli}, {Moskalenko}, {Murgia}, {Negro}, {Nuss}, {Ohno},
  {Ohsugi}, {Orienti}, {Orlando}, {Ormes}, {Paneque}, {Perkins}, {Pesce-Rollins}, {Piron}, {Pivato}, {Porter}, {Rain{\`o}}, {Rando}, {Razzano}, {Reimer}, {Reimer}, {Schmid}, {Sgr{\`o}}, {Simone}, {Siskind}, {Spada}, {Spandre}, {Spinelli}, {Stawarz}, {Takahashi}, {Thayer}, {Thompson}, {Torres}, {Tosti}, {Troja}, {Vianello}, {Wood}, {Wood}, {Zimmer}, \& {Fermi LAT Collaboration}}]{2016ApJ...826....1A}
{Ackermann}, M., {Ajello}, M., {Baldini}, L., {et~al.} 2016, \apj, 826, 1, \dodoi{10.3847/0004-637X/826/1/1}

\bibitem[{{Ait Benkhali} {et~al.}(2019){Ait Benkhali}, {Chakraborty}, \& {Rieger}}]{2019A&A...623A...2A}
{Ait Benkhali}, F., {Chakraborty}, N., \& {Rieger}, F.~M. 2019, \aap, 623, A2, \dodoi{10.1051/0004-6361/201732334}

\bibitem[{{Angioni} {et~al.}(2019){Angioni}, {Ros}, {Kadler}, {Ojha}, {M{\"u}ller}, {Edwards}, {Burd}, {Carpenter}, {Dutka}, {Gulyaev}, {Hase}, {Horiuchi}, {Krau{\ss}}, {Lovell}, {Natusch}, {Phillips}, {Pl{\"o}tz}, {Quick}, {R{\"o}sch}, {Schulz}, {Stevens}, {Tzioumis}, {Weston}, {Wilms}, \& {Zensus}}]{2019A&A...627A.148A}
{Angioni}, R., {Ros}, E., {Kadler}, M., {et~al.} 2019, \aap, 627, A148, \dodoi{10.1051/0004-6361/201935697}

\bibitem[{{Arnaud} {et~al.}(1999){Arnaud}, {Dorman}, \& {Gordon}}]{1999ascl.soft10005A}
{Arnaud}, K., {Dorman}, B., \& {Gordon}, C. 1999, {XSPEC: An X-ray spectral fitting package}, Astrophysics Source Code Library, record ascl:9910.005.
\newblock \doeprint{9910.005}

\bibitem[{{Atwood} {et~al.}(2009){Atwood}, {Abdo}, {Ackermann}, {Althouse}, {Anderson}, {Axelsson}, {Baldini}, {Ballet}, {Band}, {Barbiellini}, {Bartelt}, {Bastieri}, {Baughman}, {Bechtol}, {B{\'e}d{\'e}r{\`e}de}, {Bellardi}, {Bellazzini}, {Berenji}, {Bignami}, {Bisello}, {Bissaldi}, {Blandford}, {Bloom}, {Bogart}, {Bonamente}, {Bonnell}, {Borgland}, {Bouvier}, {Bregeon}, {Brez}, {Brigida}, {Bruel}, {Burnett}, {Busetto}, {Caliandro}, {Cameron}, {Caraveo}, {Carius}, {Carlson}, {Casandjian}, {Cavazzuti}, {Ceccanti}, {Cecchi}, {Charles}, {Chekhtman}, {Cheung}, {Chiang}, {Chipaux}, {Cillis}, {Ciprini}, {Claus}, {Cohen-Tanugi}, {Condamoor}, {Conrad}, {Corbet}, {Corucci}, {Costamante}, {Cutini}, {Davis}, {Decotigny}, {DeKlotz}, {Dermer}, {de Angelis}, {Digel}, {do Couto e Silva}, {Drell}, {Dubois}, {Dumora}, {Edmonds}, {Fabiani}, {Farnier}, {Favuzzi}, {Flath}, {Fleury}, {Focke}, {Funk}, {Fusco}, {Gargano}, {Gasparrini}, {Gehrels}, {Gentit}, {Germani}, {Giebels}, {Giglietto}, {Giommi}, {Giordano}, {Glanzman},
  {Godfrey}, {Grenier}, {Grondin}, {Grove}, {Guillemot}, {Guiriec}, {Haller}, {Harding}, {Hart}, {Hays}, {Healey}, {Hirayama}, {Hjalmarsdotter}, {Horn}, {Hughes}, {J{\'o}hannesson}, {Johansson}, {Johnson}, {Johnson}, {Johnson}, {Johnson}, {Kamae}, {Katagiri}, {Kataoka}, {Kavelaars}, {Kawai}, {Kelly}, {Kerr}, {Klamra}, {Kn{\"o}dlseder}, {Kocian}, {Komin}, {Kuehn}, {Kuss}, {Landriu}, {Latronico}, {Lee}, {Lee}, {Lemoine-Goumard}, {Lionetto}, {Longo}, {Loparco}, {Lott}, {Lovellette}, {Lubrano}, {Madejski}, {Makeev}, {Marangelli}, {Massai}, {Mazziotta}, {McEnery}, {Menon}, {Meurer}, {Michelson}, {Minuti}, {Mirizzi}, {Mitthumsiri}, {Mizuno}, {Moiseev}, {Monte}, {Monzani}, {Moretti}, {Morselli}, {Moskalenko}, {Murgia}, {Nakamori}, {Nishino}, {Nolan}, {Norris}, {Nuss}, {Ohno}, {Ohsugi}, {Omodei}, {Orlando}, {Ormes}, {Paccagnella}, {Paneque}, {Panetta}, {Parent}, {Pearce}, {Pepe}, {Perazzo}, {Pesce-Rollins}, {Picozza}, {Pieri}, {Pinchera}, {Piron}, {Porter}, {Poupard}, {Rain{\`o}}, {Rando}, {Rapposelli}, {Razzano},
  {Reimer}, {Reimer}, {Reposeur}, {Reyes}, {Ritz}, {Rochester}, {Rodriguez}, {Romani}, {Roth}, {Russell}, {Ryde}, {Sabatini}, {Sadrozinski}, {Sanchez}, {Sander}, {Sapozhnikov}, {Parkinson}, {Scargle}, {Schalk}, \& {Scolieri}}]{2009ApJ...697.1071A}
{Atwood}, W.~B., {Abdo}, A.~A., {Ackermann}, M., {et~al.} 2009, \apj, 697, 1071, \dodoi{10.1088/0004-637X/697/2/1071}

\bibitem[{{Baldini} {et~al.}(2022){Baldini}, {Bucciantini}, {Lalla}, {Ehlert}, {Manfreda}, {Negro}, {Omodei}, {Pesce-Rollins}, {Sgr{\`o}}, \& {Silvestri}}]{2022SoftX..1901194B}
{Baldini}, L., {Bucciantini}, N., {Lalla}, N.~D., {et~al.} 2022, SoftwareX, 19, 101194, \dodoi{10.1016/j.softx.2022.101194}

\bibitem[{{Ballet} {et~al.}(2023){Ballet}, {Bruel}, {Burnett}, {Lott}, \& {The Fermi-LAT collaboration}}]{2023arXiv230712546B}
{Ballet}, J., {Bruel}, P., {Burnett}, T.~H., {Lott}, B., \& {The Fermi-LAT collaboration}. 2023, arXiv e-prints, arXiv:2307.12546, \dodoi{10.48550/arXiv.2307.12546}

\bibitem[{{Brown} \& {Adams}(2012)}]{2012MNRAS.421.2303B}
{Brown}, A.~M., \& {Adams}, J. 2012, \mnras, 421, 2303, \dodoi{10.1111/j.1365-2966.2012.20451.x}

\bibitem[{{Brown} {et~al.}(2017){Brown}, {B{\r{A}}`hm}, {Graham}, {Lacroix}, {Chadwick}, \& {Silk}}]{2017PhRvD..95f3018B}
{Brown}, A.~M., {B{\r{A}}`hm}, C., {Graham}, J., {et~al.} 2017, \prd, 95, 063018, \dodoi{10.1103/PhysRevD.95.063018}

\bibitem[{{Cheng} {et~al.}(2022){Cheng}, {Huang}, {Wang}, {Huang}, \& {Liang}}]{2022ApJ...925L..19C}
{Cheng}, J.-G., {Huang}, X.-L., {Wang}, Z.-R., {Huang}, J.-K., \& {Liang}, E.-W. 2022, \apjl, 925, L19, \dodoi{10.3847/2041-8213/ac4d8e}

\bibitem[{{Cutri} {et~al.}(2013){Cutri}, {Wright}, {Conrow}, {Fowler}, {Eisenhardt}, {Grillmair}, {Kirkpatrick}, {Masci}, {McCallon}, {Wheelock}, {Fajardo-Acosta}, {Yan}, {Benford}, {Harbut}, {Jarrett}, {Lake}, {Leisawitz}, {Ressler}, {Stanford}, {Tsai}, {Liu}, {Helou}, {Mainzer}, {Gettings}, {Gonzalez}, {Hoffman}, {Marsh}, {Padgett}, {Skrutskie}, {Beck}, {Papin}, \& {Wittman}}]{2013wise.rept....1C}
{Cutri}, R.~M., {Wright}, E.~L., {Conrow}, T., {et~al.} 2013, {Explanatory Supplement to the AllWISE Data Release Products}, Explanatory Supplement to the AllWISE Data Release Products, by R. M. Cutri et al.

\bibitem[{{Di Gesu} {et~al.}(2022){Di Gesu}, {Donnarumma}, {Tavecchio}, {Agudo}, {Barnounin}, {Cibrario}, {Di Lalla}, {Di Marco}, {Escudero}, {Errando}, {Jorstad}, {Kim}, {Kouch}, {Liodakis}, {Lindfors}, {Madejski}, {Marshall}, {Marscher}, {Middei}, {Muleri}, {Myserlis}, {Negro}, {Omodei}, {Pacciani}, {Paggi}, {Perri}, {Puccetti}, {Antonelli}, {Bachetti}, {Baldini}, {Baumgartner}, {Bellazzini}, {Bianchi}, {Bongiorno}, {Bonino}, {Brez}, {Bucciantini}, {Capitanio}, {Castellano}, {Cavazzuti}, {Ciprini}, {Costa}, {De Rosa}, {Del Monte}, {Doroshenko}, {Dov{\v{c}}iak}, {Ehlert}, {Enoto}, {Evangelista}, {Fabiani}, {Ferrazzoli}, {Garcia}, {Gunji}, {Hayashida}, {Heyl}, {Iwakiri}, {Karas}, {Kitaguchi}, {Kolodziejczak}, {Krawczynski}, {La Monaca}, {Latronico}, {Maldera}, {Manfreda}, {Marin}, {Marinucci}, {Massaro}, {Matt}, {Mitsuishi}, {Mizuno}, {Ng}, {O'Dell}, {Oppedisano}, {Papitto}, {Pavlov}, {Peirson}, {Pesce-Rollins}, {Petrucci}, {Pilia}, {Possenti}, {Poutanen}, {Ramsey}, {Rankin}, {Ratheesh}, {Romani}, {Sgr{\`o}},
  {Slane}, {Soffitta}, {Spandre}, {Tamagawa}, {Taverna}, {Tawara}, {Tennant}, {Thomas}, {Tombesi}, {Trois}, {Tsygankov}, {Turolla}, {Vink}, {Weisskopf}, {Wu}, {Xie}, \& {Zane}}]{2022ApJ...938L...7D}
{Di Gesu}, L., {Donnarumma}, I., {Tavecchio}, F., {et~al.} 2022, \apjl, 938, L7, \dodoi{10.3847/2041-8213/ac913a}

\bibitem[{{Di Marco} {et~al.}(2023){Di Marco}, {Soffitta}, {Costa}, {Ferrazzoli}, {La Monaca}, {Rankin}, {Ratheesh}, {Xie}, {Baldini}, {Del Monte}, {Ehlert}, {Fabiani}, {Kim}, {Muleri}, {O'Dell}, {Ramsey}, {Rubini}, {Sgr{\`o}}, {Silvestri}, {Tennant}, \& {Weisskopf}}]{2023AJ....165..143D}
{Di Marco}, A., {Soffitta}, P., {Costa}, E., {et~al.} 2023, \aj, 165, 143, \dodoi{10.3847/1538-3881/acba0f}

\bibitem[{{Ehlert} {et~al.}(2022){Ehlert}, {Ferrazzoli}, {Marinucci}, {Marshall}, {Middei}, {Pacciani}, {Perri}, {Petrucci}, {Puccetti}, {Barnouin}, {Bianchi}, {Liodakis}, {Madejski}, {Marin}, {Marscher}, {Matt}, {Poutanen}, {Wu}, {Agudo}, {Antonelli}, {Bachetti}, {Baldini}, {Baumgartner}, {Bellazzini}, {Bongiorno}, {Bonino}, {Brez}, {Bucciantini}, {Capitanio}, {Castellano}, {Cavazzuti}, {Ciprini}, {Costa}, {De Rosa}, {Del Monte}, {Di Gesu}, {Di Lalla}, {Di Marco}, {Donnarumma}, {Doroshenko}, {Dov{\v{c}}iak}, {Enoto}, {Evangelista}, {Fabiani}, {Garcia}, {Gunji}, {Hayashida}, {Heyl}, {Iwakiri}, {Jorstad}, {Karas}, {Kitaguchi}, {Kolodziejczak}, {Krawczynski}, {La Monaca}, {Latronico}, {Maldera}, {Manfreda}, {Massaro}, {Mitsuishi}, {Mizuno}, {Muleri}, {Negro}, {Ng}, {O'Dell}, {Omodei}, {Oppedisano}, {Papitto}, {Pavlov}, {Peirson}, {Pesce-Rollins}, {Pilia}, {Possenti}, {Ramsey}, {Rankin}, {Ratheesh}, {Romani}, {Sgr{\`o}}, {Slane}, {Soffitta}, {Spandre}, {Tamagawa}, {Tavecchio}, {Taverna}, {Tawara}, {Tennant},
  {Thomas}, {Tombesi}, {Trois}, {Tsygankov}, {Turolla}, {Vink}, {Weisskopf}, {Xie}, {Zane}, {IXPE Collaboration}, {Rodi}, {Jourdain}, \& {Roques}}]{2022ApJ...935..116E}
{Ehlert}, S.~R., {Ferrazzoli}, R., {Marinucci}, A., {et~al.} 2022, \apj, 935, 116, \dodoi{10.3847/1538-4357/ac8056}

\bibitem[{{Ehlert} {et~al.}(2023){Ehlert}, {Liodakis}, {Middei}, {Marscher}, {Tavecchio}, {Agudo}, {Kouch}, {Lindfors}, {Nilsson}, {Myserlis}, {Gurwell}, {Rao}, {Aceituno}, {Bonnoli}, {Casanova}, {Ag{\'\i}s-Gonz{\'a}lez}, {Escudero}, {Husillos}, {Otero Santos}, {Sota}, {Angelakis}, {Kraus}, {Keating}, {Antonelli}, {Bachetti}, {Baldini}, {Baumgartner}, {Bellazzini}, {Bianchi}, {Bongiorno}, {Bonino}, {Brez}, {Bucciantini}, {Capitanio}, {Castellano}, {Cavazzuti}, {Chen}, {Ciprini}, {Costa}, {De Rosa}, {Del Monte}, {Di Gesu}, {Di Lalla}, {Di Marco}, {Donnarumma}, {Doroshenko}, {Dov{\v{c}}iak}, {Enoto}, {Evangelista}, {Fabiani}, {Ferrazzoli}, {Garcia}, {Gunji}, {Hayashida}, {Heyl}, {Iwakiri}, {Jorstad}, {Kaaret}, {Karas}, {Kislat}, {Kitaguchi}, {Kolodziejczak}, {Krawczynski}, {La Monaca}, {Latronico}, {Maldera}, {Manfreda}, {Marin}, {Marinucci}, {Marshall}, {Massaro}, {Matt}, {Mitsuishi}, {Mizuno}, {Muleri}, {Negro}, {Ng}, {O'Dell}, {Omodei}, {Oppedisano}, {Papitto}, {Pavlov}, {Peirson}, {Perri}, {Pesce-Rollins},
  {Petrucci}, {Pilia}, {Possenti}, {Poutanen}, {Puccetti}, {Ramsey}, {Rankin}, {Ratheesh}, {Roberts}, {Romani}, {Sgr{\'o}}, {Slane}, {Soffitta}, {Spandre}, {Swartz}, {Tamagawa}, {Taverna}, {Tawara}, {Tennant}, {Thomas}, {Tombesi}, {Trois}, {Tsygankov}, {Turolla}, {Vink}, {Weisskopf}, {Wu}, {Xie}, \& {Zane}}]{2023ApJ...959...61E}
{Ehlert}, S.~R., {Liodakis}, I., {Middei}, R., {et~al.} 2023, \apj, 959, 61, \dodoi{10.3847/1538-4357/ad05c4}

\bibitem[{{Eracleous} \& {Halpern}(1998)}]{1998ApJ...505..577E}
{Eracleous}, M., \& {Halpern}, J.~P. 1998, \apj, 505, 577, \dodoi{10.1086/306190}

\bibitem[{{Errando} {et~al.}(2024){Errando}, {Liodakis}, {Marscher}, {Marshall}, {Middei}, {Negro}, {Peirson}, {Perri}, {Puccetti}, {Rabinowitz}, {Agudo}, {Jorstad}, {Savchenko}, {Blinov}, {Bourbah}, {Kiehlmann}, {Kontopodis}, {Mandarakas}, {Romanopoulos}, {Skalidis}, {Vervelaki}, {Aceituno}, {Bernardos}, {Bonnoli}, {Casanova}, {Ag{\'\i}s-Gonz{\'a}lez}, {Husillos}, {Marchini}, {Sota}, {Kouch}, {Lindfors}, {Casadio}, {Escudero}, {Myserlis}, {Imazawa}, {Sasada}, {Fukazawa}, {Kawabata}, {Uemura}, {Mizuno}, {Nakaoka}, {Akitaya}, {Gurwell}, {Keating}, {Rao}, {Ingram}, {Massaro}, {Antonelli}, {Bonino}, {Cavazzuti}, {Chen}, {Cibrario}, {Ciprini}, {De Rosa}, {Di Gesu}, {Di Pierro}, {Donnarumma}, {Ehlert}, {Fenu}, {Gau}, {Karas}, {Kim}, {Krawczynski}, {Laurenti}, {Lisalda}, {L{\'o}pez-Coto}, {Madejski}, {Marin}, {Marinucci}, {Mitsuishi}, {Muleri}, {Pacciani}, {Paggi}, {Petrucci}, {Rodriguez Cavero}, {Romani}, {Tavecchio}, {Tugliani}, {Wu}, {Bachetti}, {Baldini}, {Baumgartner}, {Bellazzini}, {Bianchi}, {Bongiorno},
  {Brez}, {Bucciantini}, {Capitanio}, {Castellano}, {Costa}, {Del Monte}, {Di Lalla}, {Di Marco}, {Doroshenko}, {Dov{\v{c}}iak}, {Enoto}, {Evangelista}, {Fabiani}, {Ferrazzoli}, {Garcia}, {Gunji}, {Hayashida}, {Heyl}, {Iwakiri}, {Kaaret}, {Kislat}, {Kitaguchi}, {Kolodziejczak}, {La Monaca}, {Latronico}, {Maldera}, {Manfreda}, {Matt}, {Ng}, {O'Dell}, {Omodei}, {Oppedisano}, {Papitto}, {Pavlov}, {Pesce-Rollins}, {Pilia}, {Possenti}, {Poutanen}, {Ramsey}, {Rankin}, {Ratheesh}, {Roberts}, {Sgr{\`o}}, {Slane}, {Soffitta}, {Spandre}, {Swartz}, {Tamagawa}, {Taverna}, {Tawara}, {Tennant}, {Thomas}, {Tombesi}, {Trois}, {Tsygankov}, {Turolla}, {Vink}, {Weisskopf}, {Xie}, \& {Zane}}]{2024ApJ...963....5E}
{Errando}, M., {Liodakis}, I., {Marscher}, A.~P., {et~al.} 2024, \apj, 963, 5, \dodoi{10.3847/1538-4357/ad1ce4}

\bibitem[{{Ferrazzoli} {et~al.}(2024){Ferrazzoli}, {Prokhorov}, {Bucciantini}, {Slane}, {Vink}, {Cardillo}, {Yang}, {Silvestri}, {Zhou}, {Costa}, {Omodei}, {Ng}, {Soffitta}, {Weisskopf}, {Baldini}, {Di Marco}, {Doroshenko}, {Heyl}, {Kaaret}, {Kim}, {Marin}, {Mizuno}, {Pesce-Rollins}, {Sgr{\`o}}, {Swartz}, {Tamagawa}, {Xie}, {Agudo}, {Antonelli}, {Bachetti}, {Baumgartner}, {Bellazzini}, {Bianchi}, {Bongiorno}, {Bonino}, {Brez}, {Capitanio}, {Castellano}, {Cavazzuti}, {Chen}, {Ciprini}, {De Rosa}, {Del Monte}, {Di Gesu}, {Di Lalla}, {Donnarumma}, {Dov{\v{c}}iak}, {Ehlert}, {Enoto}, {Evangelista}, {Fabiani}, {Garcia}, {Gunji}, {Hayashida}, {Iwakiri}, {Jorstad}, {Karas}, {Kislat}, {Kitaguchi}, {Kolodziejczak}, {Krawczynski}, {La Monaca}, {Latronico}, {Liodakis}, {Maldera}, {Manfreda}, {Marinucci}, {Marscher}, {Marshall}, {Massaro}, {Matt}, {Mitsuishi}, {Muleri}, {Negro}, {O'Dell}, {Oppedisano}, {Papitto}, {Pavlov}, {Peirson}, {Perri}, {Petrucci}, {Pilia}, {Possenti}, {Poutanen}, {Puccetti}, {Ramsey}, {Rankin},
  {Ratheesh}, {Roberts}, {Romani}, {Spandre}, {Tavecchio}, {Taverna}, {Tawara}, {Tennant}, {Thomas}, {Tombesi}, {Trois}, {Tsygankov}, {Turolla}, {Wu}, \& {Zane}}]{2024ApJ...967L..38F}
{Ferrazzoli}, R., {Prokhorov}, D., {Bucciantini}, N., {et~al.} 2024, \apjl, 967, L38, \dodoi{10.3847/2041-8213/ad4a68}

\bibitem[{{Finke} {et~al.}(2022){Finke}, {Ajello}, {Dom{\'\i}nguez}, {Desai}, {Hartmann}, {Paliya}, \& {Saldana-Lopez}}]{2022ApJ...941...33F}
{Finke}, J.~D., {Ajello}, M., {Dom{\'\i}nguez}, A., {et~al.} 2022, \apj, 941, 33, \dodoi{10.3847/1538-4357/ac9843}

\bibitem[{{Fukazawa} {et~al.}(2015){Fukazawa}, {Finke}, {Stawarz}, {Tanaka}, {Itoh}, \& {Tokuda}}]{2015ApJ...798...74F}
{Fukazawa}, Y., {Finke}, J., {Stawarz}, {\L}., {et~al.} 2015, \apj, 798, 74, \dodoi{10.1088/0004-637X/798/2/74}

\bibitem[{Georganopoulos \& Kazanas(2003)}]{Georganopoulos_2003}
Georganopoulos, M., \& Kazanas, D. 2003, The Astrophysical Journal, 589, L5–L8, \dodoi{10.1086/375796}

\bibitem[{{Gulati} {et~al.}(2023){Gulati}, {Bhattacharya}, {Ramadevi}, {Stalin}, \& {Sreekumar}}]{2023MNRAS.521.2704G}
{Gulati}, S., {Bhattacharya}, D., {Ramadevi}, M.~C., {Stalin}, C.~S., \& {Sreekumar}, P. 2023, \mnras, 521, 2704, \dodoi{10.1093/mnras/stad716}

\bibitem[{{Guo} {et~al.}(2018){Guo}, {Zhang}, {Zhang}, \& {Liang}}]{2018RAA....18...70G}
{Guo}, S.-C., {Zhang}, H.-M., {Zhang}, J., \& {Liang}, E.-W. 2018, Research in Astronomy and Astrophysics, 18, 070, \dodoi{10.1088/1674-4527/18/6/70}

\bibitem[{{H.~E.~S.~S. Collaboration} {et~al.}(2020){H.~E.~S.~S. Collaboration}, {Abdalla}, {Adam}, {Aharonian}, {Ait Benkhali}, {Ang{\"u}ner}, {Arakawa}, {Arcaro}, {Armand}, {Ashkar}, {Backes}, {Barbosa Martins}, {Barnard}, {Becherini}, {Berge}, {Bernl{\"o}hr}, {Blackwell}, {B{\"o}ttcher}, {Boisson}, {Bolmont}, {Bonnefoy}, {Bregeon}, {Breuhaus}, {Brun}, {Brun}, {Bryan}, {B{\"u}chele}, {Bulik}, {Bylund}, {Capasso}, {Caroff}, {Carosi}, {Casanova}, {Cerruti}, {Chand}, {Chandra}, {Chen}, {Colafrancesco}, {Cury{\l}o}, {Davids}, {Deil}, {Devin}, {deWilt}, {Dirson}, {Djannati-Ata{\"\i}}, {Dmytriiev}, {Donath}, {Doroshenko}, {Drury}, {Dyks}, {Egberts}, {Emery}, {Ernenwein}, {Eschbach}, {Feijen}, {Fegan}, {Fiasson}, {Fontaine}, {Funk}, {F{\"u}{\ss}ling}, {Gabici}, {Gallant}, {Gat{\'e}}, {Giavitto}, {Glawion}, {Glicenstein}, {Gottschall}, {Grondin}, {Hahn}, {Haupt}, {Heinzelmann}, {Henri}, {Hermann}, {Hinton}, {Hofmann}, {Hoischen}, {Holch}, {Holler}, {Horns}, {Huber}, {Iwasaki}, {Jamrozy}, {Jankowsky}, {Jankowsky},
  {Jardin-Blicq}, {Jung-Richardt}, {Kastendieck}, {Katarzy{\'n}ski}, {Katsuragawa}, {Katz}, {Khangulyan}, {Kh{\'e}lifi}, {King}, {Klepser}, {Klu{\'z}niak}, {Komin}, {Kosack}, {Kostunin}, {Kraus}, {Lamanna}, {Lau}, {Lemi{\`e}re}, {Lemoine-Goumard}, {Lenain}, {Leser}, {Levy}, {Lohse}, {Lypova}, {Mackey}, {Majumdar}, {Malyshev}, {Marandon}, {Marcowith}, {Mares}, {Mariaud}, {Mart{\'\i}-Devesa}, {Marx}, {Maurin}, {Meintjes}, {Mitchell}, {Moderski}, {Mohamed}, {Mohrmann}, {Moore}, {Moulin}, {Muller}, {Murach}, {Nakashima}, {de Naurois}, {Ndiyavala}, {Niederwanger}, {Niemiec}, {Oakes}, {O'Brien}, {Odaka}, {Ohm}, {de Ona Wilhelmi}, {Ostrowski}, {Oya}, {Panter}, {Parsons}, {Perennes}, {Petrucci}, {Peyaud}, {Piel}, {Pita}, {Poireau}, {Priyana Noel}, {Prokhorov}, {Prokoph}, {P{\"u}hlhofer}, {Punch}, {Quirrenbach}, {Raab}, {Rauth}, {Reimer}, {Reimer}, {Remy}, {Renaud}, {Rieger}, {Rinchiuso}, {Romoli}, {Rowell}, {Rudak}, {Ruiz-Velasco}, {Sahakian}, {Saito}, {Sanchez}, {Santangelo}, {Sasaki}, {Schlickeiser},
  {Sch{\"u}ssler}, {Schulz}, {Schutte}, {Schwanke}, {Schwemmer}, {Seglar-Arroyo}, {Senniappan}, {Seyffert}, {Shafi}, {Shiningayamwe}, {Simoni}, {Sinha}, {Sol}, {Specovius}, {Spir-Jacob}, {Stawarz}, {Steenkamp}, {Stegmann}, {Steppa}, {Takahashi}, {Tavernier}, {Taylor}, {Terrier}, {Tiziani}, {Tluczykont}, {Trichard}, {Tsirou}, {Tsuji}, {Tuffs}, {Uchiyama}, {van der Walt}, {van Eldik}, {van Rensburg}, {van Soelen}, {Vasileiadis}, {Veh}, {Venter}, {Vincent}, {Vink}, {Voisin}, {V{\"o}lk}, {Vuillaume}, {Wadiasingh}, {Wagner}, {White}, {Wierzcholska}, {Yang}, {Yoneda}, {Zacharias}, {Zanin}, {Zdziarski}, {Zech}, {Ziegler}, {Zorn}, \& {{\.Z}ywucka}}]{2020Natur.582..356H}
{H.~E.~S.~S. Collaboration}, {Abdalla}, H., {Adam}, R., {et~al.} 2020, \nat, 582, 356, \dodoi{10.1038/s41586-020-2354-1}

\bibitem[{{Hardcastle} {et~al.}(2016){Hardcastle}, {Lenc}, {Birkinshaw}, {Croston}, {Goodger}, {Marshall}, {Perlman}, {Siemiginowska}, {Stawarz}, \& {Worrall}}]{2016MNRAS.455.3526H}
{Hardcastle}, M.~J., {Lenc}, E., {Birkinshaw}, M., {et~al.} 2016, \mnras, 455, 3526, \dodoi{10.1093/mnras/stv2553}

\bibitem[{{Harris} \& {Krawczynski}(2006)}]{2006ARA&A..44..463H}
{Harris}, D.~E., \& {Krawczynski}, H. 2006, \araa, 44, 463, \dodoi{10.1146/annurev.astro.44.051905.092446}

\bibitem[{{He} {et~al.}(2023){He}, {Sun}, {Wang}, {Rieger}, {Liu}, \& {Liang}}]{2023MNRAS.525.5298H}
{He}, J.-C., {Sun}, X.-N., {Wang}, J.-S., {et~al.} 2023, \mnras, 525, 5298, \dodoi{10.1093/mnras/stad2542}

\bibitem[{{HI4PI Collaboration} {et~al.}(2016){HI4PI Collaboration}, {Ben Bekhti}, {Fl{\"o}er}, {Keller}, {Kerp}, {Lenz}, {Winkel}, {Bailin}, {Calabretta}, {Dedes}, {Ford}, {Gibson}, {Haud}, {Janowiecki}, {Kalberla}, {Lockman}, {McClure-Griffiths}, {Murphy}, {Nakanishi}, {Pisano}, \& {Staveley-Smith}}]{2016A&A...594A.116H}
{HI4PI Collaboration}, {Ben Bekhti}, N., {Fl{\"o}er}, L., {et~al.} 2016, \aap, 594, A116, \dodoi{10.1051/0004-6361/201629178}

\bibitem[{{Hu} {et~al.}(2024){Hu}, {Yu}, {Zhang}, {Wang}, {Patra}, {Brink}, {Zheng}, {Wang}, {Kong}, {Chen}, {Zhou}, {Cao}, {Lu}, {Zhou}, {Wei}, {Huang}, {Li}, {Lou}, {Mao}, {Liang}, \& {Filippenko}}]{2024ApJ...970L..22H}
{Hu}, X.-K., {Yu}, Y.-W., {Zhang}, J., {et~al.} 2024, \apjl, 970, L22, \dodoi{10.3847/2041-8213/ad5e68}

\bibitem[{{Isobe} {et~al.}(2017){Isobe}, {Koyama}, {Kino}, {Wada}, {Nakagawa}, {Matsuhara}, {Niinuma}, \& {Tashiro}}]{2017ApJ...850..193I}
{Isobe}, N., {Koyama}, S., {Kino}, M., {et~al.} 2017, \apj, 850, 193, \dodoi{10.3847/1538-4357/aa94c9}

\bibitem[{{Isobe} {et~al.}(2020){Isobe}, {Sunada}, {Kino}, {Koyama}, {Tashiro}, {Nagai}, \& {Pearson}}]{2020ApJ...899...17I}
{Isobe}, N., {Sunada}, Y., {Kino}, M., {et~al.} 2020, \apj, 899, 17, \dodoi{10.3847/1538-4357/ab9d1c}

\bibitem[{{Kataoka} \& {Stawarz}(2005)}]{2005ApJ...622..797K}
{Kataoka}, J., \& {Stawarz}, {\L}. 2005, \apj, 622, 797, \dodoi{10.1086/428083}

\bibitem[{{Kislat} {et~al.}(2015){Kislat}, {Clark}, {Beilicke}, \& {Krawczynski}}]{2015APh....68...45K}
{Kislat}, F., {Clark}, B., {Beilicke}, M., \& {Krawczynski}, H. 2015, Astroparticle Physics, 68, 45, \dodoi{10.1016/j.astropartphys.2015.02.007}

\bibitem[{{Kouch} {et~al.}(2024){Kouch}, {Liodakis}, {Middei}, {Kim}, {Tavecchio}, {Marscher}, {Marshall}, {Ehlert}, {Di Gesu}, {Jorstad}, {Agudo}, {Madejski}, {Romani}, {Errando}, {Lindfors}, {Nilsson}, {Toppari}, {Potter}, {Imazawa}, {Sasada}, {Fukazawa}, {Kawabata}, {Uemura}, {Mizuno}, {Nakaoka}, {Akitaya}, {McCall}, {Jermak}, {Steele}, {Myserlis}, {Gurwell}, {Keating}, {Rao}, {Kang}, {Lee}, {Kim}, {Cheong}, {Jeong}, {Angelakis}, {Kraus}, {Aceituno}, {Bonnoli}, {Casanova}, {Escudero}, {Ag{\'\i}s-Gonz{\'a}lez}, {Husillos}, {Morcuende}, {Otero-Santos}, {Sota}, {Bachev}, {Antonelli}, {Bachetti}, {Baldini}, {Baumgartner}, {Bellazzini}, {Bianchi}, {Bongiorno}, {Bonino}, {Brez}, {Bucciantini}, {Capitanio}, {Castellano}, {Cavazzuti}, {Chen}, {Ciprini}, {Costa}, {De Rosa}, {Del Monte}, {Di Lalla}, {Di Marco}, {Donnarumma}, {Doroshenko}, {Dov{\v{c}}iak}, {Enoto}, {Evangelista}, {Fabiani}, {Ferrazzoli}, {Garcia}, {Gunji}, {Hayashida}, {Heyl}, {Iwakiri}, {Kaaret}, {Karas}, {Kislat}, {Kitaguchi}, {Kolodziejczak},
  {Krawczynski}, {La Monaca}, {Latronico}, {Maldera}, {Manfreda}, {Marin}, {Marinucci}, {Massaro}, {Matt}, {Mitsuishi}, {Muleri}, {Negro}, {Ng}, {O'Dell}, {Omodei}, {Oppedisano}, {Papitto}, {Pavlov}, {Peirson}, {Perri}, {Pesce-Rollins}, {Petrucci}, {Pilia}, {Possenti}, {Poutanen}, {Puccetti}, {Ramsey}, {Rankin}, {Ratheesh}, {Roberts}, {Sgr{\`o}}, {Slane}, {Soffitta}, {Spandre}, {Swartz}, {Tamagawa}, {Taverna}, {Tawara}, {Tennant}, {Thomas}, {Tombesi}, {Trois}, {Tsygankov}, {Turolla}, {Vink}, {Weisskopf}, {Wu}, {Xie}, \& {Zane}}]{2024A&A...689A.119K}
{Kouch}, P.~M., {Liodakis}, I., {Middei}, R., {et~al.} 2024, \aap, 689, A119, \dodoi{10.1051/0004-6361/202449166}

\bibitem[{{Krawczynski}(2012)}]{2012ApJ...744...30K}
{Krawczynski}, H. 2012, \apj, 744, 30, \dodoi{10.1088/0004-637X/744/1/30}

\bibitem[{{Liodakis} {et~al.}(2022){Liodakis}, {Marscher}, {Agudo}, {Berdyugin}, {Bernardos}, {Bonnoli}, {Borman}, {Casadio}, {Casanova}, {Cavazzuti}, {Rodriguez Cavero}, {Di Gesu}, {Di Lalla}, {Donnarumma}, {Ehlert}, {Errando}, {Escudero}, {Garc{\'\i}a-Comas}, {Ag{\'\i}s-Gonz{\'a}lez}, {Husillos}, {Jormanainen}, {Jorstad}, {Kagitani}, {Kopatskaya}, {Kravtsov}, {Krawczynski}, {Lindfors}, {Larionova}, {Madejski}, {Marin}, {Marchini}, {Marshall}, {Morozova}, {Massaro}, {Masiero}, {Mawet}, {Middei}, {Millar-Blanchaer}, {Myserlis}, {Negro}, {Nilsson}, {O'Dell}, {Omodei}, {Pacciani}, {Paggi}, {Panopoulou}, {Peirson}, {Perri}, {Petrucci}, {Poutanen}, {Puccetti}, {Romani}, {Sakanoi}, {Savchenko}, {Sota}, {Tavecchio}, {Tinyanont}, {Vasilyev}, {Weaver}, {Zhovtan}, {Antonelli}, {Bachetti}, {Baldini}, {Baumgartner}, {Bellazzini}, {Bianchi}, {Bongiorno}, {Bonino}, {Brez}, {Bucciantini}, {Capitanio}, {Castellano}, {Ciprini}, {Costa}, {De Rosa}, {Del Monte}, {Di Marco}, {Doroshenko}, {Dov{\v{c}}iak}, {Enoto},
  {Evangelista}, {Fabiani}, {Ferrazzoli}, {Garcia}, {Gunji}, {Hayashida}, {Heyl}, {Iwakiri}, {Karas}, {Kitaguchi}, {Kolodziejczak}, {La Monaca}, {Latronico}, {Maldera}, {Manfreda}, {Marinucci}, {Matt}, {Mitsuishi}, {Mizuno}, {Muleri}, {Ng}, {Oppedisano}, {Papitto}, {Pavlov}, {Pesce-Rollins}, {Pilia}, {Possenti}, {Ramsey}, {Rankin}, {Ratheesh}, {Sgr{\'o}}, {Slane}, {Soffitta}, {Spandre}, {Tamagawa}, {Taverna}, {Tawara}, {Tennant}, {Thomas}, {Tombesi}, {Trois}, {Tsygankov}, {Turolla}, {Vink}, {Weisskopf}, {Wu}, {Xie}, \& {Zane}}]{2022Natur.611..677L}
{Liodakis}, I., {Marscher}, A.~P., {Agudo}, I., {et~al.} 2022, \nat, 611, 677, \dodoi{10.1038/s41586-022-05338-0}

\bibitem[{{Marshall} {et~al.}(2023){Marshall}, {Liodakis}, {Marscher}, {Di Lalla}, {Jorstad}, {Kim}, {Middei}, {Negro}, {Omodei}, {Peirson}, {Perri}, {Puccetti}, {Agudo}, {Bonnoli}, {Berdyugin}, {Cavazzuti}, {Rodriguez Cavero}, {Donnarumma}, {Di Gesu}, {Jormanainen}, {Krawczynski}, {Lindfors}, {Marin}, {Massaro}, {Pacciani}, {Poutanen}, {Tavecchio}, {Kouch}, {Aceituno}, {Bernardos}, {Bonnoli}, {Casanova}, {Garcia-Comas}, {Agis-Gonzalez}, {Husillos}, {Marchini}, {Sota}, {Blinov}, {Bourbah}, {Kielhmann}, {Kontopodis}, {Mandarakas}, {Romanopoulos}, {Skalidis}, {Vervelaki}, {Borman}, {Kopatskaya}, {Larionova}, {Morozova}, {Savchenko}, {Vasilyev}, {Zhovtan}, {Casadio}, {Escudero}, {Kramer}, {Myserlis}, {Trainou}, {Imazawa}, {Sasada}, {Fukazawa}, {Kawabata}, {Uemura}, {Mizuno}, {Nakaoka}, {Akitaya}, {Masiero}, {Mawet}, {Millar-Blanchaer}, {Panopoulou}, {Tinyanont}, {Berdyugin}, {Kagitani}, {Kravtsov}, {Sakanoi}, {Antonelli}, {Bachetti}, {Baldini}, {Baumgartner}, {Bellazzini}, {Bianchi}, {Bongiorno}, {Bonino},
  {Brez}, {Bucciantini}, {Capitanio}, {Castellano}, {Cavazzuti}, {Chen}, {Ciprini}, {Costa}, {De Rosa}, {Del Monte}, {Di Gesu}, {Di Marco}, {Donnarumma}, {Doroshenko}, {Dovvciak}, {Ehlert}, {Enoto}, {Evangelista}, {Fabiani}, {Ferrazzoli}, {Garcia}, {Gunji}, {Hayashida}, {Heyl}, {Iwakiri}, {Kaaret}, {Karas}, {Kislat}, {Kitaguchi}, {Kolodziejczak}, {Krawczynski}, {La Monaca}, {Latronico}, {Maldera}, {Manfreda}, {Marin}, {Marinucci}, {Matt}, {Mitsuishi}, {Mizuno}, {Muleri}, {Ng}, {ODell}, {Oppedisano}, {Papitto}, {Pavlov}, {Pesce-Rollins}, {Petrucci}, {Pilia}, {Possenti}, {Poutanen}, {Puccetti}, {Ramsey}, {Rankin}, {Ratheesh}, {Roberts}, {Romani}, {Sgro}, {Slane}, {Soffitta}, {Spandre}, {Swartz}, {Tamagawa}, {Taverna}, {Tawara}, {Tennant}, {Thomas}, {Tombesi}, {Trois}, {Tsygankov}, {Turolla}, {Vink}, {Weisskopf}, {Wu}, {Xie}, \& {Zane}}]{2023arXiv231011510M}
{Marshall}, H.~L., {Liodakis}, I., {Marscher}, A.~P., {et~al.} 2023, arXiv e-prints, arXiv:2310.11510, \dodoi{10.48550/arXiv.2310.11510}

\bibitem[{{Marshall} {et~al.}(2024){Marshall}, {Liodakis}, {Marscher}, {Di Lalla}, {Jorstad}, {Kim}, {Middei}, {Negro}, {Omodei}, {Peirson}, {Perri}, {Puccetti}, {Laurenti}, {Agudo}, {Bonnoli}, {Berdyugin}, {Cavazzuti}, {Rodriguez Cavero}, {Donnarumma}, {Di Gesu}, {Jormanainen}, {Krawczynski}, {Lindfors}, {Madjeski}, {Marin}, {Massaro}, {Pacciani}, {Poutanen}, {Tavecchio}, {Kouch}, {Aceituno}, {Bernardos}, {Casanova}, {Garc{\'\i}a-Comas}, {Ag{\'\i}s-Gonz{\'a}lez}, {Husillos}, {Marchini}, {Sota}, {Blinov}, {Bourbah}, {Kielhmann}, {Kontopodis}, {Mandarakas}, {Romanopoulos}, {Skalidis}, {Vervelaki}, {Borman}, {Kopatskaya}, {Larionova}, {Morozova}, {Savchenko}, {Vasilyev}, {Zhovtan}, {Casadio}, {Escudero}, {Kramer}, {Myserlis}, {Trainou}, {Imazawa}, {Sasada}, {Fukazawa}, {Kawabata}, {Uemura}, {Mizuno}, {Nakaoka}, {Akitaya}, {Masiero}, {Mawet}, {Panopoulou}, {Tinyanont}, {Kagitani}, {Kravtsov}, {Sakanoi}, {Dattolo}, {Gurwell}, {Keating}, {Rao}, {Cheong}, {Jeong}, {Kang}, {Kim}, {Lee}, {Angelakis}, {Kraus},
  {Hales}, {Kameno}, {Kneissl}, {Messias}, {Nagai}, {Antonelli}, {Bachetti}, {Baldini}, {Baumgartner}, {Bellazzini}, {Bianchi}, {Bongiorno}, {Bonino}, {Brez}, {Bucciantini}, {Capitanio}, {Castellano}, {Chen}, {Ciprini}, {Costa}, {De Rosa}, {Del Monte}, {Di Marco}, {Doroshenko}, {Dov{\v{c}}iak}, {Ehlert}, {Enoto}, {Evangelista}, {Fabiani}, {Ferrazzoli}, {Garcia}, {Gunji}, {Hayashida}, {Heyl}, {Iwakiri}, {Kaaret}, {Karas}, {Kislat}, {Kitaguchi}, {Kolodziejczak}, {La Monaca}, {Latronico}, {Maldera}, {Manfreda}, {Marinucci}, {Matt}, {Mitsuishi}, {Muleri}, {Ng}, {O'Dell}, {Oppedisano}, {Papitto}, {Pavlov}, {Pesce-Rollins}, {Petrucci}, {Pilia}, {Possenti}, {Ramsey}, {Rankin}, {Ratheesh}, {Roberts}, {Romani}, {Sgr{\`o}}, {Slane}, {Soffitta}, {Spandre}, {Swartz}, {Tamagawa}, {Taverna}, {Tawara}, {Tennant}, {Thomas}, {Tombesi}, {Trois}, {Tsygankov}, {Turolla}, {Vink}, {Weisskopf}, {Wu}, {Xie}, \& {Zane}}]{2024ApJ...972...74M}
---. 2024, \apj, 972, 74, \dodoi{10.3847/1538-4357/ad5671}

\bibitem[{{Massaro} {et~al.}(2004){Massaro}, {Perri}, {Giommi}, \& {Nesci}}]{2004A&A...413..489M}
{Massaro}, E., {Perri}, M., {Giommi}, P., \& {Nesci}, R. 2004, \aap, 413, 489, \dodoi{10.1051/0004-6361:20031558}

\bibitem[{{Meisenheimer} {et~al.}(1989){Meisenheimer}, {Roser}, {Hiltner}, {Yates}, {Longair}, {Chini}, \& {Perley}}]{1989A&A...219...63M}
{Meisenheimer}, K., {Roser}, H.~J., {Hiltner}, P.~R., {et~al.} 1989, \aap, 219, 63

\bibitem[{{Meisenheimer} {et~al.}(1997){Meisenheimer}, {Yates}, \& {Roeser}}]{1997A&A...325...57M}
{Meisenheimer}, K., {Yates}, M.~G., \& {Roeser}, H.~J. 1997, \aap, 325, 57

\bibitem[{{Middei} {et~al.}(2023{\natexlab{a}}){Middei}, {Liodakis}, {Perri}, {Puccetti}, {Cavazzuti}, {Di Gesu}, {Ehlert}, {Madejski}, {Marscher}, {Marshall}, {Muleri}, {Negro}, {Jorstad}, {Ag{\'\i}s-Gonz{\'a}lez}, {Agudo}, {Bonnoli}, {Bernardos}, {Casanova}, {Garc{\'\i}a-Comas}, {Husillos}, {Marchini}, {Sota}, {Kouch}, {Lindfors}, {Borman}, {Kopatskaya}, {Larionova}, {Morozova}, {Savchenko}, {Vasilyev}, {Zhovtan}, {Casadio}, {Escudero}, {Myserlis}, {Hales}, {Kameno}, {Kneissl}, {Messias}, {Nagai}, {Blinov}, {Bourbah}, {Kiehlmann}, {Kontopodis}, {Mandarakas}, {Romanopoulos}, {Skalidis}, {Vervelaki}, {Masiero}, {Mawet}, {Millar-Blanchaer}, {Panopoulou}, {Tinyanont}, {Berdyugin}, {Kagitani}, {Kravtsov}, {Sakanoi}, {Imazawa}, {Sasada}, {Fukazawa}, {Kawabata}, {Uemura}, {Mizuno}, {Nakaoka}, {Akitaya}, {Gurwell}, {Rao}, {Di Lalla}, {Cibrario}, {Donnarumma}, {Kim}, {Omodei}, {Pacciani}, {Poutanen}, {Tavecchio}, {Antonelli}, {Bachetti}, {Baldini}, {Baumgartner}, {Bellazzini}, {Bianchi}, {Bongiorno}, {Bonino},
  {Brez}, {Bucciantini}, {Capitanio}, {Castellano}, {Ciprini}, {Costa}, {De Rosa}, {Del Monte}, {Di Marco}, {Doroshenko}, {Dov{\v{c}}iak}, {Enoto}, {Evangelista}, {Fabiani}, {Ferrazzoli}, {Garcia}, {Gunji}, {Hayashida}, {Heyl}, {Iwakiri}, {Karas}, {Kitaguchi}, {Kolodziejczak}, {Krawczynski}, {La Monaca}, {Latronico}, {Maldera}, {Manfreda}, {Marin}, {Marinucci}, {Massaro}, {Matt}, {Mitsuishi}, {Ng}, {O'Dell}, {Oppedisano}, {Papitto}, {Pavlov}, {Peirson}, {Pesce-Rollins}, {Petrucci}, {Pilia}, {Possenti}, {Ramsey}, {Rankin}, {Ratheesh}, {Romani}, {Sgr{\'o}}, {Slane}, {Soffitta}, {Spandre}, {Tamagawa}, {Taverna}, {Tawara}, {Tennant}, {Thomas}, {Tombesi}, {Trois}, {Tsygankov}, {Turolla}, {Vink}, {Weisskopf}, {Wu}, {Xie}, \& {Zane}}]{2023ApJ...942L..10M}
{Middei}, R., {Liodakis}, I., {Perri}, M., {et~al.} 2023{\natexlab{a}}, \apjl, 942, L10, \dodoi{10.3847/2041-8213/aca281}

\bibitem[{{Middei} {et~al.}(2023{\natexlab{b}}){Middei}, {Perri}, {Puccetti}, {Liodakis}, {Di Gesu}, {Marscher}, {Rodriguez Cavero}, {Tavecchio}, {Donnarumma}, {Laurenti}, {Jorstad}, {Agudo}, {Marshall}, {Pacciani}, {Kim}, {Aceituno}, {Bonnoli}, {Casanova}, {Ag{\'\i}s-Gonz{\'a}lez}, {Sota}, {Casadio}, {Escudero}, {Myserlis}, {Sievers}, {Kouch}, {Lindfors}, {Gurwell}, {Keating}, {Rao}, {Kang}, {Lee}, {Kim}, {Cheong}, {Jeong}, {Angelakis}, {Kraus}, {Antonelli}, {Bachetti}, {Baldini}, {Baumgartner}, {Bellazzini}, {Bianchi}, {Bongiorno}, {Bonino}, {Brez}, {Bucciantini}, {Capitanio}, {Castellano}, {Cavazzuti}, {Chen}, {Ciprini}, {Costa}, {De Rosa}, {Del Monte}, {Di Lalla}, {Di Marco}, {Doroshenko}, {Dov{\v{c}}iak}, {Ehlert}, {Enoto}, {Evangelista}, {Fabiani}, {Ferrazzoli}, {Garc{\'\i}a}, {Gunji}, {Hayashida}, {Heyl}, {Iwakiri}, {Kaaret}, {Karas}, {Kislat}, {Kitaguchi}, {Kolodziejczak}, {Krawczynski}, {La Monaca}, {Latronico}, {Maldera}, {Manfreda}, {Marin}, {Marinucci}, {Massaro}, {Matt}, {Mitsuishi}, {Mizuno},
  {Muleri}, {Negro}, {Ng}, {O'Dell}, {Omodei}, {Oppedisano}, {Papitto}, {Pavlov}, {Peirson}, {Pesce-Rollins}, {Petrucci}, {Pilia}, {Possenti}, {Poutanen}, {Ramsey}, {Rankin}, {Ratheesh}, {Roberts}, {Romani}, {Sgr{\`o}}, {Slane}, {Soffitta}, {Spandre}, {Swartz}, {Tamagawa}, {Taverna}, {Tawara}, {Tennant}, {Thomas}, {Tombesi}, {Trois}, {Tsygankov}, {Turolla}, {Vink}, {Weisskopf}, {Wu}, {Xie}, \& {Zane}}]{2023ApJ...953L..28M}
{Middei}, R., {Perri}, M., {Puccetti}, S., {et~al.} 2023{\natexlab{b}}, \apjl, 953, L28, \dodoi{10.3847/2041-8213/acec3e}

\bibitem[{{Migliori} {et~al.}(2011){Migliori}, {Grandi}, {Torresi}, {Dermer}, {Finke}, {Celotti}, {Mukherjee}, {Errando}, {Gargano}, {Giordano}, \& {Giroletti}}]{2011A&A...533A..72M}
{Migliori}, G., {Grandi}, P., {Torresi}, E., {et~al.} 2011, \aap, 533, A72, \dodoi{10.1051/0004-6361/201116808}

\bibitem[{{Nolan} {et~al.}(2012){Nolan}, {Abdo}, {Ackermann}, {Ajello}, {Allafort}, {Antolini}, {Atwood}, {Axelsson}, {Baldini}, {Ballet}, {Barbiellini}, {Bastieri}, {Bechtol}, {Belfiore}, {Bellazzini}, {Berenji}, {Bignami}, {Blandford}, {Bloom}, {Bonamente}, {Bonnell}, {Borgland}, {Bottacini}, {Bouvier}, {Brandt}, {Bregeon}, {Brigida}, {Bruel}, {Buehler}, {Burnett}, {Buson}, {Caliandro}, {Cameron}, {Campana}, {Ca{\~n}adas}, {Cannon}, {Caraveo}, {Casandjian}, {Cavazzuti}, {Ceccanti}, {Cecchi}, {{\c{C}}elik}, {Charles}, {Chekhtman}, {Cheung}, {Chiang}, {Chipaux}, {Ciprini}, {Claus}, {Cohen-Tanugi}, {Cominsky}, {Conrad}, {Corbet}, {Cutini}, {D'Ammando}, {Davis}, {de Angelis}, {DeCesar}, {DeKlotz}, {De Luca}, {den Hartog}, {de Palma}, {Dermer}, {Digel}, {Silva}, {Drell}, {Drlica-Wagner}, {Dubois}, {Dumora}, {Enoto}, {Escande}, {Fabiani}, {Falletti}, {Favuzzi}, {Fegan}, {Ferrara}, {Focke}, {Fortin}, {Frailis}, {Fukazawa}, {Funk}, {Fusco}, {Gargano}, {Gasparrini}, {Gehrels}, {Germani}, {Giebels}, {Giglietto},
  {Giommi}, {Giordano}, {Giroletti}, {Glanzman}, {Godfrey}, {Grenier}, {Grondin}, {Grove}, {Guillemot}, {Guiriec}, {Gustafsson}, {Hadasch}, {Hanabata}, {Harding}, {Hayashida}, {Hays}, {Hill}, {Horan}, {Hou}, {Hughes}, {Iafrate}, {Itoh}, {J{\'o}hannesson}, {Johnson}, {Johnson}, {Johnson}, {Johnson}, {Kamae}, {Katagiri}, {Kataoka}, {Katsuta}, {Kawai}, {Kerr}, {Kn{\"o}dlseder}, {Kocevski}, {Kuss}, {Lande}, {Landriu}, {Latronico}, {Lemoine-Goumard}, {Lionetto}, {Llena Garde}, {Longo}, {Loparco}, {Lott}, {Lovellette}, {Lubrano}, {Madejski}, {Marelli}, {Massaro}, {Mazziotta}, {McConville}, {McEnery}, {Mehault}, {Michelson}, {Minuti}, {Mitthumsiri}, {Mizuno}, {Moiseev}, {Mongelli}, {Monte}, {Monzani}, {Morselli}, {Moskalenko}, {Murgia}, {Nakamori}, {Naumann-Godo}, {Norris}, {Nuss}, {Nymark}, {Ohno}, {Ohsugi}, {Okumura}, {Omodei}, {Orlando}, {Ormes}, {Ozaki}, {Paneque}, {Panetta}, {Parent}, {Perkins}, {Pesce-Rollins}, {Pierbattista}, {Pinchera}, {Piron}, {Pivato}, {Porter}, {Racusin}, {Rain{\`o}}, {Rando}, {Razzano},
  {Razzaque}, {Reimer}, {Reimer}, {Reposeur}, {Ritz}, {Rochester}, {Romani}, {Roth}, {Rousseau}, {Ryde}, {Sadrozinski}, {Salvetti}, {Sanchez}, {Saz Parkinson}, {Sbarra}, {Scargle}, {Schalk}, {Sgr{\`o}}, {Shaw}, {Shrader}, \& {Siskind}}]{2012ApJS..199...31N}
{Nolan}, P.~L., {Abdo}, A.~A., {Ackermann}, M., {et~al.} 2012, \apjs, 199, 31, \dodoi{10.1088/0067-0049/199/2/31}

\bibitem[{{Peirson} \& {Romani}(2019)}]{2019ApJ...885...76P}
{Peirson}, A.~L., \& {Romani}, R.~W. 2019, \apj, 885, 76, \dodoi{10.3847/1538-4357/ab46b1}

\bibitem[{{Peng} {et~al.}(2019){Peng}, {Zhang}, {Wang}, {Wang}, \& {Zhi}}]{2019ApJ...884...91P}
{Peng}, F.-K., {Zhang}, H.-M., {Wang}, X.-Y., {Wang}, J.-F., \& {Zhi}, Q.-J. 2019, \apj, 884, 91, \dodoi{10.3847/1538-4357/ab3e6f}

\bibitem[{{Perley} {et~al.}(1997){Perley}, {Roser}, \& {Meisenheimer}}]{1997A&A...328...12P}
{Perley}, R.~A., {Roser}, H.-J., \& {Meisenheimer}, K. 1997, \aap, 328, 12

\bibitem[{{Roeser} \& {Meisenheimer}(1987)}]{1987ApJ...314...70R}
{Roeser}, H.-J., \& {Meisenheimer}, K. 1987, \apj, 314, 70, \dodoi{10.1086/165039}

\bibitem[{{Rulten} {et~al.}(2020){Rulten}, {Brown}, \& {Chadwick}}]{2020MNRAS.492.4666R}
{Rulten}, C.~B., {Brown}, A.~M., \& {Chadwick}, P.~M. 2020, \mnras, 492, 4666, \dodoi{10.1093/mnras/staa054}

\bibitem[{{Sahakyan} {et~al.}(2013){Sahakyan}, {Yang}, {Aharonian}, \& {Rieger}}]{2013ApJ...770L...6S}
{Sahakyan}, N., {Yang}, R., {Aharonian}, F.~A., \& {Rieger}, F.~M. 2013, \apjl, 770, L6, \dodoi{10.1088/2041-8205/770/1/L6}

\bibitem[{{Sambruna} {et~al.}(1999){Sambruna}, {Eracleous}, \& {Mushotzky}}]{1999ApJ...526...60S}
{Sambruna}, R.~M., {Eracleous}, M., \& {Mushotzky}, R.~F. 1999, \apj, 526, 60, \dodoi{10.1086/307981}

\bibitem[{{Schmidt}(1965)}]{1965ApJ...141....1S}
{Schmidt}, M. 1965, \apj, 141, 1, \dodoi{10.1086/148085}

\bibitem[{{Shaik} {et~al.}(2024){Shaik}, {Meyer}, {Reddy}, {Laha}, \& {Georganopoulos}}]{2024arXiv240206218S}
{Shaik}, A., {Meyer}, E.~T., {Reddy}, K., {Laha}, S., \& {Georganopoulos}, M. 2024, arXiv e-prints, arXiv:2402.06218, \dodoi{10.48550/arXiv.2402.06218}

\bibitem[{{Simkin} {et~al.}(1999){Simkin}, {Sadler}, {Sault}, {Tingay}, \& {Callcut}}]{1999ApJS..123..447S}
{Simkin}, S.~M., {Sadler}, E.~M., {Sault}, R., {Tingay}, S.~J., \& {Callcut}, J. 1999, \apjs, 123, 447, \dodoi{10.1086/313243}

\bibitem[{{Singh} {et~al.}(1990){Singh}, {Rao}, \& {Vahia}}]{1990MNRAS.246..706S}
{Singh}, K.~P., {Rao}, A.~R., \& {Vahia}, M.~N. 1990, \mnras, 246, 706

\bibitem[{Stawarz {et~al.}(2007)Stawarz, Cheung, Harris, \& Ostrowski}]{Stawarz_2007}
Stawarz, {\L}., Cheung, C.~C., Harris, D.~E., \& Ostrowski, M. 2007, The Astrophysical Journal, 662, 213, \dodoi{10.1086/517966}

\bibitem[{{Strohmayer}(2017)}]{2017ApJ...838...72S}
{Strohmayer}, T.~E. 2017, \apj, 838, 72, \dodoi{10.3847/1538-4357/aa643d}

\bibitem[{{Sun} {et~al.}(2016){Sun}, {Yang}, {Mckinley}, \& {Aharonian}}]{2016A&A...595A..29S}
{Sun}, X.-n., {Yang}, R.-z., {Mckinley}, B., \& {Aharonian}, F. 2016, \aap, 595, A29, \dodoi{10.1051/0004-6361/201629069}

\bibitem[{{Sunada} {et~al.}(2022){Sunada}, {Morimoto}, {Tashiro}, {Terada}, {Katsuda}, {Sato}, {Tateishi}, \& {Sasaki}}]{2022PASJ...74..602S}
{Sunada}, Y., {Morimoto}, A., {Tashiro}, M.~S., {et~al.} 2022, \pasj, 74, 602, \dodoi{10.1093/pasj/psac022}

\bibitem[{{Tavecchio} {et~al.}(2005){Tavecchio}, {Cerutti}, {Maraschi}, {Sambruna}, {Gambill}, {Cheung}, \& {Urry}}]{2005ApJ...630..721T}
{Tavecchio}, F., {Cerutti}, R., {Maraschi}, L., {et~al.} 2005, \apj, 630, 721, \dodoi{10.1086/432371}

\bibitem[{{Tavecchio} \& {Ghisellini}(2008)}]{2008MNRAS.385L..98T}
{Tavecchio}, F., \& {Ghisellini}, G. 2008, \mnras, 385, L98, \dodoi{10.1111/j.1745-3933.2008.00441.x}

\bibitem[{{Tavecchio} {et~al.}(2000){Tavecchio}, {Maraschi}, {Sambruna}, \& {Urry}}]{2000ApJ...544L..23T}
{Tavecchio}, F., {Maraschi}, L., {Sambruna}, R.~M., \& {Urry}, C.~M. 2000, \apjl, 544, L23, \dodoi{10.1086/317292}

\bibitem[{{Thimmappa} {et~al.}(2020){Thimmappa}, {Stawarz}, {Marchenko}, {Balasubramaniam}, {Cheung}, \& {Siemiginowska}}]{2020ApJ...903..109T}
{Thimmappa}, R., {Stawarz}, {\L}., {Marchenko}, V., {et~al.} 2020, \apj, 903, 109, \dodoi{10.3847/1538-4357/abb605}

\bibitem[{{Thimmappa} {et~al.}(2022){Thimmappa}, {Stawarz}, {Neilsen}, {Ostrowski}, \& {Reville}}]{2022ApJ...941..204T}
{Thimmappa}, R., {Stawarz}, {\L}., {Neilsen}, J., {Ostrowski}, M., \& {Reville}, B. 2022, \apj, 941, 204, \dodoi{10.3847/1538-4357/aca472}

\bibitem[{{Thomson} {et~al.}(1995){Thomson}, {Crane}, \& {Mackay}}]{1995ApJ...446L..93T}
{Thomson}, R.~C., {Crane}, P., \& {Mackay}, C.~D. 1995, \apjl, 446, L93, \dodoi{10.1086/187938}

\bibitem[{{Tingay} {et~al.}(2008){Tingay}, {Lenc}, {Brunetti}, \& {Bondi}}]{2008AJ....136.2473T}
{Tingay}, S.~J., {Lenc}, E., {Brunetti}, G., \& {Bondi}, M. 2008, \aj, 136, 2473, \dodoi{10.1088/0004-6256/136/6/2473}

\bibitem[{{Tingay} {et~al.}(2000){Tingay}, {Jauncey}, {Reynolds}, {Tzioumis}, {McCulloch}, {Ellingsen}, {Costa}, {Lovell}, {Preston}, \& {Simkin}}]{2000AJ....119.1695T}
{Tingay}, S.~J., {Jauncey}, D.~L., {Reynolds}, J.~E., {et~al.} 2000, \aj, 119, 1695, \dodoi{10.1086/301283}

\bibitem[{{Wang} {et~al.}(2022){Wang}, {Bi}, {Cao}, {Vallania}, {Wu}, {Yan}, \& {Yuan}}]{2022ChPhC..46c0003W}
{Wang}, X.-Y., {Bi}, X.-J., {Cao}, Z., {et~al.} 2022, Chinese Physics C, 46, 030003, \dodoi{10.1088/1674-1137/ac3fa9}

\bibitem[{{Wang} {et~al.}(2020){Wang}, {Zhang}, {Sun}, \& {Liang}}]{2020ApJ...893...41W}
{Wang}, Z.-J., {Zhang}, J., {Sun}, X.-N., \& {Liang}, E.-W. 2020, \apj, 893, 41, \dodoi{10.3847/1538-4357/ab7d35}

\bibitem[{{Weisskopf}(2022)}]{2022HEAD...1930101W}
{Weisskopf}, M. 2022, in AAS/High Energy Astrophysics Division, Vol.~54, AAS/High Energy Astrophysics Division, 301.01

\bibitem[{{Wilson} {et~al.}(2001){Wilson}, {Young}, \& {Shopbell}}]{2001ApJ...547..740W}
{Wilson}, A.~S., {Young}, A.~J., \& {Shopbell}, P.~L. 2001, \apj, 547, 740, \dodoi{10.1086/318412}

\bibitem[{{Wood} {et~al.}(2017){Wood}, {Caputo}, {Charles}, {Di Mauro}, {Magill}, {Perkins}, \& {Fermi-LAT Collaboration}}]{Wood2017}
{Wood}, M., {Caputo}, R., {Charles}, E., {et~al.} 2017, in International Cosmic Ray Conference, Vol. 301, 35th International Cosmic Ray Conference (ICRC2017), 824, \dodoi{10.22323/1.301.0824}

\bibitem[{{Wright} {et~al.}(2010){Wright}, {Eisenhardt}, {Mainzer}, {Ressler}, {Cutri}, {Jarrett}, {Kirkpatrick}, {Padgett}, {McMillan}, {Skrutskie}, {Stanford}, {Cohen}, {Walker}, {Mather}, {Leisawitz}, {Gautier}, {McLean}, {Benford}, {Lonsdale}, {Blain}, {Mendez}, {Irace}, {Duval}, {Liu}, {Royer}, {Heinrichsen}, {Howard}, {Shannon}, {Kendall}, {Walsh}, {Larsen}, {Cardon}, {Schick}, {Schwalm}, {Abid}, {Fabinsky}, {Naes}, \& {Tsai}}]{2010AJ....140.1868W}
{Wright}, E.~L., {Eisenhardt}, P. R.~M., {Mainzer}, A.~K., {et~al.} 2010, \aj, 140, 1868, \dodoi{10.1088/0004-6256/140/6/1868}

\bibitem[{{Xie} {et~al.}(2024){Xie}, {Wong}, {La Monaca}, {Romani}, {Heyl}, {Kaaret}, {Di Marco}, {Bucciantini}, {Liu}, {Ng}, {Di Lalla}, {Weisskopf}, {Costa}, {Soffitta}, {Muleri}, {Bachetti}, {Pilia}, {Rankin}, {Fabiani}, {Agudo}, {Antonelli}, {Baldini}, {Baumgartner}, {Bellazzini}, {Bianchi}, {Bongiorno}, {Bonino}, {Brez}, {Capitanio}, {Castellano}, {Cavazzuti}, {Chen}, {Ciprini}, {De Rosa}, {Del Monte}, {Di Gesu}, {Donnarumma}, {Doroshenko}, {Dov{\v{c}}iak}, {Ehlert}, {Enoto}, {Evangelista}, {Ferrazzoli}, {Garcia}, {Gunji}, {Hayashida}, {Iwakiri}, {Jorstad}, {Karas}, {Kislat}, {Kitaguchi}, {Kolodziejczak}, {Krawczynski}, {Latronico}, {Liodakis}, {Maldera}, {Manfreda}, {Marin}, {Marinucci}, {Marscher}, {Marshall}, {Massaro}, {Matt}, {Mitsuishi}, {Mizuno}, {Negro}, {O'Dell}, {Omodei}, {Oppedisano}, {Papitto}, {Pavlov}, {Peirson}, {Perri}, {Pesce-Rollins}, {Petrucci}, {Possenti}, {Poutanen}, {Puccetti}, {Ramsey}, {Ratheesh}, {Roberts}, {Sgr{\`o}}, {Slane}, {Spandre}, {Swartz}, {Tamagawa}, {Tavecchio},
  {Taverna}, {Tawara}, {Tennant}, {Thomas}, {Tombesi}, {Trois}, {Tsygankov}, {Turolla}, {Vink}, {Wu}, {Zane}, {IXPE Collaboration}, {Wadiasingh}, {Ho}, {Harding}, {Gendreau}, \& {Arzoumanian}}]{2024ApJ...962...92X}
{Xie}, F., {Wong}, J., {La Monaca}, F., {et~al.} 2024, \apj, 962, 92, \dodoi{10.3847/1538-4357/ad17ba}

\bibitem[{{Xue} {et~al.}(2017){Xue}, {Zhang}, {Cui}, {Liang}, \& {Zhang}}]{2017RAA....17...90X}
{Xue}, Z.-W., {Zhang}, J., {Cui}, W., {Liang}, E.-W., \& {Zhang}, S.-N. 2017, Research in Astronomy and Astrophysics, 17, 090, \dodoi{10.1088/1674-4527/17/9/90}

\bibitem[{{Yu} {et~al.}(2024){Yu}, {Zhang}, {Gan}, {Hu}, {Wu}, \& {Zhang}}]{2024ApJ...965..163Y}
{Yu}, Y.-W., {Zhang}, H.-M., {Gan}, Y.-Y., {et~al.} 2024, \apj, 965, 163, \dodoi{10.3847/1538-4357/ad2e07}

\bibitem[{{Zhang} {et~al.}(2010){Zhang}, {Bai}, {Chen}, \& {Liang}}]{2010ApJ...710.1017Z}
{Zhang}, J., {Bai}, J.~M., {Chen}, L., \& {Liang}, E. 2010, \apj, 710, 1017, \dodoi{10.1088/0004-637X/710/2/1017}

\bibitem[{{Zhang} {et~al.}(2009){Zhang}, {Bai}, {Chen}, \& {Yang}}]{2009ApJ...701..423Z}
{Zhang}, J., {Bai}, J.~M., {Chen}, L., \& {Yang}, X. 2009, \apj, 701, 423, \dodoi{10.1088/0004-637X/701/1/423}

\bibitem[{{Zhang} {et~al.}(2018){Zhang}, {Du}, {Guo}, {Zhang}, {Chen}, {Liang}, \& {Zhang}}]{2018ApJ...858...27Z}
{Zhang}, J., {Du}, S.-s., {Guo}, S.-C., {et~al.} 2018, \apj, 858, 27, \dodoi{10.3847/1538-4357/aab9b2}

\bibitem[{{Zhang} {et~al.}(2012){Zhang}, {Liang}, {Zhang}, \& {Bai}}]{2012ApJ...752..157Z}
{Zhang}, J., {Liang}, E.-W., {Zhang}, S.-N., \& {Bai}, J.~M. 2012, \apj, 752, 157, \dodoi{10.1088/0004-637X/752/2/157}

\end{thebibliography}
\bibliographystyle{aasjournal}

\clearpage

\clearpage

\begin{table}[!htp]
    \centering
        \caption{Spectal Fitting Results of Fermi-LAT Observations for Pictor A}
        \label{table_LAT}
        \begin{tabular}{lccccccc}
            \hline
            \hline
            Time range & Energy range & Model & $\Gamma_{\rm \gamma,1}$ & $\Gamma_{\rm \gamma,2}$ & $E_b$ & Flux & TS\\
            (MJD) & (GeV) & & & & (GeV) & ($\times10^{-12} \rm erg~cm^{-2}~\rm{s}^{-1}$) & \\ 
            (1) & (2) & (3) & (4) & (5) & (6) & (7) & (8)\\ 
            \hline   Full time & 0.1--500 &PL & $3.05\pm0.08$ & --{} & -- & $8.98\pm0.73$ & 669.55\\
    (54682--60501) & 0.1--500 & BPL & $3.25\pm0.15$ & $1.81\pm0.07$ & $2.46\pm0.09$ & $10.61\pm0.82$ & 773.10\\
       & 0.1--2.46 &PL & $3.12\pm0.06$ & --{} & -- & $7.68\pm0.82$ & 541.21\\
       & 2.46--500& PL & $1.97\pm0.18$ & -- & -- & $1.72\pm0.56$ & 64.06\\
    F1 (54952--55762) & 0.1--500&PL & $2.59\pm0.21$ & --{} & -- & $5.91\pm1.56$ & 50.95\\
    F2 (55762--58732) & 0.1--500&BPL & $3.13\pm0.08$ & $1.92\pm0.16$ & $2.15\pm0.27$ & $8.47\pm1.05$ & 241.27\\
    F3 (58822--59272) & 0.1--500&PL & $3.30\pm0.17$ & --{} & -- & $16.34\pm2.87$ & 156.39\\ \hline
        \end{tabular}
\tablenotetext{}{Column (1): Time range for spectral fitting; Column (2): Energy range for spectral fitting; Column (3): Spectral function utilized in fitting; Columns (4) and (5): Photon spectral indices for PL or BPL functions; Column (6): Break energy for the BPL model; Column (7): Average flux in the corresponding energy band and time period; Column (8): TS value in the corresponding energy band and time period. }
\end{table}

\begin{table}[htbp]
\centering
  \caption{Information for Each Time Bin of the Light Curve in Energy Bands 0.1--2.46 GeV and 2.46--500 GeV}
  \label{LC_detail}
\begin{tabular}{ccccccc}
\hline
\hline
Bin number & Time\_start& Time\_end& bin size & Flux & Flux error & TS \\ & (MJD) & (MJD) & (days) & ($\rm erg~cm^{-2}~\rm{s}^{-1}$) & ($\rm erg~cm^{-2}~\rm{s}^{-1}$) & \\
(1) & (2) & (3) & (4) & (5) & (6) & (7) \\
\hline
\multicolumn{7}{c}{ Energy band within 0.1--2.46 GeV}\\
\hline
1 & 54682.66 & 54772.66 & 90 & $8.59 \times 10^{-12}$ & $3.02 \times 10^{-12}$ & 9.96 \\
2 & 54772.66 & 54862.66 & 90 & $1.29 \times 10^{-11}$ & $3.57 \times 10^{-12}$ & 22.20 \\
3 & 54862.66 & 55042.66 & 180 & $8.48 \times 10^{-12}$ & $2.30 \times 10^{-12}$ & 20.29 \\
4 & 55042.66 & 55132.66 & 90 & $1.49 \times 10^{-11}$ & $3.23 \times 10^{-12}$ & 30.27 \\
5 & 55132.66 & 55222.66 & 90 & $9.92 \times 10^{-12}$ & $3.00 \times 10^{-12}$ & 16.31 \\
6 & 55222.66 & 55312.66 & 90 & $2.16 \times 10^{-11}$ & $3.40 \times 10^{-12}$ & 55.85 \\
7 & 55312.66 & 55402.66 & 90 & $1.51 \times 10^{-11}$ & $3.35 \times 10^{-12}$ & 28.32 \\
8 & 55402.66 & 55492.66 & 90 & $2.75 \times 10^{-11}$ & $3.43 \times 10^{-12}$ & 100.04 \\
9 & 55492.66 & 55582.66 & 90 & $2.20 \times 10^{-11}$ & $3.29 \times 10^{-12}$ & 62.75 \\
10 & 55582.66 & 55672.66 & 90 & $1.33 \times 10^{-11}$ & $3.45 \times 10^{-12}$ & 23.37 \\
11 & 55672.66 & 55762.66 & 90 & $1.76 \times 10^{-11}$ & $3.02 \times 10^{-12}$ & 50.97 \\
12 & 55762.66 & 55942.66 & 180 & $7.10 \times 10^{-12}$ & $2.17 \times 10^{-12}$ & 15.72 \\
13 & 55942.66 & 56122.66 & 180 & $6.68 \times 10^{-12}$ & $2.07 \times 10^{-12}$ & 15.88 \\
14 & 56122.66 & 56212.66 & 90 & $8.17 \times 10^{-12}$ & $3.14 \times 10^{-12}$ & 9.92 \\
15 & 56212.66 & 56392.66 & 180 & $7.74 \times 10^{-12}$ & $2.21 \times 10^{-12}$ & 19.80 \\
16 & 56392.66 & 57742.66 & 1350 & $2.16 \times 10^{-12}$ & $7.59 \times 10^{-13}$ & 12.49 \\
17 & 57742.66 & 58732.66 & 990 & $2.30 \times 10^{-12}$ & $1.09 \times 10^{-12}$ & 9.37 \\
18 & 58732.66 & 58822.66 & 90 & $8.56 \times 10^{-12}$ & $3.11 \times 10^{-12}$ & 11.73 \\
19 & 58822.66 & 58912.66 & 90 & $1.46 \times 10^{-11}$ & $3.20 \times 10^{-12}$ & 33.36 \\
20 & 58912.66 & 59002.66 & 90 & $2.11 \times 10^{-11}$ & $4.12 \times 10^{-12}$ & 44.33 \\
21 & 59002.66 & 59092.66 & 90 & $2.42 \times 10^{-11}$ & $4.32 \times 10^{-12}$ & 61.51 \\
22 & 59092.66 & 59182.66 & 90 & $1.79 \times 10^{-11}$ & $3.77 \times 10^{-12}$ & 37.16 \\
23 & 59182.66 & 59272.66 & 90 & $1.18 \times 10^{-11}$ & $3.96 \times 10^{-12}$ & 17.08 \\
24 & 59272.66 & 59452.66 & 180 & $7.80 \times 10^{-12}$ & $3.32 \times 10^{-12}$ & 13.54 \\
25 & 59452.66 & 59632.66 & 180 & $1.28 \times 10^{-11}$ & $2.25 \times 10^{-12}$ & 54.31 \\
26 & 59632.66 & 59812.66 & 180 & $9.78 \times 10^{-12}$ & $2.17 \times 10^{-12}$ & 33.65 \\
27 & 59812.66 & 60172.66 & 360 & $4.96 \times 10^{-12}$ & $1.31 \times 10^{-12}$ & 22.08 \\
28 & 60172.66 & 60442.66 & 270 & $4.35 \times 10^{-12}$ & $1.57 \times 10^{-12}$ & 11.46 \\
\hline
\multicolumn{7}{c}{Energy band within 2.46--500 GeV}\\
\hline
1 & 54682.66 & 55312.66 & 630 & $2.40 \times 10^{-12}$ & $1.33 \times 10^{-12}$ & 9.49 \\
2 & 55312.66 & 56482.66 & 1170 & $1.68 \times 10^{-12}$ & $8.14 \times 10^{-13}$ & 10.41 \\
3 & 56482.66 & 57112.66 & 630 & $2.37 \times 10^{-12}$ & $1.30 \times 10^{-12}$ & 9.79 \\
4 & 57112.66 & 58552.66 & 1440 & $1.74 \times 10^{-12}$ & $7.77 \times 10^{-13}$ & 9.94 \\
5 & 58552.66 & 59812.66 & 1260 & $1.57 \times 10^{-12}$ & $7.77 \times 10^{-13}$ & 9.81 \\
6 & 59812.66 & 60082.66 & 270 & $5.04 \times 10^{-12}$ & $2.50 \times 10^{-12}$ & 10.82 \\
7 & 60082.66 & 60172.66 & 90 & $8.90 \times 10^{-12}$ & $5.20 \times 10^{-12}$ & 11.17 \\
\hline
\end{tabular}
\tablenotetext{}{Column (1): Serial number of this time bin; Column (2): Time-bin start time; Column (3): Time-bin end time; Column (4): Duration of the corresponding time bin; Column (5): Average flux in corresponding time bin; Column (6): Corresponding error of the average flux; Column (7): TS value in the corresponding time bin.}
\end{table}

\begin{table}
    \begin{center}
    \caption{Results of the Spectropolarimetric Analysis in the 2--8 keV Band for Both the Nucleus and WHS}
    \label{table_IXPE}
    \begin{tabular}{ccccc}
    \hline
    \hline
    Component & Parameter & Unit & \multicolumn{2}{c}{Value} \\
    \cmidrule(r){4-5}
    & & & Nucleus & WHS \\
    \hline
    \multicolumn{5}{c}{Model = \texttt{CONSTANT}$\times$\texttt{TBABS}$\times$\texttt{POWERLAW}} \\
    \hline
    \texttt{CONSTANT} & $C_{\rm DU1-Obs.1}$ & & 1.0 (fixed) & 1.0 (fixed) \\
    & $C_{\rm DU2-Obs.1}$ & & $1.048\pm0.009$ & $1.039^{+0.094}_{-0.087}$ \\
    & $C_{\rm DU3-Obs.1}$ & & $1.021\pm0.009$ & $0.983^{+0.091}_{-0.084}$ \\
    & $C_{\rm DU1-Obs.2}$ & & $0.947\pm0.008$ & $1.129^{+0.097}_{-0.089}$ \\
    & $C_{\rm DU2-Obs.2}$ & & $1.029\pm0.009$ & $1.202^{+0.104}_{-0.096}$ \\
    & $C_{\rm DU3-Obs.2}$ & & $1.017\pm0.009$ & $1.160^{+0.102}_{-0.094}$ \\
    \texttt{TBABS} & $N_{\rm H}$ & $10^{20}$ cm$^{-2}$ & 3.62 (fixed) & 3.57 (fixed) \\
    \texttt{POWERLAW} & $\Gamma_{\rm X}$ & & $1.84\pm0.01$ & $2.07\pm0.11$ \\
    & $N_{0}$ & $10^{-3}$ ph keV$^{-1}$ cm$^{-2}$ s$^{-1}$ & $4.92\pm0.06$ & $0.11\pm0.01$ \\
    Fit Statistic & $\chi^{2}$/dof & & 233/167 & 210/167 \\
    \hline
    \multicolumn{5}{c}{Model = \texttt{CONSTANT}$\times$\texttt{TBABS}$\times$\texttt{POLCONST}$\times$\texttt{POWERLAW}} \\
    \hline
    \texttt{POLCONST} & $\Pi_{\rm X}$ & \% & $<6.6$ & $<56.4$ \\
    & $\psi_{\rm X}$ & $\degr$ & \nodata & \nodata \\
    Fit Statistic & $\chi^{2}$/dof & & 721/520 & 714/520 \\
    \hline
    \end{tabular}
    \end{center}
\end{table}

\begin{table*}
\begin{center}
\caption{Parameters of SED Fitting for both the Nucleus and WHS}
\label{table_SED}
{\scriptsize
\begin{tabular}{lcccr} % four columns, alignment for each
\hline
\hline
\multirow{2}{*}{Parameter} & \multirow{2}{*}{Symbol}  &\multirow{2}{*}{Nucleus} &\multicolumn{2}{c}{WHS} \\
&	  &	 	& Diffuse & Substructure \\
%Parameter   &Symbol     & Core \\
\hline
Electron density parameter     & $N_0$ [cm$^{-3}$]  & $6.5\times10^{5}$ & $2.6\times10^{-3}$ & $49 $  \\
Spectral index below break   & $p^{\star}_1$   & $2.2$ & $2.4$ & $2.52$  \\
Spectral index above break    & $p^{\star}_2$  & $4.42$ & $4.1$ & ~  \\
Minimum electron Lorentz factor  &$\gamma^{\star}_{\rm min}$ & $1$ & $1$ & $1$  \\
Break Lorentz factor  &$\gamma_{\rm b}$ & $1.6\times10^{3}$ & $ 2.2\times10^{5}$ & ~\\
Maximum electron Lorentz factor        &$\gamma_{\rm max}$            & $1.6\times10^{4}$ & $ 2.2\times10^{6}$ & $2.1\times 10^{8}$  \\
Radius of the blob    & $R^{\star}$ [cm/pc]  & $10^{16}$ cm & 500 pc & 30 pc \\		
Bulk Lorentz factor  & $\Gamma^{\star}$ & $5$ & $1 $ & $ 1 $   \\
Doppler boosting factor & $\delta^{\star}$ & $1.3 $ & $1$ & $1 $   \\
Magnetic Field   & $B$ [G] & $7 $ & $3.6\times10^{-4}$ & $3.5\times10^{-5}$  \\
Equipartition ratio &U$_e/$U$_B$ & $1.08   $ & $1 $ & $1.6\times10^{6}$  \\
\hline                                                             					
\end{tabular}
}
\end{center}
\tablenotetext{\star}{The parameters remain fixed during SED modeling.}
\end{table*}

\end{document}